\pdfoutput=1
\documentclass[11pt]{article}
\usepackage{cancel}
\usepackage{graphicx}
\usepackage[colorlinks=true, pdfstartview=FitV, linkcolor=blue, citecolor=blue,urlcolor=blue]{hyperref}

\usepackage{framed,color}
\definecolor{shadecolor}{rgb}{0.90, 0.90, 0.90}

\usepackage{amsbsy}
\usepackage{amssymb,amsmath}
 \usepackage{amsfonts}
\usepackage{graphicx}

\usepackage{cancel}
\usepackage{mathtools}
\usepackage{amsmath}
\usepackage{amsthm}
\usepackage{amsfonts}
\usepackage{amssymb}
\usepackage{mathrsfs}
\usepackage{graphicx}
\usepackage[T1]{fontenc}
\usepackage[OT2,T1]{fontenc}
\usepackage[latin1]{inputenc}
 \usepackage{lmodern}
\usepackage[all,cmtip]{xy}

\usepackage{bbm}
\usepackage[vcentermath]{youngtab}
\usepackage{calc}

\usepackage{setspace}
\usepackage{epigraph}
\usepackage{mathdots}
\usepackage{mathabx}
\usepackage{braket}
\usepackage{longtable}
\usepackage{float}

\def\eqref#1{(\ref{#1})}
\newtheorem{theorem}{Theorem}[section]
\newtheorem{example}{Example}[section]
\newtheorem{exercise}{Exercise}[section]

\newtheorem{lemma}{Lemma}[section]
\newtheorem{remark}{Remark}[section]

\newtheorem{proposition}{Proposition}[section]
\newtheorem{corollary}{Corollary}[section]
\newtheorem{definition}{Definition}[section]

\def\bre{\begin{remark}}
\def\ere{\end{remark}}
\def\bth{\begin{theorem}}
\def\eth{\end{theorem}}
\def\bcr{\begin{corollary}}
\def\ecr{\end{corollary}}
\def\bex{\begin{example}\small}
\def\eex{\end{example}}
\def\bexr{\begin{exercise}\small}
\def\eexr{\end{exercise}}
\def\ble{\begin{lemma}}
\def\ele{\end{lemma}}

\def\bde{\begin{definition}}
\def\ede{\end{definition}}
\def\bpr{\begin{proposition}}
\def\epr{\end{proposition}}
\def\be{\begin{equation}}
\def\ee{\end{equation}}
\def\bea{\begin{eqnarray}}
\def\eea{\end{eqnarray}}
\def\beas{\begin{eqnarray*}}
\def\eeas{\end{eqnarray*}}
\newlanguage\fakelanguage
\newcommand\cyr{\fontencoding{OT2}\fontfamily{wncyr}\selectfont
   \language\fakelanguage}
\DeclareTextFontCommand{\textcyr}{\cyr}

\numberwithin{equation}{section}

\numberwithin{theorem}{section}

\numberwithin{proposition}{section}

\numberwithin{definition}{section}

\numberwithin{remark}{section}

\numberwithin{lemma}{section}

\numberwithin{corollary}{section}

\date{}
\begin{document}
\baselineskip=14pt

\vspace{0.2cm}
\begin{center}
\begin{Large}
\fontsize{17pt}{27pt}
\selectfont

\textbf{Isomonodromic deformations along a stratum of the coalescence locus}
\end{Large}
\\
\bigskip
\begin{large} {Davide  Guzzetti}\end{large}
\\{ SISSA,  Via Bonomea, 265,  34136 Trieste $\&$  INFN, Sezione di Trieste, Italy.}
\\{ E-mail: guzzetti@sissa.it}
\\{ORCID ID: 0000-0002-6103-6563}
\bigskip
\end{center}
\begin{small}
{\bf Abstract:} We consider deformations of a differential system with Poincar\'e rank 1 at infinity and Fuchsian singularity at zero  along a stratum of a coalescence locus. We give necessary and sufficient conditions for the deformation to be strongly isomonodromic,  both as an explicit Pfaffian system (integrable deformation) and as a non linear system of PDEs on the residue matrix $A$ at the Fuchsian singularity.  This construction is complementary to that of \cite{CDG}. For the specific system here considered, the results generalize  those of \cite{JMU}, by giving up the generic conditions,  and  those of \cite{BM}, by giving up the Lidskii generic assumption.  The importance of the case here considered originates form its applications in the study of strata of Dubrovin-Frobenius manifolds and $F$-manifolds. 
\vspace{0.7cm}

\end{small}
\vskip 15pt

\tableofcontents

\vskip 1 cm 
\noindent
{\bf Notation.} For an $n\times n$ matrix $A_k$, we denote the matrix entries by either $(A_k)_{ij}$ or $A^{(k)}_{ij}$, where $i,j\in\{1,...,n\}$.   We can partition $A_k$ into $s^2$  blocks of dimension $p_a\times p_b$, where  $a,b=1,...,s$ and $p_1+...+p_s=n$. The block labelled by $a,b$, of dimension $p_a\times p_b$  will be denoted by $A^{(k)}_{[a,b]}$. 

\section{Introduction}

In the work \cite{CDG}, and in the related \cite{Eretico1,Guzz-SIGMA,Guzz-Spring18,Guzz-Degru20,guz2021}, we have 
studied  an $n\times n$ matrix differential system of the shape \eqref{27marzo2021-2} below, with an irregular singularity
 at $z=\infty$ and a Fuchsian one at $z=0$, with leading term at $\infty$ given by a diagonal matrix $\Lambda= \hbox{\rm diag}(\lambda_1,...,\lambda_n)$, whose eigenvalues $\lambda=(\lambda_1,...,\lambda_n)$ vary in a polydisc of $\mathbb{C}^n$. The polydisc contains 
 a {\it coalescence locus}, where some eigenvalues merge, namely $\lambda_j-\lambda_k\to 0$ for some $1\leq j\neq k\leq n$.  For this system,
  we have proved that a monodromy preserving deformation theory can be well defined (in an analytic way) with constant monodromy data on the whole polydisc, including the coalescence locus. This result, which generalizes the theory of Jimbo, Miwa and Ueno \cite{JMU}, is possible if the vanishing conditions $A_{jk}(\lambda)\to 0$ hold when $\lambda_j-\lambda_k\to 0$.

In this paper, we consider an $n$-dimensional differential system
\be
\label{27marzo2021-2}
\frac{dY}{dz}=\left(\Lambda +\frac{A(\lambda)}{z}\right) Y, \quad\quad \quad \lambda=(\lambda_1,\cdots,\lambda_s)\in \mathbb{D}\subset \mathbb{C}^s,
\ee
where $\mathbb{D}$ is a polydisc and 
$$ 
\Lambda=\lambda_1I_{p_1}\oplus \dots\oplus \lambda_s I_{p_s}:=\hbox{\rm diag}(\underbrace{\lambda_1,...,\lambda_1}_{p_1 },\underbrace{\lambda_2,...,\lambda_2}_{p_2},\cdots, \underbrace{\lambda_s,...,\lambda_s}_{p_s })
$$
$$ I_{p_j}=\hbox{ $p_j$-dimensional identity matrix},\quad p_1+\cdots+p_s=n.
$$ 
We can think of $\lambda$ as the parameters varying {\it within a stratum} of a coalescence locus, specified by $p_1,...,p_s$.  We would like to establish the full isomonodromy deformation theory within this stratum. 

 The deformation considered here is complementary to that of \cite{CDG}, because it occurs within  the prefixed stratum, while in \cite{CDG} the deformation takes place in a domain containing the coalescence set,  the latter being   included in the range of deformation under specific vanishing conditions on $A$.  The problem of the present paper  is therefore  different from \cite{CDG}: here  $A$ will be  any matrix and we do not suppose that the entries of $A$ corresponding to   equal eigenvalues of $\Lambda$  are zero. The deformation theory that we will develop cannot be deduced  either from \cite{CDG} or \cite{JMU}. 
   This theory  is realized in an important geometric setting, namely  at the nilpotent locus   of a Dubrovin-Frobenius manifold \cite{Dub1}, called {\it caustic} \cite{Her}. This application will be discussed in   Section \ref{3agosto2022-1}, for the type of caustics geometrically described in \cite{Reyes}.

\vskip 0.2 cm 
  In the sequel,  it will be convenient to partition   $A$  into blocks $A_{[i,j]}$, $i,j=1,...,s$, of dimension $p_i\times p_j$, inherited from $\Lambda$. We will work in the following analytic setting.

\vskip 0.2 cm 
\noindent
{\bf Assumption 1.}
\begin{itemize}
 \item {\it
 The polydisc $\mathbb{D}$ is sufficiently small so that, as $\lambda$ varies in $\mathbb{D}$, the Stokes rays defined in \eqref{27marzo2021-1} below do not cross the half-lines $\arg z=\tau+k\pi$, $k\in\mathbb{Z}$, where $\tau\in\mathbb{R}$ is  fixed, and called an {\bf admissible direction}.}

\item {\it 
$A(\lambda)$ is holomorphic in $\mathbb{D}$.}
\end{itemize}

The Stokes rays in the assumption are the rays in the universal covering $\mathcal{R}$ of $\overline{\mathbb{C}}\backslash\{0,\infty\}$ defined by 
\be 
\label{27marzo2021-1}
\Re((\lambda_i-\lambda_j)z)=0,\quad\quad \Im((\lambda_i-\lambda_j)z)<0,\quad\quad z\in\mathcal{R}.
\ee

In this paper, we establish the necessary and sufficient conditions for \eqref{27marzo2021-2} to be strongly isomonodromic on the polydisc $\mathbb{D}$. The notion of ``strong isomonodromy''  is implicit in the seminal paper \cite{JMU}, meaning that  {\it all essential monodromy data} (monodromy exponents, connection matrices, Stokes matrices, see Definition \ref{31marzo2021-1}) are 
 independent of $\lambda$. The adjective ``strong''  was introduced in \cite{Guzz-SIGMA}, to point out that a system may just be ``weakly'' isomonodromic,  namely with constant monodromy matrices, but with possibly non-constant essential monodromy data.  

  In the isomonodromy theory of \cite{JMU}, several assumptions are made to assure that the differential system is generic (the deformation is called admissible). In case of \eqref{27marzo2021-2}, the eigenvalues of $A$ are not allowed to differ by integers (so $A$ is in particular diagonalizable), and $\Lambda$ has pairwise distinct eigenvalues.  
  
  In the  paper \cite{BM}, the isomonodromy deformation theory has been extended to  rational connection with both Fuchsian and irregular singularities of any Poincar\'e rank, without several of the assumtions of \cite{JMU}. The residue matrices at the Fuchsian singularities  are not subject to restrictions, while  the leading matrix at an irregular singularity can have  any Jordan form, but  with a {\it prefixed} Jordan type $\lambda_1^{n_1}\lambda_2^{n_2}\dots\lambda_K^{n_K}$ (in a notation due to Arnol'd \cite{Arnold}). For example, in case of \eqref{27marzo2021-2}, the Jordan type is prefixed to be 
  \be
  \label{27agosto2021-1}
  \underbrace{\lambda_1\dots\lambda_1}_{p_1} \underbrace{\lambda_2\dots\lambda_2}_{p_2}\dots\underbrace{\lambda_s\dots\lambda_s}_{p_s}.
  \ee
 Besides the prefixed Jordan type, another important assumption of \cite{BM} is that  the next sub-leading matrix (in our case $A$) at an irregular singularity must be {\it Lidskii generic}, according to definition 2.1 in \cite{BM}. In our case, this means that each diagonal block of $A$ (with block partition inherited from $\Lambda$) must have distinct and nonzero eigenvalues.  Theorem 5.3 of \cite{BM} states that the deformation is isomonodromic (preserving a set of monodromy data, which include the Stokes matrices) if and only if a class of fundamental matrix solutions satisfy a certain  Pfaffian system. This result  generalizes theorem 3.1 of  \cite{JMU}. Moreover, \cite{BM} studies the generalization of the isomonodromic $\tau$-function. 
 
 \bre
 \label{24nov2021-1}
 {\rm 
 Given a  differential system $\frac{dY}{dz}=\mathfrak{A}(z,\lambda)Y$  such that the deformation $\lambda$ does not satisfying some admissibility  conditions of \cite{JMU},  by {\it generalization} of \cite{JMU} we mean: find 
 necessary and sufficient conditions  ensuring  that {\it all essential monodromy data} (strong isomonodromy) of  the differential system   are constant. These conditions are of the type: constant data  if and only if all the canonical solutions satisfy a Pfaffian system  $dY=\omega Y$,  with a very specific $\omega$ (such as \eqref{18novembre2021-1} below); or if and only if the coefficients in $\mathfrak{A}(z,\lambda)$  satisfy certain non linear PDEs, called isomonodromy deformation equations (like the equations  \eqref{21agosto2021-1} below).
 }
 \ere

  As mentioned in the beginning, an extension of the  isomonodromy deformation theory  has been achieved in \cite{CDG}  for a system such as \eqref{27marzo2021-2} when  the Jordan type of the leading matrix $\Lambda$ changes within a polydisc of $\mathbb{C}^n$. In  \cite{CDG}, $A$ is any, while 
  $$\Lambda=\hbox{\rm diag}(\lambda_1,...,\lambda_n)
  $$ has $n$ eigenvalues  $\lambda=(\lambda_1,...,\lambda_n)$ varying in a polydisc  containing a  {\it coalescence locus}   where   $\lambda_j=\lambda_k$ for some $j\neq k$. The extension  of  \cite{CDG}  can be done under  the condition  that the entries $A_{jk}(\lambda)\to 0$ for $\lambda_j-\lambda_k\to 0$.  
    
    In this paper,  the deformation parameters  $\lambda=(\lambda_1,...,\lambda_s)$  are assumed to vary inside a stratum of the coalescence locus, namely $\Lambda$ has   $s< n$  eigenvalues varying in the polydisc $\mathbb{D}$ specified before.  In this sense, the Jordan type \eqref{27agosto2021-1}  of $\Lambda$ is fixed, as in \cite{BM}. Nevertheless, we  drop any assumption on $A$, and we do not assume Lidskii generic conditions.

    We mention that in \cite{BTL} isomondormy deformations are defined for a system of type \eqref{27marzo2021-2}, with coefficients in the Lie algebra of an arbitrary complex algebraic group $G$.   In the specific case we consider here the coefficients are  $n\times n$ complex matrices, so that $G=GL_n(\mathbb{C})$. In this case, the assumptions of  \cite{BTL} require that $\Lambda$ is diagonal\footnote{\cite{BTL}  requires that $\Lambda$ is diagonalizable, so one can work in the base where it is diagonal.} with  a prefixed Jordan type \eqref{27agosto2021-1}, invariant by the deformation  $\lambda$, and  the corresponding diagonal blocks of $A$ are zero.  It is to be noticed that the  assumption $A_{[k,k]}=0$, $\forall k=1,...,s$ implies  that  theorem 3.3. and remark 3.3 of  \cite{BTL} for $G=GL_n(\mathbb{C})$ are immediately deducible from the main theorem of  \cite{CDG}, starting from  
    $\Lambda=\hbox{\rm diag}(\tilde{\lambda}_1,...,\tilde{\lambda}_n)$ and considering the coalescence 
    $$(\tilde{\lambda}_1,...,\tilde{\lambda}_n) \longmapsto (\underbrace{\lambda_1\dots\lambda_1}_{p_1} \underbrace{\lambda_2\dots\lambda_2}_{p_2}\dots\underbrace{\lambda_s\dots\lambda_s}_{p_s})
    .
    $$
   In the present paper, no assumptions on the diagonal blocks $A_{[k,k]}$ will be made, so that our results are not deducible from or 
     reducible to  theorem 3.3. and remark 3.3 of \cite{BTL} (which, as said, are obtainable from the results of  \cite{CDG}, which are 
     complementary to the present paper). It is also to be mentioned that the notion of isomonodromy  in definition 3.2 of \cite{BTL}  
     requires that only the Stokes matrices are constant, while here we require constancy of a more stringent set of monodromy data, 
     including the monodromy exponents and the central connection matrix. 
   
 \vskip 0.2 cm 
   Isomonodromic deformation equations preserving  $G$-valued Stokes matrices  were first defined in  \cite{Boalch0} for meromorphic connections on principal $G$-bundles, with $G$ a complex reductive group. The leading term at an irregular singularity is assumed to be regular semisimple in the Lie algebras $\mathfrak{g}$. In case $G=GL(n,\mathbb{C})$ and $\mathfrak{g}=Mat(n,\mathbb{C})$,  this means that its  Jordan form  has a single Jordan block for each eigenvalue, and in particular this implies  pairwise distinct eigenvalues in the diagonalizable case.    A generalization of this assumption was then given in \cite{Boalch4}, where the Jordan type is fixed (no further coalescences allowed).

   \subsubsection*{Results}
   
   Our goal is a the generalization of \cite{JMU}, in the sense of Remark \ref{24nov2021-1},  for system \eqref{27marzo2021-2}. 
   
   \vskip 0.2 cm 
     The main results of the paper are Theorems \ref{30marzo2021-2} and \ref{20agosto2021-3}. Preliminarily to them,  in Theorem \ref{27marzo2021-5} some results in the weakly isomonodromic case are given: system \eqref{27marzo2021-2} is weakly isomonodromic with an isomonodromic  fundamental matrix solution in Levelt form at $z=0$  if and only if the latter satisfies a Pfaffian system whose $\lambda$-components are  holomorphic  in $\mathbb{C}\times \mathbb{D}$. In this case, the monodromy exponents at $z=0$ are constant. This fact is mainly based on \cite{YT}.

       Theorem \ref{30marzo2021-2} states that system \eqref{27marzo2021-2}  is strongly isomonodromic if and only if a fundamental 
     matrix solution at $z=0$ in Levelt form and  the canonical solutions at $z=\infty$ (defined in the paper) all satisfy the 
     integrable Pfaffian system 
          \be
          \label{18novembre2021-1}
     dY=\left[\Bigl(\Lambda +\frac{A}{z}\Bigr)dz+\sum_{j=1}^s \Bigl(z E_{p_j} +  \widetilde{\omega}_j(\lambda)\Bigr) d\lambda_j \right]Y,
    \ee
     with
     $$ \widetilde{\omega}_j(\lambda)= \omega_j(\lambda)+\frac{\partial \mathcal{T}(\lambda)}{\partial \lambda_j}\cdot  \mathcal{T}(\lambda)^{-1},
     $$
    where $E_{p_j}:=\partial\Lambda/\partial \lambda_j$ is  the matrix with blocks $E^{(p_j)}_{[a,b]}=\delta_{aj}\delta_{bj} I_{p_j}$, for $a,b,j=1,...,s$ (all entries are  zero, except for diagonal block $ I_{p_j}$),   the matrices $ \omega_j(\lambda)$ are holomorphic in $\mathbb{C}\times \mathbb{D}$ and  univocally given in formula \eqref{31marzo2021-6}, while the block-diagonal matrix $\mathcal{T}=\mathcal{T}_1\oplus \cdots \oplus \mathcal{T}_s$ is holomorphic  invertible in $\mathbb{D}$ and reduces to Jordan form the block-diagonal part of $A(\lambda)$,  namely 
   \be
   \label{4maggio2022-1}
    \mathcal{T}_k(\lambda)^{-1} A_{[k,k]}(\lambda)~\mathcal{T}_k(\lambda)=J_k.
   \ee   
    The second part of Theorem \ref{30marzo2021-2} also says that in the strong isomonodromic case, $A$ satisfies the  non-linear system 
    \be
    \label{21agosto2021-1}
     dA=\bigl[ \sum_{j=1}^n \widetilde{\omega}_j(\lambda)d\lambda_j,A\bigr].
     \ee
      The above    \eqref{21agosto2021-1} predicts that the block-diagonal part 
       $$ 
       A_{[1,1]}\oplus \cdots \oplus A_{[s,s]} 
       $$
       is constant. 
       In particular, it has a constant Jordan form $J_1\oplus \cdots \oplus J_s$. 
       This fact is not immediately obvious and will be proved in the paper.
        It   implies that isomonodromy deformations with constant  $\mathcal{T}=\mathcal{T}_0$ are always allowed, and in such case  
    $$
    \widetilde{\omega}_j(\lambda)= \omega_j(\lambda).
    $$ 
    All the other possible isomonodromic  deformations are obtained with
     $$\mathcal{T}(\lambda)=\mathcal{T}_0 \mathfrak{B}(\lambda),
     $$
      for any $ \mathfrak{B}(\lambda)= \mathfrak{B}_1(\lambda)\oplus \cdots \oplus  \mathfrak{B}_s(\lambda)$ satisfying  
      $$ 
      [\mathfrak{B}_k(\lambda),J_k]=0,\quad k=1,...,s.
      $$      
Therefore, for a given $A_0$ at $\lambda=\lambda_0\in \mathbb{D}$, let $\mathcal{T}_0$ be the block-diagonal matrix such that 
$$\mathcal{T}_0^{-1} \cdot \bigl(A^{(0)}_{[1,1]}\oplus \cdots \oplus A^{(0)}_{[s,s]}\bigr) \cdot \mathcal{T}_0=J_1\oplus \cdots \oplus J_s
.
$$ 
Then, there are several possible strong isomonodromy deformations $A(\lambda)$, having the same constant block-diagonal part $A^{(0)}_{[1,1]}\oplus \cdots \oplus A^{(0)}_{[s,s]}$, but different off-diagonal blocks $A_{[i,j]}(\lambda)$, $1\leq i\neq j \leq s$,  obeying  equation \eqref{21agosto2021-1}  with  the different  possible choices of $\mathcal{T}(\lambda)= \mathcal{T}_0 \mathfrak{B}(\lambda)$. These different deformations are related by a $\lambda$-dependent gauge transformation, as will be explained after Theorem \ref{30marzo2021-2}. 

In the 3-dimensional case, when the only non-trivial case is  $\Lambda=\hbox{\rm diag}(\lambda_1,\lambda_2,\lambda_2)$ (up to permutations), the  above freedom in the coefficients $\widetilde{\omega}_j$ implies the existence of  the particular isomonodromy deformation with constant $A$ and non constant
$\mathcal{T}(\lambda)$, and conversely of the deformation with non constant $A(\lambda)$ and constant $\mathcal{T}$. Since a gauge transformation can reduce to  constant $A$,  the 3-dimensional case is rigid. See Section \ref{1settembre2021-1} for details. The first non-trivial case occurs for $n=4$ and $s=3$. Already in simplified situations, we will show in Section \ref{20settembre2022-9} that the isomonodromic problem is at least as transcendental as a Painlev\'e equation.

\bre[{\bf Important Remark}]
\label{10agosto2022-1}
{\rm
 That strong isomonodromy deformations of \eqref{27marzo2021-2} exist with constant $\mathcal{T}$, and that they can always be reached by a gauge transformation,  does not at all mean that these deformations are trivial. Indeed, even in case $\mathcal{T}$ is constant, $A=A(\lambda)$ is in general a {\it highly non trivial} and {\it highly transcendental} matrix function, which must satisfy the extremely difficult system \eqref{21agosto2021-1} with the $\widetilde{\omega}_j(\lambda)=\omega_j(\lambda)$. Only in case $n=3$ (and $s=1,2$)   one can reduce to the deformations with constant $A$, but  starting form  $n\geq 4$,  solutions to the deformation equations \eqref{21agosto2021-1} are transcendental. See Section \ref{20settembre2022-9} for a 4-dimensional  example}. 
\ere
    Theorem \ref{20agosto2021-3}  is  the converse to the second part of Theorem \ref{30marzo2021-2}. It says that system \eqref{27marzo2021-2} is strongly isomonodromic if $A$ is not partially resonant (Definition \ref{20agosto2021-7}) and  satisfies the Frobenius integrable  system \eqref{21agosto2021-1}, with
$$
       \widetilde{\omega}_j(\lambda)= \omega_j(\lambda)+ \mathcal{D}_j(\lambda),
 $$
 where the $\mathcal{D}_j=\mathcal{D}_{[1,1]}^{(j)}\oplus\cdots\oplus  \mathcal{D}_{[s,s]}^{(j)}$ are holomorphic block-diagonal matrices, arbitrary\footnote{So the case $\mathcal{D}_j=0$ for all $j=1,...,s$ is possible.}
  except for the differential constraint  $$
\partial_j\mathcal{D}_k-\partial_k\mathcal{D}_j=  [\mathcal{D}_j(\lambda),~  \mathcal{D}_k(\lambda)],
 $$
 which is required by the integrability  of \eqref{21agosto2021-1}. 
 Hence,  
 $$
 dT=\bigl(\sum_j\mathcal{D}_j(\lambda)d\lambda_j\bigr) T.
 $$
 is integrable. 
  It admits holomorphic fundamental matrix solutions $\mathcal{T}(\lambda)=\mathcal{T}_1\oplus \cdots \oplus \mathcal{T}_s$ such that \eqref{4maggio2022-1} holds.

         
         \vskip 0.2 cm 
           The above theorems  \ref{30marzo2021-2} and  \ref{20agosto2021-3}  generalize to the non-generic case \eqref{27marzo2021-2} the strategy and the results of   \cite{JMU}.  

\bre
{\rm As a corollary, if $\Lambda$ has pairwise distinct eigenvalues, \eqref{27marzo2021-2} is strongly isomonodromic if and only if  \eqref{21agosto2021-1} holds with $ \widetilde{\omega}_j(\lambda)= \omega_j(\lambda)+ \mathcal{D}_j(\lambda)$, with $ \mathcal{D}_j(\lambda)$ diagonal satisfying $\partial_j\mathcal{D}_k-\partial_k\mathcal{D}_j=0$ (here $j,k=1,...,n$).  In this case   $\mathcal{T}(\lambda)$ above is {\it any} diagonal matrix.
}
\ere

An important application of the isomonodromy deformation theory here developed will be given in Section \ref{3agosto2022-1} for  the {\it caustic} of a semisimple Dubrovin-Frobenius manifold $M$  \cite{Dub1,Dub2} of dimension $n$. The caustic is a hypersurface $\mathcal{K}\subset M$  of codimension 1, such that the multiplication defined on the tangent bundle is nilpotent.  Following \cite{Her} and \cite{Reyes},  a generic point of $\mathcal{K}$ has a neighbourhood in $M$ were local coordinates $(t_1,t_2,u_3,...,u_n)$ are defined, such that $\mathcal{K}$ corresponds to $t_2=0$ and the vector $\partial/\partial t_2|_{t_2=0}$ is nilpotent.  The flat sections of the Dubrovin deformed connection (defined in Section \ref{3agosto2022-1}), expressed  in these coordinates and restricted at the caustic, are solutions of a Pfaffian system   exactly of the type \eqref{18novembre2021-1} (see system \eqref{29luglio2022-5}), with deformation parameters given by the coordinates $(t_1,u_3,...,u_n)$ on $\mathcal{K}$. It has  non trivial $\mathcal{T}$ depending on the flat metric defined of $M$,  and 
$$
\Lambda:=\hbox{diag}(\underbrace{t_1,t_1},\underbrace{u_3,...,u_n}_{n-2~distinct}).
$$   
Its $z$-component is always strongly isomonodromic in the sense described in this paper. Indeed, we will show that in case of a caustic  the deformation equations \eqref{21agosto2021-1}, concretely realized by system \eqref{12agosto2022-8},  are sufficient conditions for  strong isomonodromy.  The results are summarized in Proposition \ref{6agosto2022-3} and Remark \ref{13agosto2022-2}.  Moreover, the deformation theory developed in this paper allows us to predict some properties of the caustic (see Corollary \ref{12agosto2022-2} and point 3) of Remark \ref{13agosto2022-1}).

 \subsubsection*{Some further remarks}

The  integrability conditions of a   Pfaffian system \eqref{18novembre2021-1} with  {\it given } $\widetilde{\omega}_j(\lambda)= \omega_j(\lambda)$ of the specific form  \eqref{31marzo2021-6} are the non-linear 
``deformation equations'' \eqref{21agosto2021-1} and their compatibility  conditions. This is an elementary  computation 
 and is not new (see also the proof of part II of Theorem \ref{30marzo2021-2} here). For distinct eigenvalues these deformation equations are  a particular 
case of the JMMS equations introduced in  \cite{JMMS} (in particular section  4 and the appendix A. See also \cite{Hard}), while in case 
of repeated  eigenvalues, but no further coalescences,  \eqref{18novembre2021-1} and  \eqref{21agosto2021-1} fit into the more general deformations equations
 studied in  \cite{Boalch3} (
 see  also \cite{Boalch4}).
 
   The purpose of the present paper is not to  give deformations equations (integrability conditions) for a Pfaffian system, but  to  derive the Pfaffian system   \eqref{18novembre2021-1} as the   necessary 
 and sufficient condition for {\it all} the essential monodromy data   of \eqref{27marzo2021-2}  to be  constant; \eqref{21agosto2021-1} is consequently the 
 integrability condition of \eqref{18novembre2021-1}. Moreover,  in case there are no partial resonances, we show that     \eqref{21agosto2021-1} is also a 
 sufficient condition for {all} the essential monodromy data to be constant. This, in the spirit of Remark \ref{24nov2021-1}. 

\vskip 0.2cm

 To conclude,  we  make two more general comments. The first is that the main difficulty to generalize \cite{JMU} to 
 non generic cases    (in the sense of Remark \ref{24nov2021-1}) 
 is to find a suitable canonical representations for a class of fundamental matrix solutions (like the Levelt form at a 
 Fuchsian singularity and the solutions having a canonical asymptotics in Stokes sectors), and to deal with {\it the
  change of those representations}  when the Jordan type of the leading matrix at an irregular singularity changes, namely
   some eigenvalues merge. To our knowledge, this is an extremely difficult problem, which is far from being solved. 
 In the literature, we either find  attempts to deal with coalescences of eigenvalues with a change of Jordan type, but with suitable analyticity  and semisimplicity  assumptions, such as in  the work \cite{CDG}, or the Jordan types are fixed such as in \cite{BM} and in the present paper.

 The second comment -- which in a sense expands Remark \ref{24nov2021-1} -- is that there are two approaches in order to describe, {\it from the analytic viewpoint}, the isomonodromy deformations of a differential system
\be
\label{10dic2019-1}
\frac{dY}{dz}=\mathfrak{A}(z,\lambda)Y,
\ee
where $\mathfrak{A}(z,\lambda)$ is rational in $z$ and analytic in a domain of $\lambda$ (such as $\mathbb{D}$ here). 
One approach starts by proving the existence of fundamental matrix solutions  of \eqref{10dic2019-1}, holomorphic in the deformation parameters (under certain assumptions) in the  $\lambda$-domain, and  characterized by a  certain canonical form, such as Levelt form at Fuchsian singularities and canonical asymptotics at irregular ones.  Then, one must show  that these solutions satisfy a Pfaffian
system 
\be
\label{9dic2019-1}
dY = \omega Y,
\ee
 with a specific $\omega(z,\lambda)$, possibly determined by $\mathfrak{A}(z,\lambda)$, if and only if the deformation is isomonodromic, namely preserves a certain class of monodromy data (such as monodromy matrices or essential monodromy data of the above mentioned solutions). This is the approach approach of \cite{JMU} and the approach we mainly follow in our paper.

The other approach starts by
assuming that we are given a Pfaffian system (\ref{9dic2019-1}), satisfying the Frobenius integrability condition $d\omega=\omega\wedge \omega$,  and such that the $dz$ component of $\omega$ gives a differential system (\ref{10dic2019-1}), namely 
 $$\omega(z,\lambda)\Bigr|_{\lambda \rm ~fixed}=\mathfrak{A}(z,\lambda)dz.
 $$
 This implies that the monodromy matrices of a fundamental matrix solution $Y(z,\lambda)$ of  (\ref{9dic2019-1}) are constant,  so that system (\ref{10dic2019-1}) is weakly  isomonodromic. 
 Then, this approach  proceeds by showing if, depending on the specific $\omega(z,\lambda)$,  the Pfaffian system admits fundamental matrix solutions with a  canonical structure, whose corresponding  essential monodromy data are constant.  This is, for example, the approach of \cite{Bolibruch0,Bolibruch,Bolibruch1,YT}. 

\vskip 0.3 cm 
\noindent {\bf Acknowledgments.} I thank P. Boalch for several useful remarks and for pointing out some references. The author is member of   the European Union's H2020 research and innovation programme under the Marie Sk\l{l}odowska-Curie grant No. 778010 {\it IPaDEGAN}. He is also a member of 
 the Project 'Mathematical Methods in Non
Linear Physics' (MMNLP), Commissione Scientifica Nazionale 4 - Fisica
Teorica (CNS4) of the Istituto Nazionale di Fisica Nucleare (INFN).

\section{Preliminaries}
   
It is a standard result \cite{Wasow,AnosBol}  that, for each fixed $\lambda$, system \eqref{27marzo2021-2} admits a fundamental solution with {\it Levelt form}\footnote{This is an improper Levelt form, obtained by a permutation $Y\mapsto YP$, $P$ a suitable permutation matrix,  from a proper Levelt form \eqref{10agosto2022-2} of Section \ref{2maggio2021-1}.} at $z=0$:
\be
\label{27marzo2021-3} 
Y^{(0)}(z,\lambda):= G^{(0)}(\lambda) \widehat{Y}^{(0)}(z,\lambda)  z^{D^{(0)} }z^{L^{(0)}(\lambda)},
\ee
\be
\label{29dic2021-1}
J^{(0)}=D^{(0)} + S^{(0)}, \quad \quad L^{(0)}=S^{(0)} +R^{(0)},
\ee
where:
\be
\label{28marzo2021-2}
\widehat{Y}^{(0)}(z,\lambda)=I+\sum_{k=1}^\infty F_k^{(0)} (\lambda) z^k\quad \hbox{ is convergent for finite $|z|$};
 \ee
 the matrix $J^{(0}(\lambda)$ is a Jordan form of $A(\lambda)$;  the eigenvalues  $\mu_1(\lambda),\cdots, \mu_n(\lambda)$ of  $A(\lambda)$ are uniquely decomposed as 
$$
 \mu_j(\lambda)=d^{(0)}_j+\rho^{(0)}_j(\lambda), \quad \hbox{ with $0\leq \Re \rho^{(0)}_j <1$ and $d^{(0)}_j\in\mathbb{Z}$};
$$
and
$$  \mathrm{diag}(S^{(0)})=\hbox{\rm diag}(\rho^{(0)}_1,...,\rho^{(0)}_n),
  \quad 
  D^{(0)}=\hbox{\rm diag}(d^{(0)}_1,...,d^{(0)}_n)
 .
 $$
  The matrix $R^{(0)}$ is nilpotent, with entries
$$ 
(R^{(0)}(\lambda))_{ij}\neq 0 \hbox{ only if } \mu_i(\lambda)-\mu_j(\lambda)\in \mathbb{N}\backslash \{0\}.
$$
The invertible matrix $G^{(0)} $ puts $A$ in Jordan form. In general, the dependence of $Y^{(0)}$ on $\lambda$ is not holomorphic in $\mathbb{D}$. 
\vskip 0.2 cm 
\noindent
{\bf Assumption 2.} {\it $A(\lambda)$ is holomorphically similar to a Jordan form $J^{(0}(\lambda)$  in $\mathbb{D}$, namely  there is $G^{(0)}=G^{(0)}(\lambda) $  holomorphically invertible  in $\mathbb{D}$ such that 
$$J^{(0)}(\lambda)= G^{(0)}(\lambda) ^{-1} A(\lambda)~G^{(0)}(\lambda).
$$
}

The matrix $A(\lambda)$ is said to be {\it resonant} at $\lambda\in\mathbb{D}$ if there exist $i\neq j \in\{1,...,n\}$ such that 
$\mu_i(\lambda)-\mu_j(\lambda)\in\mathbb{Z}\backslash\{0\}$.  If the eigenvalues do not depend on $\lambda$ in $\mathbb{D}$, we simply say that $A$ is resonant.

\vskip 0.2 cm 
If there are no  resonances, Assumption 2 guarantees that   $Y^{(0)}(z,\lambda)$ can be taken holomorphic on $\mathcal{R}\times \mathbb{D}$. Otherwise,  in addition to Assumption 2 we need to require  that if $
\mu_i(\lambda)-\mu_j(\lambda)=\ell_{ij}\in \mathbb{Z}\backslash\{0\}$ for some $\lambda$, then the {\it resonance persists all over $\mathbb{D}$}, namely
\be
\label{27marzo2021-6} 
 \quad \mu_i(\lambda )-\mu_j(\lambda)=\ell_{ij}\in \mathbb{Z}\backslash\{0\} \quad  \forall~\lambda \in \mathbb{D}.
\ee
Then, holomorphy of \eqref{27marzo2021-3} follows from its standard formal computation (see \cite{Wasow}).
The reason for \eqref{27marzo2021-6} is that if it does not hold, then $R^{(0)}(\lambda)$ may have an extremely wild behaviour in $\lambda$.

\section{Weak isomonodromic deformations}

\ble[Isospectrality] 
\label{27marzo2021-9}
Let $A(\lambda)$ be holomorphic on $\mathbb{D}$. If \eqref{27marzo2021-2} has for each $\lambda\in\mathbb{D}$ a fundamental matrix solution $Y(z;\lambda)$ whose monodromy 
 $$
Y(z;\lambda)\longmapsto Y(ze^{2\pi i };\lambda)=Y(z;\lambda) M,
$$
  is the same for all $\lambda\in\mathbb{D}$  (i.e. the monodromy matrix $M$ is constant), then the eigenvalues of $A(\lambda)$ are constant on $\mathbb{D}$.
In particular,  \eqref{27marzo2021-6} holds in case of resonances.
  If moreover Assumption 2 holds, then  a Levelt form   $Y^{(0)}$ is holomorphic on $\mathcal{R}\times \mathbb{D}$. 
 
 \ele
 
 Notice that in Lemma \ref{27marzo2021-9} it is not assumed that $Y(z;\lambda)$ depends holomorphically on $\lambda$. 

\begin{proof}
There exists for each $\lambda$ an invertible connection matrix $C(\lambda)$ such that 
$$
Y(z;\lambda)= Y^{(0)}(z,\lambda) C(\lambda)
 = G^{(0)}(\lambda)\widehat{Y}^{(0)}(z,\lambda) z^{D^{(0)}(\lambda)} z^{L^{(0)}(\lambda)} C(\lambda)
.
$$
By assumption,  $M=C(\lambda)^{-1}e^{2\pi i L^{(0)}(\lambda)}~C(\lambda)$ does not depend on $\lambda$, so the eigenvalues $\rho_j^{(0)}$ of $L^{(0)}$ are constant. Since $A(\lambda)$ is holomorphic, its eigenvalues $\mu_j(\lambda)=d_j^{(0)}+\rho_j^{(0)}$  are continuous. It follows that both the integers $d_j^{(0)}$  and the eigenvalues $\mu_j$ are constant.  
Clearly, \eqref{27marzo2021-6} holds,  so that holomorphy follows from the  formal computation (see \cite{Wasow}) of \eqref{27marzo2021-3} and \eqref{28marzo2021-2}. 
\end{proof}

\bde
System  \eqref{27marzo2021-2} is {\bf weakly isomonodromic} in $\mathbb{D}$ if there exists a fundamental matrix solution $Y^{\rm hol}(z,\lambda)$ {\it depending holomorphically} on  $(z,\lambda)\in \mathcal{R}\times \mathbb{D}$, with  $\lambda$-independent monodromy matrix  $M$, defined by 
$$ 
Y^{\rm hol}(z,\lambda)\longmapsto Y^{\rm hol}(z,\lambda) M,\quad z\longmapsto ze^{2\pi i }.
$$ 
\ede

\bpr  Let  $A(\lambda)$ be holomorphic in $\mathbb{D}$.  
System \eqref{27marzo2021-2} is  weakly isomonodromic in $\mathbb{D}$ if and only if it is the $z$-component of an integrable Pfaffian system 
\be
\label{27marzo2021-4}
dY = \omega(z,\lambda) Y,\quad\quad\quad  \omega(z,\lambda)=\left(\Lambda +\frac{A}{z}\right)dz+\sum_{j=1}^s \omega_j(z,\lambda)d\lambda_j.
\ee
with $\omega(z,\lambda)$ holomorphic in $\overline{\mathbb{C}}\backslash\{0,\infty\}\times \mathbb{D}$, satisfying the integrability condition 
$$d\omega=\omega\wedge \omega.
$$ 
\epr 
The proof is standard.

\bth
\label{27marzo2021-5}
\begin{shaded}
Let $A(\lambda)$ be holomorphic in $\mathbb{D}$. System \eqref{27marzo2021-2} is weakly isomonodromic in $\mathbb{D}$ with holomorphic fundamental matrix solution  $Y^{\rm hol}$ coinciding with a fundamental solution in Levelt form $   Y^{(0)}$, 
if and only     
$$ 
\hbox{the coefficients $\omega_j(z,\lambda)$ are holomorphic in $\mathbb{C}\times \mathbb{D}$}.
$$  
\vskip 0.2 cm 
\noindent
In this case, the following facts hold.  
\begin{itemize}
\item $D^{(0)}$ and $L^{(0)}$ are constant,  or equivalenlty $J^{(0)}$ and $R^{(0)}$ are constant.

\item   $A(\lambda)$ is holomorphically similar to $J^{(0)}$ through a fundamental matrix solution  $G^{(0)}(\lambda)$ of 
$$ 
dG= \sum_{j=1}^s \omega_j(0,\lambda)d\lambda_j ~G.
$$
\end{itemize}
\end{shaded}  
\eth
 
 The matrices $\omega_j(z,\lambda)$ in Theorem \ref{27marzo2021-5} may have isolated singularity in $z=\infty$. 
The requirement $Y^{\rm hol}= Y^{(0)}$ is equivalent to the requirement that  
$ 
Y^{\rm hol} = Y^{(0)} C$,
  for  $C$ constant invertible matrix. 

\vskip 0.2 cm 
 Assumption 2  is not explicitly written in the statement of Theorem \ref{27marzo2021-5}.   
 If \eqref{27marzo2021-2} is weakly isomonodromic in $\mathbb{D}$ and we assume that there is a holomorphic fundamental matrix solution  in Levelt form  $  Y^{(0)}$, then Assumption 2  is a necessary condition, so it is automatically assumed by requiring that $Y^{(0)}$ is holomorphic. Conversely, if the coefficients $\omega_j(z,\lambda)$ are holomorphic in $\mathbb{C}\times \mathbb{D}$, then it follows from Proposition \ref{28marzo2021-3} below that   $A(\lambda)$ is holomorphically similar to $J^{(0)}$, namely  Assumption 2  is satisfied. 

\vskip 0.2 cm 
 Theorem \ref{27marzo2021-5}  holds also for 
\be
\label{29dic2021-4}
\begin{aligned}
\frac{dY}{dz}=\mathfrak{A}(z,t) Y,  \quad\quad\quad \mathfrak{A}(z,t):=\Lambda +\frac{A(t)}{z-a}
\\
\noalign{\medskip}
t:=(\lambda,a),\quad  \lambda=(\lambda_1,\cdots,\lambda_s)\in \mathbb{D}\subset \mathbb{C}^{s},\quad a\in\mathbb{C}.
\end{aligned}
\ee
where one deformation parameter is the pole $z=a$. 
In this case, \eqref{27marzo2021-4} is replaced by 
\be
\label{29dic2021-6}
\omega(z,t) = \mathfrak{A}(z,t) dz+\sum_{j=1}^s \omega_j(z,t)d\lambda_j+\left(\omega_0(z,t)-\frac{A(t)}{z-a}\right)da.
\ee
The coefficients $\omega_j(z,\lambda)$ and $\omega_0(z,t)$  are holomorphic in $\mathbb{C}\times \mathbb{D}$. 
We stress that the above holds for any $A(t)$, including the non-diagonalizable and resonant cases. 
Moreover, $G^{(0)}(t)$ is a fundamental matrix solution of 
\be
\label{29dic2021-2} 
dG= \left(\sum_{j=1}^s \omega_j(a,t)d\lambda_j +\omega_0(a,t)da+\varphi(t)da\right)G,
\ee
with 
\be
\label{29dic2021-3}
\varphi=G^{(0)}\left(F_1^{(0)}+[F_1^{(0)},J^{(0)}]+R_1^{(0)}\right) (G^{(0)})^{-1}.
\ee
Here, $F_1^{(0)}$ appears in the Taylor expansion \eqref{28marzo2021-2}, $J^{(0)}$ in \eqref{29dic2021-1}, and $R_1^{(0)}$ is the first term  in the decomposition of $R^{(0)}=\sum_{\ell=1}^m R_\ell ^{(0)}$, where 
$$ 
(R_\ell^{(0)})_{ij}\neq 0 \hbox{ only if } \mu_i-\mu_j=\ell  \in \mathbb{N}\backslash \{0\}.
$$ 
Notice that \eqref{29dic2021-2}  is linear, because $F_1^{(0)}$ and $R_1^{(0)}$ are obtained by the standard formal computation yielding \eqref{27marzo2021-3}, which is  done using the differential system
$ d\widetilde{Y}/dz= (G^{(0)})^{-1} \mathfrak{A}(z,t) G^{(0)}  \widetilde{Y}$, after the gauge transformation $Y=G^{(0)}\widetilde{Y}$. Explicit computation shows that  $G^{(0)}$ cancels in \eqref{29dic2021-3}, namely $\varphi$ is only determined by  $ \mathfrak{A}(z,t)$. 

\vskip 0.2 cm 
{\small

Expressions \eqref{29dic2021-6}, \eqref{29dic2021-2}, \eqref{29dic2021-3} can be obtained following the same steps of the proofs of Propositions  \ref{28marzo2021-3} and \ref{27marzo2021-8} below. Just notice that for example in \eqref{29dic2021-7} one has
$$
\omega =d(G^{(0)}\widehat{Y}^{(0)})\cdot (G^{(0)}\widehat{Y}^{(0)})^{-1}+ G^{(0)}\widehat{Y}^{(0)}
\frac{D^{(0)}+(z-a)^{D^{(0)}}L^{(0)} (z-a)^{-D^{(0)}}}{(z-a)}(G^{(0)}\widehat{Y}^{(0)})^{-1} dz
$$
$$
- G^{(0)}\widehat{Y}^{(0)}
\frac{D^{(0)}+(z-a)^{D^{(0)}}L^{(0)} (z-a)^{-D^{(0)}}}{z-a}(G^{(0)}\widehat{Y}^{(0)})^{-1} da.
$$ 
Then  use the definitions of the monodromy exponents \eqref{29dic2021-1}, which imply that 
$$
D^{(0)}+(z-a)^{D^{(0)}}L^{(0)} (z-a)^{-D^{(0)}}=J^{(0)}+\sum_{\ell=1}^m R_\ell^{(0)} (z-a)^\ell.
$$
}
\subsection{Proof of Theorem \ref{27marzo2021-5}}

Theorem \ref{27marzo2021-5} follows from  Propositions \ref{28marzo2021-3} and \ref{27marzo2021-8} below.  These propositions apply also to system \eqref{29dic2021-4}, with form \eqref{29dic2021-6} and \eqref{29dic2021-2}.

By its definition $D^{(0)}$ is locally constant   on subsets of $\mathbb{D}$. It may have  jump discontinuities on $\mathbb{D}$, so that $dD^{(0)}$ is not well defined. 
  We will sometimes write $dD^{(0)}=0$ with abuse of notation when we want to indicate that $D^{(0)}$ is constant on the whole $\mathbb{D}$. 

\bpr
\label{28marzo2021-3}
\begin{shaded}
Let $A(\lambda)$ be holomorphic in $\mathbb{D}$ and let \eqref{27marzo2021-2} be the $z$-component of an integrable Pfaffian system \eqref{27marzo2021-4} whose   coefficients $\omega_j(z,\lambda)$  are holomorphic in $\mathbb{C}\times \mathbb{D}$. 

\noindent
Then, there is a fundamental matrix solution $Y^{(0)}(z,\lambda)$ of \eqref{27marzo2021-4} in Levelt form \eqref{27marzo2021-3},    with 
\be
\label{20agosto2021-1} 
dD^{(0)}= dL^{(0)}=0, \quad \hbox{ or equivalently} \quad dJ^{(0)}=dR^{(0)}=0\quad \hbox{ on $\mathbb{D}$}.
\ee
Moreover, the matrix  $A(\lambda)$ is holomorphically similar to $J^{(0)}$ through  a fundamental matrix solution $G^{(0)}(\lambda)$ of
$$ 
dG= \sum_{j=1}^s \omega_j(0,\lambda)d\lambda_j ~G.
$$
\end{shaded}
\epr

\begin{proof} Proposition  \ref{28marzo2021-3} is a particular case  of the main results of \cite{YT} on fundamental matrix solutions of Pfaffian systems at a logarithmic (Fuchsian) singularity.
\end{proof}

 The converse of the above is the following
\bpr
\label{27marzo2021-8}
\begin{shaded}
Let $A(\lambda)$ be holomorphic in $\mathbb{D}$ and  let Assumption 2 hold. 

\vskip 0.2 cm 
a) Suppose that  \eqref{27marzo2021-2} is weakly isomonodromic.  
If  there is a fundamental matrix solution  of   \eqref{27marzo2021-4}  in Levelt form $Y^{(0)}(z,\lambda)$, then the  coefficients $\omega_j(z,\lambda)$  of \eqref{27marzo2021-4} are holomorphic in $\mathbb{C}\times \mathbb{D}$ and \eqref{20agosto2021-1} holds. 

\vskip 0.2 cm 
b) Conversely, if system  \eqref{27marzo2021-2} has  a fundamental solution $Y^{(0)}(z,\lambda)$  in Levelt form \eqref{27marzo2021-3} such that  \eqref{20agosto2021-1} holds, then the system is weakly isomonodromic, and the corresponding Pfaffian system \eqref{27marzo2021-4} has coefficients $\omega_j(z,\lambda)$   holomorphic in $\mathbb{C}\times \mathbb{D}$. 

\vskip 0.2 cm 
In both cases a) and b),  $G^{(0)}(\lambda)$ is a fundamental matrix solution of 
 $$ 
dG= \sum_{j=1}^s \omega_j(0,\lambda)d\lambda_j ~G.
$$
\end{shaded}
\epr

\begin{proof}
a)  Being  \eqref{27marzo2021-2} weakly isomonodromic, there is an isomonodromic $Y^{\rm hol}(z,\lambda)$, with constant monodromy matrix $M$,  satisfying  \eqref{27marzo2021-4}. By Lemma \ref{27marzo2021-9}, a solution $Y^{(0)}(z,\lambda)$  of \eqref{27marzo2021-2}  exists holomorphic in $\mathcal{R}\times \mathbb{D}$, with 
$$ 
D^{(0)} \hbox{ constant}.
$$
 By the assumption in a),   $Y^{(0)}$ also satisfies \eqref{27marzo2021-4}.  Being solutions of \eqref{27marzo2021-2},  $Y^{\rm hol}$ and $Y^{(0)}$  are related by a holomorphic  connection matrix $C(\lambda)$:
 $$ 
 Y^{\rm hol}(z,\lambda)=Y^{(0)}(z,\lambda)C(\lambda).
 $$
 Since both $dY^{\rm hol}=\omega Y^{\rm hol}$ and $dY^{(0)}=\omega Y^{(0)}$ hold, then 
 $$
 dC=0.
 $$
 Let us rewrite
$$ 
 Y^{\rm hol}(z,\lambda)=Y^{(0)}(z,\lambda)C= G^{(0)}(\lambda)\widehat{Y}^{(0)}(z,\lambda) z^{D^{(0)}}  Cz^{\mathcal{L}^{(0)}},$$
 $$ \mathcal{L}^{(0)}(\lambda):=C^{-1}~L^{(0)}(\lambda)~C.
 $$
 Since $dM=d(\exp\{2\pi i \mathcal{L}^{(0)}\}) =0$ by assumption, we have $d\mathcal{L}^{(0)}=0$, and then 
 $$ 
 dL^{(0)}=0.
 $$ 
Therefore, we find
\begin{align}
\omega=&~dY^{\rm hol}\cdot (Y^{\rm hol})^{-1}= d(Y^{(0)} \cdot  (Y^{(0)})^{-1})
\\
\noalign{\medskip}
\label{29dic2021-7}
= &~ d(G^{(0)}\widehat{Y}^{(0)})\cdot (G^{(0)}\widehat{Y}^{(0)})^{-1}+ G^{(0)}\widehat{Y}^{(0)}
\frac{D^{(0)}+z^{D^{(0)}}L^{(0)} z^{-D^{(0)}}}{z}(G^{(0)}\widehat{Y}^{(0)})^{-1} dz
\end{align}
Now, the definition of $D^{(0)}$ and $L^{(0)}$ implies that $D^{(0)}+z^{D^{(0)}}L^{(0)} z^{-D^{(0)}} $ is holomorphic at $z=0$ and 
\be
\label{28marzo2021-1}
 \lim_{z\to 0} (D^{(0)}+z^{D^{(0)}}L^{(0)} z^{-D^{(0)}} )= J^{(0)}, 
\ee 
 so that 
 $$ 
 \omega =dG^{(0)}\cdot (G^{(0)})^{-1}+ \hbox{\rm reg}(z,\lambda) +\left(\frac{A(\lambda)}{z}  +  \hbox{\rm reg}_1(z,\lambda)\right) dz  .
 $$ 
 Here 
 $$
\hbox{\rm reg}(z,\lambda) =O(z), \hbox{ for } z\to 0
$$
 is a 1-form in $dz$ and  $d\lambda_1,...,d\lambda_s$, holomorphic in $\mathbb{C}\times \mathbb{D}$. 
Moreover, $\hbox{\rm reg}_1(z,\lambda)$ is a holomorphic matrix  in $\mathbb{C}\times \mathbb{D}$  with behaviour
$$
\hbox{\rm reg}_1(z,\lambda) =O(1), \hbox{ for } z\to 0.
$$ 
 We conclude that 
 $$ 
 \omega = \sum_{j=1}^s \omega_j(z,\lambda) d\lambda_j + \left(\frac{A(\lambda)}{z}  +  \hbox{\rm reg}_1(z,\lambda)\right)dz ,
 $$ 
 where both $\hbox{\rm reg}_1(z,\lambda)$  and   the matrices $\omega_j(z,\lambda)$ are  holomorphic in $\mathbb{C}\times \mathbb{D}$, of order $O(1)$ for $z\to 0$. In particular, 
 $$ 
dG^{(0)} \cdot (G^{(0)})^{-1}= \sum_{j=1}^s \omega_j(0,z)d\lambda_j.
$$

 b)  Suppose that $dD^{(0)}=dL^{(0)}=0$,  so that for $z\longmapsto z e^{2\pi i}$ the monodromy $Y^{(0)}\longmapsto Y^{(0)} e^{2\pi i L^{(0)}}$ is constant. This  implies that  Lemma \ref{27marzo2021-9} holds, so that $Y^{(0)}(z,\lambda)$ is holomorphic in $\mathcal{R}\times \mathbb{D}$. We prove that $Y^{(0)}$ satisfies a Pfaffian system. We define 
 $$ 
 \omega(z,\lambda):=dY^{(0)}(z,\lambda)\cdot (Y^{(0)}(z,\lambda))^{-1}.
 $$ 
 This is single valued with respect to $z$, because the monodromy of  $Y^{(0)}$ is constant. The structure of $\omega(z,\lambda) =\omega_0(z,\lambda) dz +\sum_{j=1}^s \omega_j(z,\lambda) d\lambda_j$  is computable from \eqref{27marzo2021-3}:
 $$ 
 dY^{(0)}\cdot (Y^{(0)})^{-1}=
 $$
 $$=
  dG^{(0)}\cdot G^{(0)}+G^{(0)}d\widehat{Y}^{(0)}\cdot (G^{(0)} \widehat{Y}^{(0)})^{-1} + 
 G^{(0)}\widehat{Y}^{(0)}\frac{D^{(0)}+ z^{D^{(0)} } L^{(0)}z^{-D^{(0)} }}{z}(G^{(0)}\widehat{Y}^{(0)})^{-1}
 $$
 $$ 
 =d G^{(0)}\cdot G^{(0)} +\hbox{\rm reg}( z,\lambda) +\left(\frac{A_1}{z}+\hbox{\rm reg}_1(z,\lambda)\right)dz.
 $$ 
 In the last step, we have used \eqref{28marzo2021-2} and \eqref{28marzo2021-1}. Here $\hbox{\rm reg}( z,\lambda)$ stands for  a matrix valued 1-form in the $d\lambda_j$'s, holomorphic in $\mathbb{C}\times \mathbb{D}$, and of order $O(z)\to 0$ as $z\to 0$, while $\hbox{\rm reg}_1(z,\lambda)$ is a matrix  holomorphic in  $\mathbb{C}\times \mathbb{D}$ of order $O(1)$ as $z\to 0$ (and we know that it must be $\Lambda$).   
 In conclusion, we have found that  
  $$
  \sum_{j=1}^s \omega_j(z,\lambda)d\lambda_j:=d G^{(0)}\cdot G^{(0)}+\hbox{\rm reg}( z,\lambda),
  $$ 
  and in particular
  $
  d G^{(0)}\cdot (G^{(0)})^{-1}= \sum_{j=1}^s \omega_j(0,\lambda)d\lambda_j
 .
 $
\end{proof}

\section{Canonical solutions of \eqref{27marzo2021-2} at $z=\infty$}

Let us partition $A$ in blocks $A_{[i,j]}$, $i,j=1,...,s$, of dimension $p_i\times p_j$, inherited from $\Lambda$.   
Let 
$$\mathcal{T}(\lambda)=\mathcal{T}_1(\lambda)\oplus \cdots \oplus  \mathcal{T}_s(\lambda),
$$ 
 be a block diagonal matrix such that 
 \be
 \label{28marzo2021-4}
 \mathcal{T}_k(\lambda)^{-1} A_{[k,k]}(\lambda)~\mathcal{T}_k(\lambda)=J_k(\lambda) \quad \hbox{Jordan form},\quad \quad k=1,...,s.
 \ee
 It has structure  $\mathcal{T}_k(\lambda)=\mathcal{T}_k^0(\lambda)\mathfrak{B}_k(\lambda)$,  where $\mathcal{T}_k^0(\lambda)$ is a chosen  matrix satisfying \eqref{28marzo2021-4} and $\mathfrak{B}_k(\lambda)$ is any matrix such that  $[\mathfrak{B}_k,J_k]=0$. 
 
 \vskip 0.2 cm 
\noindent
{\bf Assumption 3.} {\it $A_{[1,1]}(\lambda)\oplus \cdots \oplus A_{[s,s]}(\lambda)$ is holomorphically reducible to Jordan form 
$$
J(\lambda)=J_1(\lambda)\oplus \cdots \oplus J_s(\lambda).
$$
 This means that each  $\mathcal{T}_k(\lambda)$ is holomorphic on $\mathbb{D}$, and so is each $J_k(\lambda)$.}

 \vskip 0.2 cm 
 We can  arrange each  $J_k$    into  $h_k\leq p_k$  Jordan blocks $J_1^{(k)}$, ..., $
J_{h_k}^{(k)}$
\be
\label{20agosto2021-2}
J_k= J_1^{(k)}\oplus \cdots \oplus
J_{h_k}^{(k)}.
\ee    
Each block $J_j^{(k)}$, $1\leq j\leq h_k$,  has dimension $r_j\times r_j$, with $r_j\geq 1$, $r_1+\cdots+r_{h_k}=p_k$.  Each $J_j^{(k)}$ has only one eigenvalue $\mu_j^{(k)}$, with structure,
$$
 J_j^{(k)}(\lambda)=\mu_j^{(k)}(\lambda) I_{r_j} +H_{r_j},
 \quad
 \quad
 I_{r_j}=\hbox{ $r_j\times r_j$ identity matrix},
 $$
  $$ \hbox{\rm $H_{r_j}=0$ ~if $r_j=1$}, \quad \quad 
 H_{r_j}=\left[
 \begin{array}{cccccc}
 0& 1 &  &  \cr
                   &0 &1      &        & &\cr 
                    & & \ddots & \ddots & &\cr
                    &          &         & 0 &1\cr
                    &           &         &  & 0 
 \end{array}
\right]~\hbox{  if $r_j\geq 2$}.
$$
Note that $\mu_1^{(k)}$, ..., $\mu_{h_k}^{(k)}$ are not necessarily distinct.
 The decomposition  $\mu_j^{(k)}=d_j^{(k)}+\rho_j^{(k)}$, with $d_j^{(k)}\in\mathbb{Z}$ and $0\leq \Re \rho_j^{(k)} <1$, induces the decomposition 
\be
\label{31marzo2021-3}
J_k= D_k +S_k,\quad k=1,...,s. 
\ee
where $D_k$ is diagonal with eigenvalues $d_j^{(k)}$ and $S_k$ is Jordan with eigenvalues $\rho_j^{(k)}$.  We let
$$ D:=D_1\oplus \cdots \oplus D_s,\quad S:=S_1\oplus \cdots \oplus S_s \quad \hbox{ so that } J=D+S.
$$

\vskip 0.2 cm 
 If Assumption 3 holds,  the gauge 
 $$ 
 Y(z,\lambda)=\mathcal{T}(\lambda) \widehat{X}(z,\lambda) 
 $$
 transforms system \eqref{27marzo2021-2} into 
 \be
 \label{8settembre2021-1}
  \frac{d\widehat{X}}{dz}=\left(\Lambda +\frac{\mathcal{A}(\lambda)}{z}\right) \widehat{X},\quad \quad 
 \mathcal{A}:=\mathcal{T}^{-1} A \mathcal{T} \equiv 
 \begin{pmatrix} 
 J_1 & \mathcal{A}_{[1,2]} & \cdots &  \mathcal{A}_{[1,s]}
 \\
  \mathcal{A}_{[2,1]} & J_2 & \cdots &  \mathcal{A}_{[2,s]}
 \\
 \vdots & \vdots & \ddots& \vdots 
 \\
  \mathcal{A}_{[s,1]} & \mathcal{A}_{[s,2]} &\cdots & J_s
 \end{pmatrix}
 \ee
We can then  apply to \eqref{8settembre2021-1} the computations of section 4.1 of \cite{CDG},  which allow to find formal solutions of \eqref{27marzo2021-2}  depending holomorphically on $\lambda\in\mathbb{D}$, with structure 
 \be
 \label{21maggio2021-3}
 Y_F(z,\lambda)= \mathcal{T}(\lambda) \Bigl( I+\sum_{j=1}^\infty F_j(\lambda) z^{-j}\Bigr) z^{D(\lambda)} z^{L(\lambda)} e^{\Lambda z}.
 \ee
   Here  
 \be
 \label{31marzo2021-4}
L:=L_1\oplus \cdots \oplus L_s 
, \quad
\quad 
 L_k:=S_k + R_k ,\quad \quad R_k\hbox{ is nilpotent}.
\ee
Each $ R_k$ has possibly non zero blocks  
\be
\label{31marzo2021-2}
[R_k]_{{\rm block} ~a,b} \neq 0 \quad \hbox{ only if } \quad \mu_b^{(k)}(\lambda)-\mu_a^{(k)}(\lambda) =\ell_{ba}\in \mathbb{N}\backslash \{0\},\quad a\neq b=1,...,h_k.
\ee
The diagonal matrix $D(\lambda)$ is locally constant, from its very definition, and may have discrete jumps as $\lambda$ varies in  $\mathbb{D}$. 
The computation of the $F_k(\lambda)$ and $R=R_1\oplus \cdots \oplus R_s$  follows exactly the procedure of proposition 4.1 of \cite{CDG}. 

In case there are no resonances in  $A_{[k,k]}(\lambda)$, then $R_k(\lambda)=0$. If there are no resonances in all the blocks  $A_{[k,k]}$, $\forall k=1,...,s$,   then
 $$R=0 \quad \quad\Longrightarrow \quad  L(\lambda)=S(\lambda),$$
  and 
 $$ 
 Y_F(z,\lambda)= \mathcal{T}(\lambda) \Bigl( I+\sum_{j=1}^\infty F_j(\lambda) z^{-j}\Bigr) z^{D(\lambda)} z^{S(\lambda)} e^{\Lambda z}= \mathcal{T}(\lambda) \Bigl( I+\sum_{j=1}^\infty F_j(\lambda) z^{-j}\Bigr) z^{J(\lambda)}  e^{\Lambda z}
 $$
 Then, by Assumption 3,  $Y_F(z,\lambda)$ depends holomorphically on $\lambda$.
 
In case of resonance of some $A_{[k,k]}(\lambda)$,  a sufficient condition for the $F_j(\lambda)$'s and  $L(\lambda)$ to depend holomorphically on $\lambda$   is that  when it happens that  
$
\mu_b^{(k)}(\lambda)-\mu_a^{(k)}(\lambda) =\ell_{ba}\in \mathbb{N}\backslash \{0\}$ for some value of $\lambda$, then 
the resonance persists all over $\mathbb{D}$, namely
 \be
 \label{28marzo2021-5}
\mu_b^{(k)}(\lambda)-\mu_a^{(k)}(\lambda) =\ell_{ba}\in \mathbb{N}\backslash \{0\}\quad \forall~ \lambda\in\mathbb{D}.
\ee
In this case,  being $D$ locally constant in $\mathbb{D}$,    $Y_F(z,\lambda)$ locally depends holomorphically on $\lambda$. 

\bde
\label{20agosto2021-7}
In the terminology  introduced in \cite{sabbah}, if there exists $k$ such that $A_{[k,k]}$  is resonant, we say that $A$ has a {\bf partial resonance}.
 \ede
 
  \vskip 0.2 cm
  
 A formal solution \eqref{21maggio2021-3} with given $\mathcal{T}$, $L$, $D$  and $\Lambda$ is uniquely determined  only if all the matrices $A_{[1,1]}(\lambda),\dots,  A_{[s,s]}(\lambda)$  are non-resonant (see corollary 4.1 of \cite{CDG}). 

\bre
{\rm 
In case $\Lambda={\rm diag}(\lambda_1,...,\lambda_n)$ has pairwise distinct eigenvalues, then 
$$Y_F(z,\lambda)= \mathcal{T}(\lambda) \Bigl( I+\sum_{j=1}^\infty F_j(\lambda) z^{-j}\Bigr) z^{{\rm diag}(A(\lambda))} e^{\Lambda z},
$$
and $ \mathcal{T}(\lambda)$ is an arbitrary invertible diagonal matrix. One can choose it  to be the identity matrix. 
}
\ere

\subsubsection*{Stokes Matrices}
 
 Consider an admissible direction $\tau$ as in Assumption 1 and the following   $\lambda$-independent sectors in $\mathcal{R}$ of central angular opening $\pi+2\delta$:
 $$
 \mathcal{S}_\nu : \quad (\tau+(\nu-1)\pi)-\delta<\arg z< (\tau+\nu\pi)+\delta,\quad \nu\in\mathbb{Z},\quad \delta>0. 
 $$
 If Assumption 1 holds, there is a sufficiently small $\delta$ such that $\mathcal{S}_\nu\cap \mathcal{S}_{\nu+1}$ does not contain Stokes rays as $\lambda$ varies in $\mathbb{D}$. From \cite{Sh4}, we know that to a prefixed formal solution \eqref{21maggio2021-3} there correspond  actual solutions  satisfying 
  \begin{align}
  \label{21maggio2021-4}
  &Y_\nu(z,\lambda)= \mathcal{T}(\lambda)\widehat{Y}_\nu(z,\lambda)z^{D(\lambda)} z^{L(\lambda)} e^{\Lambda z},
 \\
 \noalign{\medskip} 
 \label{22maggio2021-2}
 &
 \widehat{Y}_\nu(z,\lambda)\sim    I+\sum_{j=1}^\infty F_j(\lambda) z^{-j}, \quad z\to \infty \hbox{ in }\mathcal{S}_\nu.
 \end{align}
 For short, we will improperly write 
 $$
Y_\nu(z,\lambda)\sim Y_F(z,\lambda), \quad z\to \infty \hbox{ in }\mathcal{S}_\nu.
 $$
They are uniquely determined by the above asymptiotic behaviour (as proved in theorem 6.2 of \cite{CDG}). When  Assumption 3 and \eqref{28marzo2021-5} hold, they are holomorphic in $\mathcal{R}\times \mathbb{D}$ . In this case, the holomorphic {\bf Stokes matrices} $\mathbb{S}_\nu(\lambda)$  are defined by 
 $$ 
 Y_{\nu+1} (z,\lambda)=  Y_\nu (z,\lambda)\mathbb{S}_\nu(\lambda).
 $$
 
 \section{More on the Levelt form}
 \label{2maggio2021-1}
 
 This technical section can be skipped at first reading. It introduces  details needed especially in the proof in  the Appendix. The reader not interested in the Appendix  may just read the last sentence of this section, starting with  ``In conclusion,...''.
 
  Consider a $N\times N$  system $Y^\prime =\mathfrak{A}(z)Y$, such that $\mathfrak{A}(z)$ has a Fuchsian singularity in $z=a$, for   $a\in \mathbb{C}$, or  in $z=\infty$.
  The residue matrix of $\mathfrak{A}(z)$ at  $z=a$ (or $z=\infty$) has a Jordan form 
  $$J=J_1\oplus\cdots\oplus J_r,
  $$ 
  with 
  $$ J_j=\mu_j I_{m_j}+H_{m_j},\quad m_1+...+m_r=N,
  $$
   {\small $$ \hbox{\rm $H_{m_j}=0$ ~if $m_j=1$}, \quad \quad 
 H_{m_j}=\left[
 \begin{array}{cccccc}
 0& 1 &  &  \cr
                   &0 &1      &        & &\cr 
                    & & \ddots & \ddots & &\cr
                    &          &         & 0 &1\cr
                    &           &         &  & 0 
 \end{array}
\right]~\hbox{  if $m_j\geq 2$}.
$$
}
We can arrange the Jordan form so that the eigenvalues   $\mu_1,...,\mu_r$ of  $J$  have real parts forming a non increasing sequence if $z=a$ is the singularity:
\be
\label{4ottobre2021-1} 
\Re \mu_1 \geq \Re \mu_2\geq \dots \geq \Re \mu_r;
\ee
 or  a non decreasing sequence in case $z=\infty$ is the singularity:
 \be
\label{4ottobre2021-2}
\Re \mu_1 \leq \Re \mu_2\leq \dots \leq \Re \mu_r.
\ee
    We also write  $\mu_j=\rho_j+d_j$, with $0\leq \Re \rho_i<1$ and $d_j\in\mathbb{Z}$, and   
   $$ 
   J=D+S, 
   $$ 
   where $D$ is the diagonal matrix of integers $d_j$.

 The differential system can be reduced to normal form by a standard procedure \cite{Wasow}, and this allows to find a fundamental matrix solution in Levelt form 
 \be
 \label{3agosto2022-2}
 Y(z)=\mathcal{G}(\zeta) \zeta^D \zeta^L, 
 \ee 
 where
  $\zeta=z-a$ if $a$ is the singularity,  or $\zeta=z$ if $\infty$ is the singularity. Here, $\mathcal{G}(\zeta)$ is holomorphic at $z=a$ (or at $z=\infty$). In case the matrix coefficient $\mathfrak{A}(z)$ is holomorphic only  in a sector centered at the singularity, and admits  there  an asymptotic expansion, then $ \mathcal{G}(\zeta)$ is holomorphic in that sector, with asymptotic expansion there \cite{Wasow}. 
  Moreover, the monodromy exponent $L$ is 
  $$ 
  L=S+R, 
   $$
 where the  matrix $R$ is  nilpotent and obtained by the formal computation of the normal form. 
 
  Consider the block partition of $R$ inherited from $J$. 
     For the singularity  $z=a$, it  possibly has a non-trivial block in position $(i,j)$, with $1\leq i\neq j \leq r$, if  $\mu_i -\mu_j = d_i - d_j \geq 1$ is  integer.   For the singularity  $z=\infty$, $R$ possibly has a non-trivial block in position $(j,i)$  if  $\mu_i -\mu_j = d_i - d_j \geq  1$ is integer.    It follows from the ordering \eqref{4ottobre2021-1} or \eqref{4ottobre2021-2} that  $R$  only  has possibly non zero blocks in the upper triangular part of its block partition  ($R$ is upper triangular  if  $J$ is diagonal). The diagonal  blocks  of $R$ are zero (the  diagonal is zero if  $J$ is diagonal).

\vskip 0.2 cm 
\noindent
{\bf Examples.}
 The solution   \eqref{27marzo2021-3} is an example for $a=0$. The solutions $Y_\nu(z,\lambda)$ in \eqref{21maggio2021-4} contains the matrix factor $\mathcal{T}(\lambda)\widehat{Y}_\nu(z,\lambda)z^{D(\lambda)} z^{L(\lambda)}$, which  is an example  with  $\zeta=z$ and $\mathcal{G}(\zeta)$  holomorphic at $z=\infty$ in a sector  $ \mathcal{S}_\nu$: indeed, it   is a fundamental solution in  Levelt form for the Fuchsian  system (4.1)   at  $z=\infty$ of  the paper \cite{CDG}. 
\vskip 0.2 cm

   Notice once more that, with the given ordering  \eqref{4ottobre2021-1} or \eqref{4ottobre2021-2},     for  $1\leq i<j\leq r$ we have $\mu_i\neq \mu_j$ and $\rho_i=\rho_j$  whenever $\mu_i-\mu_j \equiv d_i-d_j\neq 0$ is a non-zero  integer,  and correspondingly $R$  possibly has a non-zero block in position $(i,j)$. Therefore, possibly acting by a permutation  $L\longmapsto P^{-1}L P$ if necessary (which means changing $Y\longmapsto YP$ by a permutation matrix $P$),  we can always do the above construction in such a way that  $L$ admits another partition into blocks 
$$L=L_1\oplus \cdots \oplus L_\ell,  \quad\quad \hbox{with $\ell\leq r$},
$$
where   each block $L_q$ is  upper triangular, it has only one eigenvalue $\sigma_q$  equal to some $\rho_i=\rho_j=...$  from the set $\{\rho_1,...,\rho_r\}$, satisfying $0\leq \Re  \sigma_q <1$, and $\sigma_p\neq \sigma_q$ for $1\leq p\neq q\leq \ell$, and  the corresponding diagonal matrix $D$ of integer parts of the eigenvalues of $J$  is split into blocks $D={D}_1\oplus\cdots\oplus 
{D}_\ell$, with 
$$
{D}_q=\hbox{diag}({d}_{q,1},{d}_{q,2},....),  \quad \quad q=1,...,\ell,
$$ 
 where for each $q$ the integers  form a non-increasing finite sequence $${d}_{q,1}\geq {d}_{q,2}\geq...$$ in case the singularity is $z=a$;  or  a non-decreasing  finite sequence $${d}_{q,1}\leq {d}_{q,2}\leq...$$ in case $z=\infty$ is the singularity. 
We can therefore rewrite 
    $$
 L= S+R,\quad
 \quad\hbox{with }\quad 
 {S}={S}_1\oplus\cdots\oplus {S}_\ell,\quad {R}={R}_1\oplus\cdots\oplus {R}_\ell.
 $$
Each block $L_q$, $1\leq q \leq \ell$, consists of  sub-blocks, according to  the structure (for some integer $k_q$):
\be
\label{5ottobre2021-1} 
L_q={S}_q+{R}_q,\quad 
\quad  {S}_q=\left(\begin{array}{cccc}
 {S}_1^{(q)} &  &&\\
                                & {S}_2^{(q)} && \\
                                &     &\ddots &\\
                                &    &       &  {S}_{\kappa_q}^{(q)}                   
 \end{array}
 \right) 
 ,\quad
 {R}_q=\left(
 \begin{array}{cccc}
 \boldsymbol{0} &  *   & *&* \\
                                & \boldsymbol{0}&* &* \\
                                &     &\ddots &* \\
                                &    &       & \boldsymbol{0}                     
 \end{array}
 \right),
\ee
 where each matrix $ {S}_i^{(q)}$   is a Jordan matrix with the same eigenvalue  $\sigma_q$  on the diagonal and 1's on the second upper diagonal: 
$$
  {S}_i^{(q)}=\left(
 \begin{array}{cccc}
\sigma_q &  1 & 0&0 \\
                                &\sigma_q &1 &0 \\
                                &     &\ddots &1 \\
                                &    &       & \sigma_q                     
 \end{array}
 \right) ,\quad i=1,2,...,k_q;
 $$
while in  $R_q$ the $ \boldsymbol{0}$ are zero diagonal blocks (corresponding to the blocks ${S}_i^{(q)}$), and  each  $*$ is an off-diagonal block which is possibly non zero (now the block partition of $R_q$ in \eqref{5ottobre2021-1}  is inherited from that of $S_q$).

\vskip 0.2 cm 
One can also decompose  the above $L$ as 
\be
\label{30aplile2021-1}
L=\Sigma +N, \quad \hbox{  $\Sigma$ diagonal and $N$ nilpotent},
\ee
with 
$$ 
\Sigma= \sigma_1 I_1\oplus\cdots  \oplus  \sigma_\ell I_\ell, 
\quad 
N=N_1\oplus\cdots  \oplus N_\ell.
$$
Here $ I_1,..., I_\ell$ are identity matrices, each $I_q$ having the dimension of $L_q$. 
It follows that 
\be
\label{30aplile2021-2}
[\Sigma, N]=0.
\ee
Therefore,
$$ 
z^D z^L= z^D z^\Sigma z^N= z^\Delta z^N,
$$
where
\be
\label{9agosto2022-1}
 \Delta:= D+\Sigma
\ee
is a diagonal matrix, whose eigenvalues are the eigenvalues of $J$. The above properties allow to write 
$$ 
z^\Delta z^N= z^{\Delta} \sum_{k=1}^{\overline{k}} \frac{N^k}{k!}(\ln z)^k \quad \hbox{ finite sum},
$$
 where $\overline{k}$ depends on the order of nilpotency of $N$.

 In conclusion, a Levelt form \eqref{3agosto2022-2} can  be always chosen so that $D$ and $L$ (and so $S$ and $R$)  satisfy 
the above properties \eqref{30aplile2021-1}, \eqref{30aplile2021-2} and \eqref{9agosto2022-1}, 
namely
\be
\label{10agosto2022-2}
 Y(\zeta)=\mathcal{G}(\zeta) \zeta^D \zeta^L
=\mathcal{G}(\zeta)
\zeta^\Delta \zeta^N.
\ee
This can always be achieved by a permutation matrix $P$, by changing a fundamental matrix solution in Levelt form $Y$   to another fundamental solution $YP$ with  Levelt form  having the desired properties. In the Appendix, we will need the above choice of Levelt form. 
 
 \bre
 {\rm
 In this section we have given the analytic construction of the ``proper'' Levelt form, just starting from the analytic structure of fundamental solutions at a Fuchsian singularity. For the geometric viewpoint, see 
 \cite{AnosBol}.
 }\ere
 
 \section{Strong Isomonodromy Deformations}
 \label{21settembre2022-1}
 
We define a {\bf central connection matrix} $C_0(\lambda)$ associated with  $Y_0(z,\lambda)$ in \eqref{21maggio2021-4} with $\nu=0$, and with a fundamental solution $Y^{(0)}(z,\lambda)$ in Levelt form \eqref{27marzo2021-3}  at $z=0$, by 
$$ 
Y_0(z,\lambda)=Y^{(0)}(z,\lambda)C_0(\lambda).
$$ 
Notice that 
$$ 
Y_\nu(z,\lambda)=Y^{(0)}(z,\lambda)~C_0~\mathbb{S}_0\cdots \mathbb{S}_{\nu-1}.
$$

 \bde 
 \label{31marzo2021-1}
  Let Assumption 1 hold and let    system \eqref{27marzo2021-2} be weakly isomonodromic in
  $\mathbb{D}$ with holomorphic fundamental matrix solution $Y^{\rm hol}=Y^{(0)}$ in Levelt form, so that Theorem \ref{27marzo2021-5} holds,  Assumption 2 is satisfied and the essential monodromy data
  $$L^{(0)}, \quad D^{(0)} \quad \hbox{ are constant}.
  $$  
  If also Assumption  3 holds, system \eqref{27marzo2021-2} is said to be  {\bf strongly isomonodromic} on $\mathbb{D}$ when also the remaining  essential monodromy data are constant, namely 
$$
d\mathbb{S}_\nu= 0, \quad dL=0, \quad D \hbox{ is constant }, \quad dC_0=0.
$$ 
\ede

\bre{\rm
For a strongly isomonodromic system, the relations \eqref{28marzo2021-5}, if any,  are satisfied by definition, so that the fundamental matrices $Y_\nu(z,\lambda)$ are holomorphic on $\mathcal{R}\times \mathbb{D}$. 
}
\ere

\bth
\label{30marzo2021-2}
\begin {shaded}
~
\vskip 0.1 cm 
\noindent
 {\bf Part I.}  System \eqref{27marzo2021-2} is strongly isomonodromic in $\mathbb{D}$ if and only if the fundamental matrix solutions $Y^{(0)}(z,\lambda)$ and  $Y_\nu(z,\lambda)$ satisfy   for every $\nu\in\mathbb{Z}$ the integrable  Pfaffian system \eqref{27marzo2021-4} of the specific form
\be
\label{30marzo2021-1}
dY = \omega(z,\lambda) Y,\quad\quad\quad  
\omega(z,\lambda)=\left(\Lambda +\frac{A}{z}\right)dz+\sum_{j=1}^s \Bigl(z E_{p_j} + \widetilde{\omega}_j(\lambda)\Bigr)d\lambda_j,
\ee
where
$$
\widetilde{\omega}_j(\lambda)=\omega_j(\lambda)+\frac{\partial \mathcal{T}(\lambda)}{\partial \lambda_j}\cdot  \mathcal{T}(\lambda)^{-1}
,
$$
and   $\omega_j(\lambda)$ has blocks
\be
\label{31marzo2021-6}
\omega^{(j)}_{[a,a]}(\lambda)=0,\quad \quad\quad \omega^{(j)}_{[a,b]}(\lambda)
= 
\frac{A_{[a,b]}(\lambda)~(\delta_{aj}-\delta_{bj})}{\lambda_a-\lambda_b},\quad \quad a\neq b=1,...,s,
\ee
while $$ \mathcal{T}(\lambda)= \mathcal{T}_1(\lambda)\oplus \cdots \oplus  \mathcal{T}_s(\lambda)$$ is a holomorphic matrix reducing  to Jordan form the diagonal blocks of $A$ as in \eqref{28marzo2021-4}. 
Strong isomonodromy deformations  with  constant  $\mathcal{T}$  are allowed. In this case, 
$$
\widetilde{\omega}_j(\lambda)=\omega_j(\lambda).
$$

\vskip 0.3 cm 
\noindent
{\bf Part II.} If system \eqref{27marzo2021-2} is strongly isomonodromic in $\mathbb{D}$, then 
\be
\label{20agosto2021-6}
 \frac{\partial A}{\partial \lambda_j}=[\widetilde{\omega}_j(u), A],\quad j=1,...,s.
 \ee
 In particular, 
 $$
\frac{\partial A_{[1,1]}}{\partial \lambda} =\frac{\partial A_{[2,2]}}{\partial \lambda} \dots=\frac{\partial A_{[s,s]}}{\partial \lambda} =0
,
$$
 and so  the block-diagonal part of $A$  and the Jordan forms $J_k$ in \eqref{28marzo2021-4} are constant on $\mathbb{D}$.
\end{shaded}
\eth

   For a strong isomonodromy deformation, PART II says 
    that $A_{[1,1]}\oplus \cdots \oplus A_{[s,s]}$ is constant, so it can be reduced  to Jordan form by a constant block-diagonal matrix $\mathcal{T}_0$.  If $\mathcal{T}=\mathcal{T}_1\oplus\cdots \oplus \mathcal{T}_s$ is another matrix satisfying \eqref{28marzo2021-4}, but  not constant, then it has the structure 
 $$ 
 \mathcal{T}(\lambda)=\mathcal{T}_0 \mathfrak{B}(\lambda), 
\quad\quad 
 \mathfrak{B}(\lambda)=\mathfrak{B}_1(\lambda)\oplus\cdots \oplus \mathfrak{B}_s(\lambda),
\quad \hbox{ with }\quad [\mathfrak{B}_j(\lambda),J_j]=0.
$$
 
 The isomonodromic fundamental matrix solutions $Y_\nu(z,\lambda)$ which satisfy the Pfaffian system  
 \eqref{30marzo2021-1} have structure \eqref{21maggio2021-4} with constant $\mathcal{T}=\mathcal{T}_0$ if and only if system  \eqref{30marzo2021-1} is of the specific form with coefficients  $
\widetilde{\omega}_j(\lambda)=\omega_j(\lambda)
$.  

In other words, if a differential system 
$$ 
\frac{dY}{dz}=\left(\Lambda_0+\frac{A_0}{z}\right)Y
$$
is given at $\lambda=\lambda_0$, where $\Lambda_0$ has repeated eigenvalues, then  it can have different 
isomonodromy deformations \eqref{30marzo2021-1}, differing by the specific
 $\sum_j \widetilde{\omega}_j(\lambda) d\lambda_j$, namely by the specific $\mathcal{T}(\lambda)$.   
 For all these deformations, the diagonal blocks are constant and equal to those of $A_0$, but the off-diagonal
  blocks of $A(\lambda)$ are different for different deformations,  satisfying different systems  \eqref{20agosto2021-6}, with different\footnote{This allows, in case $n=3$ and $s=2$, to have isomonodromy deformations with constant $A$, see Section \ref{1settembre2021-1}. For $n\geq 4$, it is not possible to reach a constant $A$ by a gauge transformation, see Remark \ref{10agosto2022-1}.}   matrix 
  coefficients $\widetilde{\omega}_j$ depending on the choice of $\mathcal{T}(\lambda)$.

Two isomonodromy deformations with different $\mathcal{T}$ are related by a  gauge transformation. Suppose that $Y$ satisfies \eqref{30marzo2021-1} with  matrices $A(\lambda)$ and  $
\widetilde{\omega}_j(\lambda)=\omega_j(\lambda)+\partial_j \mathcal{T} \cdot \mathcal{T}^{-1}
$, where $\omega_j$  is in \eqref{31marzo2021-6}. Consider the gauge transformation 
$$ 
Y=\mathcal{T}(\lambda)  \widecheck{\mathcal{T}}^{-1}(\lambda)~  \widecheck{Y},
$$ 
where  $\widecheck{\mathcal{T}}$ is another matrix Jordanizing the block-diagonal part of $A(\lambda)$. 
Then,  $\widecheck{Y}$ satisfies a system \eqref{30marzo2021-1} of the form
$$ 
d\widecheck{Y}=\left[\left(\Lambda +\frac{\widecheck{A}}{z}\right)dz+\sum_{j=1}^s \Bigl(z E_{p_j} + \widecheck{\omega}_j\Bigr)d\lambda_j+d\widecheck{\mathcal{T}}\cdot \widecheck{\mathcal{T}}^{-1} \right]\widecheck{Y}.
$$  
where
\be 
\label{13ottobre2021-1}
\widecheck{A}(\lambda):=\widecheck{\mathcal{T}}\bigr(\mathcal{T}^{-1}A(\lambda)\mathcal{T}\bigr)\widecheck{\mathcal{T}}^{-1},\quad\quad  \widecheck{\omega}_j(\lambda)=\widecheck{\mathcal{T}}\bigl(\mathcal{T}^{-1}\omega_j(\lambda)\mathcal{T}\bigr)\widecheck{\mathcal{T}}^{-1}.
\ee
Notice that  $ \widecheck{\omega}_j$ is the same as in definition \eqref{31marzo2021-6} with $A$ replaced by $\widecheck{A}$ (the block-diagonal parts of $A$ and  $\widecheck{A}$ are the same).

  \bre
  \label{1settembre2021-2}
  {\rm Suppose that we have a deformation with $\widetilde{\omega}_j=\omega_j$ for all $j$. Then the dependence of $Y_\nu$  on $\lambda$ is 
    $$
  Y_\nu(z,\lambda)= \mathcal{T}~\widehat{Y}_\nu(z;\lambda_2-\lambda_1,...,\lambda_s-\lambda_1)~z^{D} z^{L} ~e^{\Lambda z}.
  $$
   Indeed, formula  \eqref{31marzo2021-6} implies that 
$$ 
\sum_{j=1}^s \omega_j(\lambda)=0, \quad \Longrightarrow \quad \sum_{j=1}^s \frac{\partial A}{\partial \lambda_j}=0,
 \quad \Longrightarrow \quad
A=A(\lambda_2-\lambda_1,...,\lambda_s-\lambda_1).
$$
  Moreover, let us write \eqref{21maggio2021-4} as
  $$Y_\nu(z,\lambda)=H_\nu(z,\lambda) e^{\Lambda z}
  \quad \hbox{
   with}\quad  H_\nu:=\mathcal{T}\widehat{Y}_\nu z^{D} z^{L}.
   $$
     Then 
   $ 
   \frac{\partial H_\nu e^{\Lambda z}}{\partial \lambda_j}= \frac{\partial H_\nu}{\partial \lambda_j} e^{\Lambda z} + H_\nu \cdot z E_{p_j}  e^{\Lambda z}
   $.  Since  $Y_\nu$ satisfies \eqref{30marzo2021-1} with $\widetilde{\omega}_j=\omega_j$, we also have $ 
   \frac{\partial H_\nu e^{\Lambda z}}{\partial \lambda_j}= (z E_{p_j}+\omega_j(\lambda))H_\nu e^{\Lambda z}
   $. 
   Thus,  
   $$
   \frac{\partial H_\nu}{\partial \lambda_j}=z[E_{p_j},H_\nu]+\omega_j(\lambda) H_\nu,
   \quad \Longrightarrow \quad 
  \sum_{j=1}^s \frac{\partial H_\nu}{\partial \lambda_j}=0,
  $$
  so that $H_\nu=H_\nu(z;\lambda_2-\lambda_1,...,\lambda_s-\lambda_1)$. 
  Notice also  that $z^Dz^L$  commutes with $E_{p_j}$ and that in the   strong isomonodromic case  $d\mathcal{T}=dD=dL=0$, so that we also obtain 
  $$
   \frac{\partial \widehat{Y}_\nu}{\partial \lambda_j}=z[E_{p_j},\widehat{Y}_\nu]+\omega_j(\lambda) \widehat{Y}_\nu,
   \quad \Longrightarrow \quad 
  \sum_{j=1}^s \frac{\partial \widehat{Y}_\nu}{\partial \lambda_j}=0,
  $$
This concludes. 
  }
  \ere

\centerline{**********************}
\begin{proof} [Proof of Theorem \ref{30marzo2021-2}] Let system \eqref{27marzo2021-2} be strongly isomonodromic in $\mathbb{D}$ (Definition \ref{31marzo2021-1}), so that $Y^{\rm hol}=Y^{(0)}$ in Levelt form is holomorphic, and by Assumption  3  all the $Y_\nu$ are holomorphic. Thus, we can take differentials. We define 
$$
\omega(z,\lambda):= dY^{(0)}\cdot (Y^{(0)})^{-1}\underset{dC^{(0)}=0}=dY_0\cdot (Y_0)^{-1}\underset{\rm all ~d\mathbb{S}_\nu=0}=dY_\nu\cdot (Y_\nu)^{-1},\quad \forall \nu\in\mathbb{Z}.
 $$
This is single valued for the counter-clockwise loop $z\mapsto z e^{2\pi i }$, because the monodromy  $e^{2\pi i L^{(0)}}$ of $Y^{(0)}$ and the monodromy $e^{2\pi i L} \bigl(\mathbb{S}_\nu \mathbb{S}_{\nu+1}\bigr)^{-1}$ of $Y_\nu$ are constant. Its singularities may only be located at  $z=0,\infty$. 
We find the structure of $\omega$ at $z=0$ and $z=\infty$ respectively.

\vskip 0.2 cm 
\noindent
Structure at $z=0$. Let us decompose the differential as $d=d_z+d_\lambda$, the former being the component on $dz$, the latter on $d\lambda_1,...,d\lambda_s$.  Firstly, we compute
\begin{align*} 
d_zY^{(0)}\cdot (Y^{(0)})^{-1}&=G^{(0)}d_z\widehat{Y}^{(0)}\cdot (\widehat{Y}^{(0)})^{-1}+ G^{(0)}\widehat{Y}^{(0)} \frac{D^{(0)}+z^{D^{(0)}} L z^{-D^{(0)}}}{z} (G^{(0)}\widehat{Y}^{(0)})^{-1}dz
\\
&= \left(\frac{A}{z}+\hbox{\rm reg}_1(z,\lambda)\right)dz,\quad\quad \hbox{\rm reg}_1(z,\lambda)=O(1),\quad z\to 0,
\end{align*}
where we have used \eqref{28marzo2021-1}. Here, $\hbox{\rm reg}_1(z,\lambda)$ is holomorphic for $z\in\mathbb{C}$ and  $\lambda\in\mathbb{D}$. Then, we compute 
\begin{align*} 
d_\lambda  Y^{(0)}\cdot (Y^{(0)})^{-1}&\underset{dD^{(0)}=dL^{(0)}=0}=d_\lambda G^{(0)} \cdot (G^{(0)})^{-1}+G^{(0)} d_\lambda\widehat{Y}^{(0)}\cdot (\widehat{Y}^{(0)})^{-1} (G^{(0)})^{-1}
\\
&= d_\lambda G^{(0)} \cdot (G^{(0)})^{-1}+\hbox{\rm reg}(z,\lambda),\quad\quad \hbox{\rm reg}(z,\lambda)=O(z)\to 0,\quad z\to 0,
\end{align*}
where $\hbox{\rm reg}(z,\lambda)$ is holomorphic for $z\in\mathbb{C}$ and  $\lambda\in\mathbb{D}$. 

\vskip 0.2 cm 
\noindent
Structure at $z=\infty$. Firstly, we compute 
\begin{align*}
d_z Y_\nu \cdot Y_\nu^{-1}=& \mathcal{T} d_z \widehat{Y}_\nu \cdot \widehat{Y}_\nu^{-1}\mathcal{T}^{-1}+
\\
&+\Bigl(\mathcal{T}  \widehat{Y}_\nu\frac{D+z^D L z^{-D}}{z}  \widehat{Y}_\nu^{-1}\mathcal{T}^{-1}+ 
\mathcal{T}  \widehat{Y}_\nu~ z^D z^L~ \Lambda ~z^{-L} z^{-D}~ \widehat{Y}_\nu^{-1}\mathcal{T}^{-1}\Bigr)dz.
\end{align*}
Due to the block structure of $D$ and $L$ and diagonality of $\Lambda$, we have 
 $ z^D z^L~ \Lambda ~z^{-L} z^{-D}=\Lambda$, while by \eqref{31marzo2021-3}, \eqref{31marzo2021-4}, \eqref{31marzo2021-2} we have $ (D+z^D L z^{-D})/z = J/z+O(z^{-2})$. Hence,  
$$
d_z Y_\nu \cdot Y_\nu^{-1}= \Bigl(\Lambda +\widetilde{\hbox{\rm reg}}\left(z^{-1},\lambda\right)\Bigr)dz ,
\quad\quad
\widetilde{\hbox{\rm reg}}\left(z^{-1},\lambda\right)=O\left(\frac{1}{z}\right)\to 0,\quad z\to \infty,
$$
being $\widetilde{\hbox{\rm reg}}(1/z,\lambda)$ analytic for $z\in\overline{\mathbb{C}}\backslash\{0\}$ and $\lambda\in\mathbb{D}$.  Then, we compute 
$$
d_\lambda Y_\nu \cdot Y_\nu^{-1}\underset{dD=dL=0}= d_\lambda \mathcal{T} \cdot \mathcal{T}^{-1}+
\mathcal{T} d_\lambda \widehat{Y}_\nu \cdot Y_\nu^{-1} \mathcal{T}^{-1}+ z\cdot \mathcal{T} \widehat{Y}_\nu ~z^D z^{L} d\Lambda z^{-L}z^{-D} (\mathcal{T} \widehat{Y}_\nu)^{-1}.
$$
As before, from the diagonality of $d\Lambda$ we receive  $z^D z^{L} d\Lambda z^{-L}z^{-D}=d\Lambda$, so that   
 $$
 d_\lambda Y_\nu \cdot Y_\nu^{-1}
 =
  d_\lambda \mathcal{T} \cdot \mathcal{T}+z d\Lambda 
+ \mathcal{T} [F_1,d\Lambda] \mathcal{T}^{-1} +  \widehat{\hbox{\rm reg}}\left(z^{-1},\lambda\right) ,
$$
where  $\widehat{\hbox{\rm reg}}\left(z^{-1},\lambda\right)$  is a 1-form in $d\lambda_1,...,d\lambda_n$, analytic for $z\in\overline{\mathbb{C}}\backslash\{0\}$ and $\lambda\in\mathbb{D}$, with behaviour \
$$
\widehat{\hbox{\rm reg}}\left(z^{-1},\lambda\right)=O\left(\frac{1}{z}\right)\to 0,\quad z\to \infty,
$$
Notice that $d\Lambda=E_{p_1}d\lambda_1+\dots +E_{p_s} d\lambda_s$. 
By Liouville theorem and the above  behaviours at $z=0,\infty$ we conclude that 
\begin{equation}
\label{31marzo2021-5}
\omega =\left(\Lambda +\frac{A}{z}\right)dz + \sum_{j=1}^s\Bigl(z E_{p_j} + [\mathcal{T} F_1 \mathcal{T}^{-1}, E_{p_j} ] +\frac{\partial \mathcal{T}}{\partial \lambda_j}\mathcal{T}^{-1}\Bigr)d\lambda_j.
\end{equation}
It remains to show that $[\mathcal{T} F_1(\lambda)\mathcal{T}^{-1},E_{p_j}]$ equals \eqref{31marzo2021-6}. The computations of section 4.1 of \cite{CDG}, Proposition 4,1, yield the off-diagonal blocks
\be
\label{8settembre2021-4}
F^{(1)}_{[i,j]}=\frac{\mathcal{A}_{[i,j]}}{\lambda_j-\lambda_i},\quad 1\leq i\neq j \leq s.
\ee
They suffice to evaluate $ [\mathcal{T} F_1 \mathcal{T}^{-1}, E_{p_j} ] $, since the diagonal blocks do not contribute. From the definition of  $\mathcal{A}$ in \eqref{8settembre2021-1}  and \eqref{8settembre2021-4} we receive 
$$ 
F^{(1)}_{[i,j]}=\frac{ \mathcal{T}_i^{-1}A_{[i,j]}\mathcal{T}_j}{\lambda_j-\lambda_i}
\quad
\Longrightarrow
\quad
(\mathcal{T} F_1 \mathcal{T}^{-1})_{[i,j]}
=
 \mathcal{T}_iF^{(1)}_{[i,j]}\mathcal{T}_j^{-1}
 =
  \frac{A_{[i,j]}}{\lambda_j-\lambda_i}.
$$
Using the last formula, we  conclude that 
  $$ [\mathcal{T} F_1(\lambda)\mathcal{T}^{-1},E_{p_k}]_{[a,a]}=0,  
  \quad
  \quad 
 [\mathcal{T} F_1(\lambda)\mathcal{T}^{-1},E_{p_k}]_{[a,b]}= \frac{A_{[a,b]}(\lambda)~(\delta_{ak}-\delta_{bk})}{\lambda_a-\lambda_b},\quad a\neq b.
 $$ 

It remains to show that isomonodromy deformations with   constant $\mathcal{T}$ are possible. This will be proved  after Lemma \ref{1april2021-1}.

\vskip 0.2 cm 
$\bullet$ Conversely, we assume that all the fundamental matrices $Y^{(0)}$ and $Y_\nu$, $\nu\in \mathbb{Z}$, of system \eqref{27marzo2021-2} also satisfy 
$$
dY = \Bigl[\left(\Lambda +\frac{A}{z}\right)dz+\sum_{j=1}^s \Bigl(z E_{p_j} + \widetilde{\omega}_j(\lambda)\Bigr)d\lambda_j\Bigr] Y,
$$
 with holomorphic $\widetilde{\omega}_j(\lambda)$ and $A(\lambda)$.  
 In particular, this means that $dY^{(0)}\cdot (Y^{(0)})^{-1}$ and $dY_\nu\cdot Y_\nu^{-1}$ depend homomorphically on $\lambda$. 

Since $\omega$  has Fuchsian singularity at $z=0$, by Proposition  \ref{28marzo2021-3} we know that indeed it has holomorphic solution $Y^{(0)}$ in Levelt form, and\footnote{This can also be seen directly by taking 
$$
d_\lambda Y^{(0)} \cdot (Y^{(0)})^{-1} 
=
 d_\lambda (G^{(0)}\widehat{Y}^{(0)})\cdot   (G^{(0)}\widehat{Y}^{(0)})^{-1}+
$$
$$+
(G^{(0)}Y^{(0)}) \Bigl(
d\Delta^{(0)} \ln z + z^{\Delta^{(0)}} \sum_{k=1}^{\overline{k}_0} \frac{d (N^{(0)})^k}{k!} (\ln z)^k ~ z^{-N^{(0)}}z^{-\Delta^{(0)}}
\Bigr)(G^{(0)}Y^{(0)})^{-1}.
$$
Since $\omega$ does not contain terms in $\ln z$, it follows that $d\Delta^{(0)}=d N^{(0)}=0$, so that $D^{(0)}$ and $L^{(0)}$ are constant. Here, $\Delta^{(0)}$ and $N^{(0)}$ are the analogous of $\Delta$ and $N$ in \eqref{30aplile2021-1}.} 
$$ 
D^{(0)},\quad L^{(0)}\quad \hbox{ are constant}
.
$$

The fact that $dC_0= d\mathbb{S}_\nu=0$ is straightforward. Indeed, since all fundamental solutions satisfy $dY=\omega Y$, we have 
$$ dY_{\nu+1}\cdot Y_{\nu+1}^{-1}=dY_{\nu}\cdot Y_{\nu}^{-1}  \quad \Longleftrightarrow\quad  d\mathbb{S}_\nu=0.
$$
$$ dY_{0}\cdot Y_{0}^{-1}=dY^{(0)}\cdot( Y^{(0)})^{-1}  \quad \Longleftrightarrow\quad  d C_0=0.
$$

Finally, we show that $D$ and $L$ are constant. Recall from Section \ref{2maggio2021-1}  that we can write $Y_\nu=\mathcal{T} \widehat{Y}_\nu z^\Delta z^N$, with diagonal $\Delta$ and nilpotent $N$.   By assumption 
\begin{align*} 
 \sum_{j=1}^s\bigl(z E_{p_j} +\widetilde{\omega}_j(\lambda)\bigr)d\lambda_j &=
  d_\lambda Y_\nu \cdot Y_\nu^{-1}
 = d\mathcal{T} \cdot \mathcal{T}^{-1} +\mathcal{T} d_\lambda \widehat{Y}_\nu \cdot \widehat{Y}_\nu^{-1}\mathcal{T}^{-1} + 
\\
&
~ + 
\mathcal{T} \widehat{Y}_\nu\Bigl(
 \ln z~d\Delta+z^\Delta \sum_{k=1}^{\overline{k}}\frac{dN^k}{k!}
(\ln z)^k ~z^{-N}z^{-\Delta} \Bigr) ~(\mathcal{T} \widehat{Y}_\nu)^{-1}
+
\\
&
~ +z\cdot \mathcal{T} \widehat{Y}_\nu d\Lambda (\mathcal{T} \widehat{Y}_\nu )^{-1}.
\end{align*}
Since logarithmic terms cannot occur, necessarily $d\Delta=dN=0$, so that $D$ and $L$ are constant. From  the dominant terms at $z=\infty$  in the above computation we receive 
$$ 
\widetilde{\omega}_j(\lambda)=  [\mathcal{T} F_1 \mathcal{T}^{-1}, E_{p_j} ] +\frac{\partial \mathcal{T}}{\partial \lambda_j}\mathcal{T}^{-1}.
$$

\vskip 0.2 cm 
\noindent
$\bullet$ PART II.   Suppose the system is strongly isomonodromic. By PART I, the matrices $Y^{(0)}$ and $Y^{(\nu)}$ solve a Pfaffian system $dY=\omega Y$ where $\omega$ has structure 
$$
\omega =\left(\Lambda +\frac{A}{z}\right)dz + \sum_{j=1}^s\Bigl(z E_{p_j} +\widetilde{\omega}_j(\lambda)\Bigr)d\lambda_j.
 $$
 We write for short
 $$
 \omega=\sum_{\alpha=0}^s \varphi_\alpha(x)dx^\alpha,\quad\quad 
 (x^0,x^1,...,x^s):=(z,\lambda_1,...,\lambda_s).
 $$
  Thus, $\omega$ is  integrable, i.e. $d\omega=\omega\wedge \omega$, which explicitly is  
\be
\label{4novembre2021-1}
\frac{\partial \varphi_\beta}{\partial x^\alpha}+\varphi_\beta \varphi_\alpha = \frac{\partial \varphi_\alpha}{\partial x^\beta}+\varphi_\alpha \varphi_\beta,\quad \alpha\neq \beta=0,1,...,s, 
\ee
For $\beta=0$ and $\alpha=j\in\{1,...,s\}$, \eqref{4novembre2021-1} is 
$$ 
\frac{\partial}{\partial \lambda_j}\left(\Lambda +\frac{A}{z}\right)+ \left(\Lambda +\frac{A}{z}\right)\Bigl(zE_{p_j} + \widetilde{\omega}_j(\lambda)\Bigr)=
\frac{\partial}{\partial z} \Bigl(zE_{p_j} + \widetilde{\omega}_j(\lambda)\Bigr) +\Bigl(zE_{p_j} + \widetilde{\omega}_j(\lambda)\Bigr)\left(\Lambda +\frac{A}{z}\right).
$$
Expanding, we see that the equality is true if and only if the coefficients of $z^{-1}$ and $z^0$ are respectively equal, namely
\begin{align}
\label{31marzo2021-8}
& \frac{\partial A}{\partial \lambda_j}= [\widetilde{\omega}_j(\lambda),A],
\\
\noalign{\medskip}
\label{31marzo2021-9}
&
[\Lambda, \widetilde{\omega}_j(\lambda)]=[E_{p_j},A].
\end{align}
The  equations \eqref{31marzo2021-9} have general solution 
$$
 \bigl[\widetilde{\omega}_j(\lambda)\bigr]_{\rm block ~a,a}\hbox{ arbitrary},\quad 
\bigl[\widetilde{\omega}_j(\lambda)\bigr]_{\rm block~ a,b}=
\frac{A_{[a,b]}(\lambda)~(\delta_{aj}-\delta_{bj})}{\lambda_a-\lambda_b},\quad \quad a\neq b=1,...,s.
$$
Thus, 
\be
\label{31marzo2021-7}
 \widetilde{\omega}_j(\lambda)=\omega_j(\lambda)+ \mathcal{D}_j(\lambda),
\ee 
where $\omega_j(\lambda)$ is \eqref{31marzo2021-6}, while $ \mathcal{D}_j(\lambda)$  is an arbitrary block-diagonal matrix, which in our case must be 
$$
\mathcal{D}_j=\frac{\partial \mathcal{T}}{\partial \lambda _j} \cdot  \mathcal{T}^{-1}.
$$ 

The integrability condition \eqref{4novembre2021-1} for $\alpha=k$, $\beta=j$, with $j\neq k\in \{1,...,s\}$ is 
$$ 
\frac{\partial}{\partial \lambda_k}\Bigl(zE_{p_j} + \widetilde{\omega}_j(\lambda)\Bigr)+ \Bigl(zE_{p_j} + \widetilde{\omega}_j(\lambda)\Bigr)\Bigl(zE_{p_k} + \widetilde{\omega}_k(\lambda)\Bigr)=\hbox{ the same with $j,k$ exchanged}$$
This is true if and only if 
\begin{align} 
\label{1aprile2021-4}
& [E_{p_j},  \widetilde{\omega}_k(\lambda)]=[E_{p_k},  \widetilde{\omega}_j(\lambda)],
\\
\noalign{\medskip}
\label{31marzo2021-10}
& \frac{\partial  \widetilde{\omega}_j(\lambda)}{\partial \lambda_k} +  \widetilde{\omega}_j(\lambda) \widetilde{\omega}_k(\lambda)= \frac{\partial  \widetilde{\omega}_k(\lambda)}{\partial \lambda_j} +  \widetilde{\omega}_k(\lambda) \widetilde{\omega}_j(\lambda),\quad\quad 1\leq j\neq k\leq s.
\end{align}
The  equalities \eqref{1aprile2021-4} are automatically satisfied\footnote{
First, notice that each $\mathcal{D}_j$ commutes with $\Lambda$ and each $E_{p_k}$, so that 
$$
[\Lambda, \widetilde{\omega}_j(\lambda)]=[E_{p_j},A],\quad\quad  [E_{p_j},  \widetilde{\omega}_k(\lambda)]=[E_{p_k},  \widetilde{\omega}_j(\lambda)],\quad\quad1\leq j\neq k\leq s,
$$
are equivalent to 
$$[\Lambda, {\omega}_j(\lambda)]=[E_{p_j},A],\quad\quad  [E_{p_j}, 
{\omega}_k(\lambda)]=[E_{p_k},  {\omega}_j(\lambda)],\quad\quad 1\leq j\neq k\leq s.
$$
Then, we show that 
$$
[\Lambda, {\omega}_j(\lambda)]=[E_{p_j},A],\quad\Longrightarrow\quad  [E_{p_j}, 
{\omega}_k(\lambda)]=[E_{p_k},  {\omega}_j(\lambda)],\quad\quad 1\leq j\neq k\leq s.
$$
Indeed, the $[\Lambda, {\omega}_j(\lambda)]=[E_{p_j},A]$ imply that the matrices $\omega_j$ are as in \eqref{31marzo2021-6}. Substituting \eqref{31marzo2021-6} we obtain the blocks 
$$ 
\Bigl([E_{p_j},\omega_k]-[E_{p_k},\omega_j]\Bigr)_{[a,b]}= \delta_{ja}\omega^{(k)}_{[a,b]}-
\omega^{(k)}_{[a,b]}\delta_{jb}-\delta_{ka}\omega^{(j)}_{[a,b]}+\omega^{(j)}_{[a,b]}\delta_{kb}
$$
$$
=\frac{A_{[a,b]}}{\lambda_a-\lambda_b}\Bigl(
\delta_{ja}(\delta_{ka}-\delta_{kb})-\delta_{jb}(\delta_{ka}-\delta_{kb})-\delta_{ka}(\delta_{ja}-\delta_{jb})+\delta_{kb}(\delta_{ja}-\delta_{jb})
\Bigr)=0.
$$} by \eqref{31marzo2021-7} and  \eqref{31marzo2021-6}, while \eqref{31marzo2021-10} implies the Frobenius  integrability of \eqref{31marzo2021-8}. 
Notice that the block-diagonal part of \eqref{31marzo2021-10} is (see Lemma \ref{21ottobre2022-1} for details)
$$
\frac{\partial  \mathcal{D}_j(\lambda)}{\partial \lambda_k} +   \mathcal{D}_j(\lambda)  \mathcal{D}_k(\lambda)= \frac{\partial   \mathcal{D}_k(\lambda)}{\partial \lambda_j} +   \mathcal{D}_k(\lambda)  \mathcal{D}_j(\lambda),\quad j,k=1,...,s,
$$
and admits in particular the holomorphic solution $\mathcal{D}_j=\partial \mathcal{T}/\partial \lambda _j \cdot  \mathcal{T}^{-1}$ for some holomorphic $ \mathcal{T}(\lambda)=\mathcal{T}_1(\lambda)\oplus \cdots \oplus \mathcal{T}_s(\lambda)$, in accordance with the required structure \eqref{31marzo2021-5}, which necessarily holds in case of strong isomonodromic deformations. 

Let $A_{[k,k]}$ denote as usual a diagonal-block of $A$, and let $\mathcal{D}^{(j)}_{[k,k]}$ be a diagonal-block of $\mathcal{D}_j$, $k=1,...,s$. The block diagonal part of \eqref{31marzo2021-8} now reduces to 
\be
\label{1april2021-2}
\frac{\partial A_{[k,k]}}{\partial \lambda_j} =[ \mathcal{D}^{(j)}_{[k,k]}, A_{[k,k]}], \quad k=1,...,s.
\ee
If $\mathcal{D}=0$, namely $\mathcal{T}$ is constant, the above implies constancy of the  block diagonal part of  $A$. If $\mathcal{T}$ is not constant, we need the following technical Lemma, proved in the Appendix. 

\ble
\label{1april2021-1} 
Let $\mathcal{T}=\mathcal{T}_1\oplus \cdots \oplus \mathcal{T}_s$ be a matrix yielding a Jordan form  
$J=J_1\oplus \cdots \oplus J_s= \mathcal{T}^{-1}\Bigl(
A_{[1,1]}\oplus \cdots \oplus  A_{[s,s]}\Bigr) \mathcal{T}
$, 
where each $J_k$ is as in \eqref{20agosto2021-2}. 
If the deformation is strongly isomonodromic, then 
  $$
  \bigl[\mathcal{T}^{-1}  d \mathcal{T}, J\bigr]=0.
  $$
\ele

 In case of strong isomonodromic deformation, we have  $\mathcal{D}_j=\partial \mathcal{T}/\partial \lambda _j \cdot  \mathcal{T}^{-1}$. Using Lemma \ref{1april2021-1} and \eqref{1april2021-2}, we  prove that $A_{[1,1]},\dots,A_{[s,s]}$ are constant. Indeed
\begin{align*} \frac{\partial A_{[k,k]}}{\partial \lambda_j} &=[ \mathcal{D}^{(j)}_{[k,k]}, A_{[k,k]}]
\\
&
= \frac{\partial \mathcal{T}_k}{\partial \lambda_j} \mathcal{T}_k^{-1}A_{[k,k]}-A_{[k,k]} \frac{\partial \mathcal{T}_k}{\partial \lambda_j} \mathcal{T}_k^{-1}
\\
&
= \frac{\partial \mathcal{T}_k}{\partial \lambda_j} J_k \mathcal{T}_k^{-1} -\mathcal{T}_kJ_k \Bigl(\mathcal{T}_k^{-1} \frac{\partial \mathcal{T}_k}{\partial \lambda_j} \Bigr)\mathcal{T}_k^{-1}
\\
&
\underset{\rm  Lemma ~ \ref{1april2021-1}}=
 \frac{\partial \mathcal{T}_k}{\partial \lambda_j} J_k \mathcal{T}_k^{-1} -\mathcal{T}_k \mathcal{T}_k^{-1} \frac{\partial \mathcal{T}_k}{\partial \lambda_j} J_k \mathcal{T}_k^{-1}=0.
 \end{align*}
 This proves PART II. 
 
By the constancy of the diagonal blocks of $A$ for a strong isomonodromy deformation,  it is possible  to consider deformations with   
 $ \mathcal{T}$ constant.   Conversely, if $\mathcal{T}$ is constant, so that all $\mathcal{D}_j=0$, then \eqref{1april2021-2} implies that all the $A_{[k,k]}$ are constant.  
 \end{proof}

 It is convenient to point out from the proof above that  $\bigl[\Lambda,\widetilde{\omega}_j\bigr]=\bigl[ E_{p_j}, A\bigr]
$ is equivalent to
 \be
 \label{21maggio2021-7}
  \widetilde{\omega}_j(\lambda)=\omega_j(\lambda) +\mathcal{D}_j(\lambda)
 \ee
 where $\omega_j(\lambda)$ is \eqref{31marzo2021-6} and 
\be
 \label{22maggio2021-1}
\mathcal{D}_j= \mathcal{D}_{[1,1]}^{(j)}\oplus\cdots\oplus  \mathcal{D}_{[s,s]}^{(j)}
\quad \hbox{arbitrary  block-diagonal matrix}.
\ee
It is also convenient to state the following
 
 \ble
 \label{21ottobre2022-1}
 Assume that the matrices $\widetilde{\omega}_j(\lambda)$ are defined by 
 $$[\Lambda, \widetilde{\omega}_j]=[E_{p_j},A],\quad j=1,...,s,
 $$
   so that they have structure \eqref{21maggio2021-7}. Let $\partial_j:=\frac{\partial}{\partial\lambda_j}$. The following facts hold. 
 
 a)  The system 
 $$
 \partial_i\widetilde{\omega}_j-\partial_j \widetilde{\omega}_i=[\widetilde{\omega}_i,\widetilde{\omega}_j],\quad\quad i,j=1,...,s,
 $$
 is equivalent to 
 $$
 \left\{
 \begin{aligned}
 &
 \partial_i{\omega}_j-\partial_j {\omega}_i=[{\omega}_i,{\omega}_j]+[\omega_i,\mathcal{D}_j]+[\mathcal{D}_j,\omega_i]& \hbox{block off-diagonal,}
 \\
 \noalign{\medskip}
 &
 \partial_i\mathcal{D}_j-\partial_j\mathcal{D}_i=[\mathcal{D}_i,\mathcal{D}_j] &  \hbox{block diagonal.}
 \end{aligned}
 \right.
 $$

 b) Assume moreover that 
\be
\label{21maggio2021-2-bis}
 dA=\bigl[ \sum_{j=1}^n \widetilde{\omega}_j(\lambda)d\lambda_j,A\bigr],
\ee
 then 
$$
\left\{
\begin{aligned}
& \partial_i\widetilde{\omega}_j-\partial_j \widetilde{\omega}_i-[\widetilde{\omega}_i,\widetilde{\omega}_j]
=
\partial_i\mathcal{D}_j-\partial_j\mathcal{D}_i-[\mathcal{D}_i,\mathcal{D}_j] ,
\\
\noalign{\medskip}
&
 \partial_i{\omega}_j-\partial_j {\omega}_i
 -
 \Bigl(
 [{\omega}_i,{\omega}_j]+[\omega_i,\mathcal{D}_j]+[\mathcal{D}_j,\omega_i]
 \Bigr)
=
0
;
\end{aligned}
\right.
$$
and
 $$
 dA^D=[
\mathcal{D},A^D],
 $$
 where
 \be
\label{10giugno2021-1}
A^D(\lambda):=A_{[1,1]}(\lambda)\oplus \cdots \oplus A_{[s,s]}(\lambda)
\ee

 \ele
 Notice that part b) above implies that if \eqref{21maggio2021-2-bis} holds, then  the system $
 \partial_i\widetilde{\omega}_j-\partial_j \widetilde{\omega}_i=[\widetilde{\omega}_i,\widetilde{\omega}_j]$ is equivalent to $\partial_i\mathcal{D}_j-\partial_j\mathcal{D}_i=[\mathcal{D}_i,\mathcal{D}_j]$, $i,j=1,...,s$.  
 
 \begin{proof}
a) Given any matrix $\widetilde{\omega}_j$, we can write it as $\widetilde{\omega}_j=\omega_j+\mathcal{D}_j$, a sum of a block off-diagonal term $\omega_j$ + block diagonal term $\mathcal{D}_j$. Then
$$
\partial_i\widetilde{\omega}_j-\partial_j \widetilde{\omega}_i-[\widetilde{\omega}_i,\widetilde{\omega}_j]
=
$$
$$
=\underbrace{\Bigl\{
 \underbrace{\partial_i{\omega}_j-\partial_j {\omega}_i
 }_{(*1*)}-
 \Bigl(
 \underbrace{[{\omega}_i,{\omega}_j]+[\omega_i,\mathcal{D}_j]+[\mathcal{D}_j,\omega_i]}_{(*2*)}
 \Bigr)
\Bigr\}
}_{(\bullet)}
+
\Bigl\{\underbrace{
\partial_i\mathcal{D}_j-\partial_j\mathcal{D}_i-[\mathcal{D}_i,\mathcal{D}_j] }_{(*3*)}
\Bigr\}
$$
Clearly, $(*1*)$ is block off-diagonal and $(*3*)$ is block diagonal. 
It is easy to see that in $(*2*)$, also $[\omega_i,\mathcal{D}_j]+[\mathcal{D}_j,\omega_i]$ is block off-diagonal. Indeed
$$
[\omega_i,\mathcal{D}_j]_{[a,a]}=\omega^{(i)}_{[a,a]}\mathcal{D}^{(j)}_{[a,a]}-\mathcal{D}^{(j)}_{[a,a]}\omega^{(i)}_{[a,a]}\underset{\omega^{(i)}_{[a,a]}=0}=0.
$$
If the matrices $\widetilde{\omega}_j$ are defined by $[\Lambda, \widetilde{\omega}_j(\lambda)]=[E_{p_j},A]$ for $j=1,...,s$, namely if $\omega_j$ is as in \eqref{31marzo2021-6}, then the whole $(\bullet)$ is block off-diagonal. Indeed,
$$[{\omega}_i,{\omega}_j]_{[a,a]}=\sum_{b\neq a} (\omega^{(i)}_{[a,b]}\omega^{(j)}_{[b,a]}-\omega^{(j)}_{[a,b]}\omega^{(i)}_{[b,a]})
$$
$$
\underset{\eqref{31marzo2021-6}}= \sum_{b\neq a} \frac{A_{[a,b]}A_{[b,a]}}{(\lambda_a-\lambda_b)(\lambda_b-\lambda_a)}
((\delta_{ia}-\delta_{ib})(\delta_{jb}-\delta_{ja})-(\delta_{ja}-\delta_{jb})(\delta_{ib}-\delta_{ia}))=0.
$$
This proves a).

b) We consider the blocks of   $(\bullet)$ and  substitute  $\widetilde{\omega}_k=\omega_k+\mathcal{D}_k$ as in   \eqref{21maggio2021-7}. Then,  where derivatives of the blocks $\omega_{[a,b]}^{(k)}$  occur, we express the $\omega_{[a,b]}^{(k)}$ in terms of $A$ using  \eqref{31marzo2021-6}, and then we re-substitute $\partial_kA_{[a,b]}=[\omega_k+\mathcal{D}_k,A]_{[a,b]}$. 
After a lengthy computation we receive
$$
(\bullet)=
0.
$$
This proves the first part of b). The blocks of \eqref{21maggio2021-2-bis} are 
$$
\frac{\partial A_{[a,b]}}{\partial \lambda_j}=\sum_{c=1}^s \left(\omega^{(j)}_{[a,c]}A_{[c,b]}-A_{[a,c]}\omega^{(j)}_{[c,b]}\right)
+
\mathcal{D}_a^{(j)} A_{[a,b]}-A_{[a,b]}\mathcal{D}_b^{(j)}.
$$
For $a=b$, the structure \eqref{31marzo2021-6} implies that 
$$
\sum_{c=1}^s \left(\omega^{(j)}_{[a,c]}A_{[c,a]}-A_{[a,c]}\omega^{(j)}_{[c,a]}\right)=0,
\quad\Longrightarrow \quad 
 \frac{\partial A_{[a,a]}}{\partial \lambda_j}=[\mathcal{D}_a^{(j)}, A_{[a,a]}].
$$

\end{proof}
 
 We state the converse of Part II of Theorem \ref{30marzo2021-2}. 
 \bth
\label{20agosto2021-3}
\begin{shaded}

  Let  $A$ and $\widetilde{\omega}_1,...,\widetilde{\omega}_1$ satisfy the system
 \begin{align}
 \label{21maggio2021-1}
& \bigl[\Lambda,\widetilde{\omega}_j\bigr]=\bigl[ E_{p_j}, A\bigr],\quad j=1,...,s, \quad (\hbox{so that } 
\widetilde{\omega}_j=\omega_j+\mathcal{D}_j \hbox{ as in \eqref{21maggio2021-7}});
\\
\noalign{\medskip}
 \label{21maggio2021-2}
 &dA=\bigl[ \sum_{j=1}^n \widetilde{\omega}_j(\lambda)d\lambda_j,A\bigr];
 \\
\noalign{\medskip}
\label{21ottobre2022-2}
&
\partial_i\mathcal{D}_j-\partial_j\mathcal{D}_i=[\mathcal{D}_i,\mathcal{D}_j],\quad 1\leq i\neq j \leq s;
 \end{align}
  Then, the following facts hold.

 \begin{itemize}
 
 \item[1.] System \eqref{21maggio2021-2} is Frobenius integrable. 
 \item[2.] The connection 
 \be
 \label{21maggio2021-8}
\omega (z,\lambda)=\left(\Lambda +\frac{A}{z}\right)dz + \sum_{j=1}^s\Bigl(z E_{p_j} +\widetilde{\omega}_j(\lambda)\Bigr)d\lambda_j.
\ee
with matrices \eqref{31marzo2021-6}, \eqref{21maggio2021-7} and \eqref{22maggio2021-1} is Frobenius integrable. The Pfaffian system $dY=\omega Y$  has a fundamental matrix solution $Y^{(0)}(z,\lambda)$ in Levelt form \eqref{27marzo2021-3}, with constant exponents as in \eqref{20agosto2021-1}.

\item[3.]  
Assumption 3 holds and $A^D(\lambda)$ defined in \eqref{10giugno2021-1} 
 admits constant Jordan form 
 $$\mathcal{T} A^D\mathcal{T}^{-1}=J \equiv J_1\oplus \cdots\oplus J_s,
 $$
  where $\mathcal{T}(\lambda)
 $ is  a holomorphic invertible matrix solution  of the integrable Pfaffian system
 \be
 \label{21maggio2021-10}
 dT=\mathcal{D}(\lambda) T,
 \ee 
where $\mathcal{D}(\lambda):=\sum_{j=1}^n \mathcal{D}_j(\lambda) d\lambda_j$. Such a $\mathcal{T}(\lambda)
 $ can be chosen block diagonal as $J$. 
\item[4.]  
If there are no partial resonances (i.e.  
   all the  blocks $A_{[k,k]}$, $k=1,...,s$, are non-resonant), then every  $Y_\nu(z,\lambda)$   in \eqref{21maggio2021-4} with $\mathcal{T}$ as in point 3.   satisfies  
   $$
  dY_\nu=\omega(z,\lambda) Y_\nu
  ,
  $$ 
 so that   system \eqref{27marzo2021-2} is strongly isomonodromic. 
  
  \end{itemize}
  \end{shaded}

 \eth

\bre
{\rm  
We can choose the solution $\mathcal{D}=0$ of \eqref{21ottobre2022-2}, and in this case $\mathcal{T}$  at point 3. is constant. 
}
\ere
\bre 
{\rm In the proof of Theorem \ref{20agosto2021-3}, we will see that for every $\nu\in\mathbb{Z}$ there exists a holomorphic matrix valued one-form  $K_\nu(\lambda)$ such that each $Y_\nu(z,\lambda)$ in \eqref{21maggio2021-4}, with $\mathcal{T}$ as at point 2., satisfies the system
 $$
  dY_\nu=\omega(z,\lambda) Y_\nu +Y_\nu K_\nu(\lambda);
  $$
 If   
   all the  blocks $A_{[k,k]}$, $k=1,...,s$ are non-resonant, then  we will see that $K_\nu=0$.}
\ere

\bcr
\label{4agosto2022-1}
\begin{shaded}
If $\Lambda$ has pairwise distinct eigenvalues, system \eqref{27marzo2021-2} is strongly isomonodromic if and only if   \eqref{21maggio2021-1}-\eqref{21maggio2021-2} are satisfied. 
\end{shaded}
\ecr
\begin{proof}[Proof of Corollary \ref{4agosto2022-1}]
At point 3. the non-resonance condition always holds if $\Lambda$ has pairwise distinct eigenvalues $\lambda_1,...,\lambda_n$, so that $A_{[k,k]}=A_{kk}$, $k=1,...,n$.
\end{proof}

\begin{proof}[Proof of Theorem \ref{20agosto2021-3}]  
{~}

\vskip 0.2 cm

1. A computation gives 
$$
\frac{\partial^2 A}{\partial \lambda_i\partial\lambda_j}
-
\frac{\partial^2 A}{\partial \lambda_j\partial\lambda_i}
\underset{\eqref{21maggio2021-2}}=\bigl[
\partial_i\widetilde{\omega}_j-\partial_j\widetilde{\omega}_i-[\widetilde{\omega}_i,\widetilde{\omega}_j],A
\bigr],\quad 1\leq i\neq j \leq s.
$$
From Lemma \ref{21ottobre2022-1}, if \eqref{21ottobre2022-2} holds, we receive $
\partial_i\widetilde{\omega}_j-\partial_j\widetilde{\omega}_i-[\widetilde{\omega}_i,\widetilde{\omega}_j]=0
$, which is \eqref{31marzo2021-10}. Therefore $\partial_i\partial_jA- \partial_j\partial_iA=0$, so that \eqref{21maggio2021-2} is integrable.

 \vskip 0.2 cm 
 2. Equations \eqref{1aprile2021-4} are  automatically satisfied by the structure of the  $\widetilde{\omega}_j$ as in \eqref{21maggio2021-7}. In 1. we have seen that  \eqref{31marzo2021-10} holds. 
Now, \eqref{1aprile2021-4}-\eqref{31marzo2021-10} and \eqref{21maggio2021-1}-\eqref{21maggio2021-2} are the Frobenius integrability conditions of \eqref{21maggio2021-8}. The last statement at point 2. is proved by Proposition \ref{28marzo2021-3}.

\vskip 0.2 cm 

3. By \eqref{21ottobre2022-2}, system \eqref{21maggio2021-10} is integrable and admits a holomorphic fundamental matrix solution, call it  $\mathfrak{T}(\lambda)$. 
Any fundamental matrix  solution is $T(\lambda)=\mathfrak{T}(\lambda)T_0$, for an invertible constant matrix $T_0$. 
Take any such  $T(\lambda)$. Using \eqref{21maggio2021-10}, we receive
 $$ 
 d(T^{-1} A^D T)= T^{-1} \Bigl(dA^D+[A^D,\mathcal{D}]\Bigr)T.
 $$
By Lemma \ref{21ottobre2022-1}, $dA^D+[A^D,\mathcal{D}]=0$. Consequently, 
$$ 
d(T^{-1} A^D T)=0.
$$
So, $T^{-1} A^D T$ is constant. Therefore, there is a choice of  $T_0$ such that 
$$ 
\mathcal{T}^{-1}(\lambda) A^D(\lambda) \mathcal{T}(\lambda)=J \quad \hbox{ Jordan constant}.
$$
Since $A^D$ is block diagonal, we can take a block diagonal  fundamental matrix solution $\mathfrak{T}(\lambda)=\mathfrak{T}_1(\lambda)\oplus \cdots \oplus \mathfrak{T}_s(\lambda)$. Therefore, also $T_0$ can be taken block diagonal, and so is $\mathcal{T}(\lambda)$.

\vskip 0.2 cm 

\begin{figure}
\centerline{\includegraphics[width=0.6\textwidth]{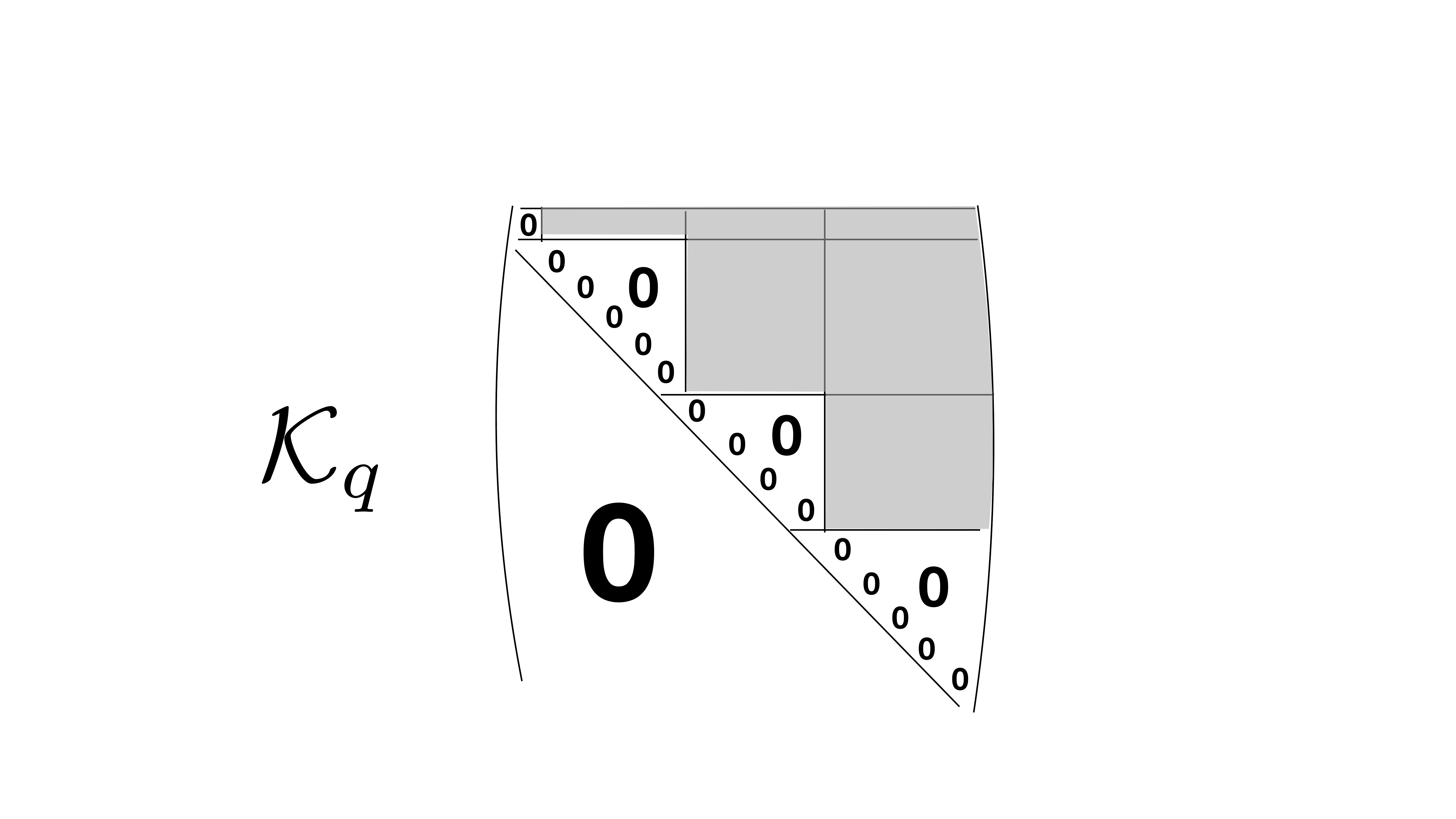}}
\caption{The structure of the sub-blocks $\mathcal{K}_q$ of $K^{(\nu)}_{[a,a]}$ corresponding to $\mathcal{N}_q$,  $\mathcal{D}_q$ of figure \ref{2maggio2021-4} and  $\Sigma_q=\sigma_q I_q$. They are split into sub-blocks.}
\label{22maggio2021-14}
\end{figure}

4. Since Assumption 3 holds by  point 3. and $J$ is constant,  there are fundamental matrices $Y_\nu(z,\lambda)$ as in \eqref{21maggio2021-4} satisfying 
$$ 
\frac{dY_\nu}{dz}=\left(\Lambda+\frac{A(\lambda)}{z}\right)Y_\nu, 
$$ 
which are holomorphic in $\mathcal{R}(\overline{\mathbb{C}}\backslash\{0,\infty\})\times \mathbb{D}$. Therefore, 
$$ 
\varphi_\nu(z,\lambda):=d_\lambda Y_\nu -\sum_{j=1}^s \bigl(zE_{p_j} + \widetilde{\omega}_j\bigr)d\lambda_j~ Y_\nu,
$$
is well defined, where $d_\lambda$ is the differential w.r.t. $\lambda_1,...,\lambda_s$. Using $d_\lambda \partial_z=\partial_z d_\lambda$ and \eqref{21maggio2021-2} we obtain 
$$ 
\frac{\partial \varphi_\nu}{\partial z}= \left(\Lambda+\frac{A(\lambda)}{z}\right) \varphi_\nu + \sum_{j=1}^s\Bigl(
\bigl[A,E_{p_j}]+[\Lambda,\widetilde{\omega}_j\bigr]
\Bigr) d\lambda_j ~Y_\nu.
$$
By \eqref{21maggio2021-1}, the above reduces to 
$$
\frac{\partial \varphi_\nu}{\partial z}= \left(\Lambda+\frac{A(\lambda)}{z}\right) \varphi_\nu
.
$$
Therefore, there is on $\mathbb{D}$ a holomorphic matrix 1-form  $K_\nu(\lambda)$ (not necessarily invertible) such that 
$$ 
\varphi_\nu=Y_\nu K_\nu.
$$
We rewrite \eqref{21maggio2021-4} with the specific Levelt form as in Section \ref{2maggio2021-1}:
$$
Y_\nu(z,\lambda)= \mathcal{T}(\lambda)\widehat{Y}_\nu(z,\lambda)z^{\Delta} z^{N(\lambda)} e^{\Lambda z}
.
$$
From point 2. (constancy of $J$)  we know already that $\Delta$ is constant. Nevertheless, we will indicate $d\Delta$ in the following, also if it is zero.  We 
compute the structure of 
$$ 
K_\nu= Y_\nu^{-1} d_\lambda Y_\nu - Y_\nu^{-1} \sum_{j=1}^s \bigl(zE_{p_j} + \widetilde{\omega}_j\bigr)d\lambda_j~ Y_\nu.
$$
We have
$$ 
Y_\nu^{-1} d_\lambda Y_\nu =
$$
{\small
$$
=e^{-\Lambda z}
\left(
z^{-N} z^{-\Delta} (\mathcal{T} \widehat{Y}_\nu)^{-1} d_\lambda( \mathcal{T} \widehat{Y}_\nu)
z^\Delta z^N+ z^{-N}   d\Delta ~ \ln z~ z^N + z^{-N} \sum_{k=1}^{\overline{k}} \frac{dN^k}{k!}( \ln z)^k z^N +z d\Lambda
\right)e^{\Lambda z}
.
$$
}
Notice that $d\Lambda= \sum_j E_{p_j} d\lambda_j$, so that  
\begin{align*}
e^{\Lambda z} K_\nu e^{-\Lambda z} 
=&
z^{-N} z^{-\Delta} \left[
 (\mathcal{T} \widehat{Y}_\nu)^{-1} \left(
 d_\lambda (\mathcal{T} \widehat{Y}_\nu) -\sum_j (zE_{p_j} +\widetilde{\omega}_j) d\lambda_j ~ \mathcal{T} \widehat{Y}_\nu
 \right)\right]z^\Delta z^N
\\
&+
z^{-N} \left( d\Delta \ln z +\sum_{k=1}^{\overline{k}} \frac{dN^k}{k!}( \ln z)^k \right)z^N +\sum_j zE_{p_j} d\lambda_j.
\end{align*}
Notice that $[\Delta,d\Delta]=0$ (also in case $d\Delta\neq 0$, because $\Delta$ is diagonal), and recall that  $[\Delta,\Lambda]=[\Delta,d\Lambda]=0$ and  $[N,\Lambda]=[N,d\Lambda]=0$, so that  
\begin{align*}
z^\Delta z^N e^{\Lambda z} K_\nu e^{-\Lambda z} z^{-N} z^{-\Delta}
=& (\mathcal{T} \widehat{Y}_\nu)^{-1} \left(
 d_\lambda (\mathcal{T} \widehat{Y}_\nu) -\sum_j (zE_{p_j} +\widetilde{\omega}_j) d\lambda_j ~ \mathcal{T} \widehat{Y}_\nu \right)+
 \\
 &+d\Delta \ln z + z^\Delta\left(
 \sum_{k=1}^{\overline{k}} \frac{dN^k}{k!}( \ln z)^k 
 \right) z^{-\Delta} + \sum_j zE_{p_j} d\lambda_j.
 \end{align*}
The asymptotic expansion \eqref{22maggio2021-2} holds in a sector $\mathcal{S}_\nu$  of amplitude greater than $\pi$. In such a sector, we obtain 
\begin{align*}
z^\Delta z^N e^{\Lambda z} K_\nu e^{-\Lambda z} z^{-N} z^{-\Delta}
=&- \sum_j zE_{p_j} d\lambda_j +\mathcal{T}^{-1}d\mathcal{T} +\mathcal{T}^{-1}\left(\sum_j [\mathcal{T}
F_1\mathcal{T}^{-1}, E_{p_j}]d\lambda_j\right)\mathcal{T} +
\\
&-\mathcal{T}^{-1}\left(\sum_j  {\omega}_jd\lambda_j \right)\mathcal{T}- \mathcal{T}^{-1}
\mathcal{D} \mathcal{T}+\sum_j zE_{p_j} d\lambda_j+
\\
&+d\Delta \ln z + z^\Delta\left(
 \sum_{k=1}^{\overline{k}} \frac{dN^k}{k!}( \ln z)^k 
 \right) z^{-\Delta} +O\left(\frac{1}{z}\right).
\end{align*}
In the computations in the proof of Theorem \ref{30marzo2021-2} we have seen that   $[\mathcal{T}
F_1\mathcal{T}^{-1}, E_{p_j}]= \omega_j$. By   point 2., $d\mathcal{T}=\mathcal{D}(\lambda) \mathcal{T}$. Therefore
$$
z^\Delta z^N e^{\Lambda z} K_\nu e^{-\Lambda z} z^{-N} z^{-\Delta}
=d\Delta \ln z + z^\Delta\left(
 \sum_{k=1}^{\overline{k}} \frac{dN^k}{k!}( \ln z)^k 
 \right) z^{-\Delta} +O\left(\frac{1}{z}\right).
$$
The off-diagonal blocks of the r.h.s. are of order $O(1/z)$, because $N$ and $\Delta$ are block-diagonal. The  l.h.s is 
$$ 
e^{(\lambda_a-\lambda_b)z}~z^{\Delta_a} z^{N_{[a,a]}} K^{(\nu)}_{[a,b]}  z^{-N_{[b,b]}}z^{-\Delta_b},
\quad
\quad
a\neq b =1,...,s.$$
Since $e^{(\lambda_a-\lambda_b)z}$ diverges exponentially in a subsector of $\mathcal{S}_\nu$, while the r.h.s. does not, necessarily 
 $$ K^{(\nu)}_{[a,b]} =0, \quad
\quad
a\neq b =1,...,s.
$$
The diagonal blocks are
$$ 
z^{\Delta_a} z^{N_{[a,a]}} K^{(\nu)}_{[a,a]}  z^{-N_{[a,a]}}z^{-\Delta_a}=
d\Delta_a \ln z + z^{\Delta_a}\left(
 \sum_{k=1}^{\overline{k}} \frac{dN_{[a,a]}^k}{k!}( \ln z)^k 
 \right) z^{-\Delta_a} +O\left(\frac{1}{z}\right).
$$
Namely
$$ 
z^{\Delta_a}  K^{(\nu)}_{[a,a]}z^{-\Delta_a}+ z^{\Delta_a}([N_{[a,a]},K^{(\nu)}_{[a,a]}]\ln z +\hbox{ terms in $(\ln z)^r$ with  $r\geq 2$})z^{-\Delta_a}
=
$$
$$
= d\Delta_a\ln z +z^{\Delta_a}( dN_{[a,a]}\ln z + \hbox{ terms in $(\ln z)^r$, $r\geq 2$})z^{-\Delta_a} +O(1/z).
$$
Therefore, it is necessary that 
\be
\label{22maggio2021-12}
z^{\Delta_a}  K^{(\nu)}_{[a,a]}z^{-\Delta_a}=O(1/z).
\ee
Recall that $\Delta=D+\Sigma$. Proceeding as in the Appendix, we see that $N_{[a,a]}$ has   diagonal block structure, and correspondingly so has $K_{[a,a]}^{(\nu)}$. Let for simplicity $\mathcal{K}:=K_{[a,a]}^{(\nu)}$ for a fixed $a$. Then $\mathcal{K}=\mathcal{K}_1\oplus\cdots\oplus \mathcal{K}_\ell$, as in the Appendix. Since the integer diagonal entries in $\mathcal{D}_q$, $q=1,...,\ell$ are an increasing sequence as in figure \ref{2maggio2021-4},  each diagonal sub-block $\mathcal{K}_q$ has structure as in figure \ref{22maggio2021-14}. 

In case of no resonance in $A_{[a,a]}$, then $\mathcal{D}_q= d_q I_q$ in figure \ref{2maggio2021-4}, where $d_q$ is integer  and $I_q$ is an identity matrix of suitable dimension. Thus, the diagonal  sub-blocks of  \eqref{22maggio2021-12} reduce to 
$$ 
  \mathcal{K}_q=O(1/z),\quad q=1,...,\ell.
  $$ 
  This implies that $ K^{(\nu)}_{[a,a]}=0.$
\end{proof}

\section{The 3-dimensional case}
\label{1settembre2021-1}

Let  $n=3$. 
The isomonodromy problem  of the case with no coalescences, namely $\Lambda=\hbox{\rm diag}((\lambda_1,\lambda_2,\lambda_3)$, $\lambda_i\neq \lambda_j$, is highly transcendental. For example, for a certain class of matrices $A(\lambda)$, the isomonodromy deformation equations \eqref{21agosto2021-1} are equivalent to the sixth Painlev\'e equation \cite{Hard,Dub2,MazzoIrr,Boalch1,DG}. Coalescences of  pairs  of eigenvalues correspond to  the   fixed singularities  of the sixth Painlev\'e equation.

 The opposite situation is the trivial case  $\Lambda=\hbox{\rm diag}(\lambda_1,\lambda_1,\lambda_1)$. All the fundamental matrix solutions of system \eqref{27marzo2021-2} are 
$$ 
Y(z,\lambda_1)=e^{z\lambda_1} z^{A(\lambda_1)} C(\lambda_1),\quad \det C(\lambda_1)\neq 0.
$$
The system is isomonodromic if and only if $A$ is constant. 

The only non-trivial case  we need to consider along a stratum of the coalescence locus is, up to permutation,  
\be
\label{12maggio2022-5}
\frac{dY}{dz}=\left(
\begin{pmatrix} \lambda_1 & 0 & 0 
\\
0 & \lambda_2 & 0
\\
0 & 0 & \lambda_2
\end{pmatrix} +
\frac{A}{z}
\right)Y.
\ee
The gauge $Y=e^{\lambda_2 z} \widetilde{Y} $ and the change of variable $\zeta=x z$, where $x:=\lambda_1-\lambda_2$,  yield 
\be
\label{12maggio2022-6} 
\frac{d\widetilde{Y}}{d\zeta}=\left(
\begin{pmatrix} 1 & 0 & 0 
\\
0 & 0 & 0
\\
0 & 0 & 0
\end{pmatrix} +
\frac{A}{\zeta}
\right)\widetilde{Y}.
\ee
Suppose that $A$ is   constant. Then the above admits fundamental matrix solutions with constant essential monodromy data. Thus, the fundamental matrix solutions  of the starting system also have constant data, if $x$ varies in a sufficiently small domain away from $x=0$. Therefore, the starting system is strongly isomonodromic.  

We can obtain this result  also from the point of view of Theorem \ref{30marzo2021-2}. Since the diagonal blocks  $A_{[1,1]}\equiv A_{11}$ and $A_{[2,2]}$  are always constant in the isomonodromic case, we are allowed to consider an isomonodromic deformation with 
 $$ 
 \mathcal{T}(\lambda)=\mathcal{T}_0 \mathfrak{B}(\lambda), 
\quad\quad 
 \mathfrak{B}(\lambda)=\mathfrak{B}_1(\lambda)\oplus \mathfrak{B}_2(\lambda),
\quad \quad [\mathfrak{B}_j(\lambda),J_j]=0,\quad j=1,2,
$$
where   $\mathcal{T}_0 $ is constant. We can also choose $\mathfrak{B}(\lambda)\equiv \mathfrak{B}(\lambda_1-\lambda_2)$. This implies that 
$$ 
\frac{\partial \mathcal{T}}{\partial \lambda_2}= - \frac{\partial \mathcal{T}}{\partial \lambda_1}.
$$
 Moreover
$$ 
\omega_1=\frac{1}{\lambda_1-\lambda_2} \begin{pmatrix}
0 & A_{12} & A_{13}
\\
A_{21} & 0 & 0 
\\
A_{31} & 0 & 0 
\end{pmatrix},\quad \quad \omega_2=-\omega_1.
$$
Since $\widetilde{\omega}_1+\widetilde{\omega}_2=0$, the same arguments of Remark \ref{1settembre2021-2} imply that 
$$ 
A=A(x),\quad \quad   x=\lambda_1-\lambda_2,
$$
so that $\omega_j=\omega_j(x)$. 
Therefore, the gauge $Y=e^{\lambda_2 z} \widetilde{Y} $ transforms  the Pfaffian system \eqref{30marzo2021-1} into 
$$
d\widetilde{Y}=\left\{
\left[\begin{pmatrix} x & 0 & 0 
\\
0 & 0 & 0
\\
0 & 0 & 0
\end{pmatrix}+
\frac{A}{z}\right]dz
+
\Bigl(zE_1+\widetilde{\omega}_1(x)\Bigr)dx
\right\}
\widetilde{Y}
,
\quad\quad \widetilde{\omega}_1(x)=\omega_1(x)+\frac{d\mathcal{T}(x)}{dx}\mathcal{T}(x)^{-1},
$$
where $E_1=\hbox{\rm diag}(1,0,0)$. 
 Its integrability condition \eqref{20agosto2021-6}   reduces to 
\be
\label{1settembre2021-4}
\frac{dA}{dx}=\Bigl[\omega_1(x)+\frac{d\mathcal{T}(x)}{dx}\mathcal{T}(x)^{-1}~,~A\Bigr] .
\ee
Now, a suitable choice of $\mathcal{T}(x)$  can be made such that 
\be
\label{1settembre2021-3}
\Bigl[\omega_1(x)+\frac{d\mathcal{T}(x)}{dx}\mathcal{T}(x)^{-1}~,~A\Bigr]=0.
\ee
This choice of $\mathcal{T}$ is   obtained by setting $A=A_0$ constant, calculating by linear algebra a constant $\mathcal{T}_0$ which Jordanizes $A_{11}^{(0)}\oplus A_{[2,2]}^{(0)}$:
$$
\mathcal{T}_0^{-1}\bigl(A_{11}^{(0)}\oplus A_{[2,2]}^{(0)}\bigr)\mathcal{T}_0=J_1\oplus J_2,
$$
  and the general $\mathfrak{B}(x)$ such that   $[\mathfrak{B}_j(x),J_j]=0$. The so obtained  $ \mathcal{T}(x)=\mathcal{T}_0 \mathfrak{B}(x)$ must be  substituted into \eqref{1settembre2021-3}, which can  be solved as a system of differential equations  for the entries of $\mathfrak{B}(x)$. 

\vskip 0.2 cm 
\noindent
{\bf Example.} We consider for simplicity matrices of the form 
$$ 
A=\begin{pmatrix}
0 & A_{12} & A_{13}
\\
A_{21} & 0 & A_{23}
\\
A_{31} & A_{32} & 0
\end{pmatrix}.
$$
In the isomonodromic case, $A_{23}$ and $A_{32}$ are constant, due to constancy of the diagonal blocks of $A$. 
To simplify computations, we  suppose that $A_{23}A_{32}\neq 0$, so that the block $A_{[2,2]}$ is diagonalizable, with eigenvalues $\pm \sqrt{A_{32}A_{23}} $.   Thus, the general form of $\mathcal{T}(x)$ is 
\be
\label{13ottobre2021-2}
\mathcal{T}(x)
=
\underbrace{
\begin{pmatrix}
1 & 0 & 0
\\
0 & \frac{A_{23}}{\sqrt{A_{23}A_{32}} }& - \frac{A_{23}}{\sqrt{A_{23}A_{32}}}
\\
0 & 1& 1
\end{pmatrix}
}_{=:~\mathcal{T}_0^{\rm particular}}
\cdot
\underbrace{
 \begin{pmatrix}
a(x) & 0 & 0
\\
0 & b(x)&0
\\
0 & 0& c(x)
\end{pmatrix}
}_{\mathfrak{B}(x)},
\ee
where $ \mathcal{T}_0^{\rm particular}$ is a particular choice for a constant  matrix diagonalizing the block-diagonal part of $A$. 
Equation \eqref{1settembre2021-3} has solution 
\be
\label{13ottobre2021-3}
b(x) = b_0x^\rho a(x),\quad c(x) = c_0 x^{-\rho }a(x), \quad\quad b_0,c_0\in\mathbb{C}\backslash\{0\},\quad \rho:=\sqrt{A_{32}A_{23}}.
\ee
With this choice of $\mathcal{T}(x)$, equation \eqref{1settembre2021-4} becomes 
\be 
\label{13ottobre2021-5}
\frac{dA}{dx}=0\quad \Longrightarrow \quad A \hbox{ is constant}.
\ee
Notice that this introduces the integration constants $A_{12},A_{13},A_{21},A_{31}$.
If instead  we  choose  a constant $\mathcal{T}$, namely 
\be
\label{13ottobre2021-4}
\mathcal{T}=\mathcal{T}_0:=\begin{pmatrix}
1 & 0 & 0
\\
0 & \frac{A_{23}}{\sqrt{A_{23}A_{32}} }& - \frac{A_{23}}{\sqrt{A_{23}A_{32}}}
\\
0 & 1& 1
\end{pmatrix}\cdot \begin{pmatrix}
a_0 & 0 & 0
\\
0 & b_0&0
\\
0 & 0& c_0
\end{pmatrix}\quad\hbox{constant},
\ee
  then equation \eqref{1settembre2021-4} has solution given by a non-constant $A(x)$. Indeed, now $A$ must satisfy  
$$ 
\frac{dA(x)}{dx}=[\omega_1(x),A(x)]\equiv  \begin{pmatrix}
0 & \frac{A_{32}}{x}A_{13}(x) & \frac{A_{23}}{x}A_{12}(x) 
\\
- \frac{A_{23}}{x}A_{31}(x) & 0&0
\\
-\frac{A_{32}}{x}A_{21}(x) & 0& 0
\end{pmatrix}
.
$$
Since $A_{23}$ and $A_{32}$ are  constant, the above is a linear system with Fuchsian singularity at $x=0$, with general solution 
\begin{align*}
& A_{12}(x) = c_1x^{\rho} + c_2x^{-\rho}, & A_{13}(x) = \sqrt{\frac{A_{23}}{A_{32}}}(c_1x^\rho - c_2x^{-\rho}),
\\
& A_{21}(x) = c_3x^{\rho} + c_4x^{-\rho},
& 
A_{31}(x) = -\sqrt{\frac{A_{32}}{A_{23}}}(c_3x^\rho - c_4x^{-\rho}).
\end{align*}
Here, $c_1,c_2,c_3,c_4$ are the integration constants. 
It is a computation to check that the above $A(x)$ has structure
$$ 
A(x)=\mathcal{T}_0\bigl( \mathcal{T}^{-1}(x) A_0 \mathcal{T}(x)\bigr)\mathcal{T}_0^{-1},
$$
as predicted by the discussion leading to \eqref{13ottobre2021-1}, where $A_0$ is a constant, i.e. a solution of  \eqref{13ottobre2021-5}, $\mathcal{T}_0$ is \eqref{13ottobre2021-4} and  $ \mathcal{T}(x)$ is \eqref{13ottobre2021-2} with the functions \eqref{13ottobre2021-3}.

\bre{\rm 
 There is a precise correspondence, that we describe below,  between constant monodromy data of \eqref{12maggio2022-5} and of \eqref{12maggio2022-6}.  We consider the case when  $A$ is {\it constant}. This  can always be achieved by a gauge transformation, as explained before.  In order to simplify computations, we further assume that $A$ is diagonalizable and  non-resonant, so that \eqref{27marzo2021-3} is
$$
Y^{(0)}(z,\lambda)= G^{(0)}(x) \Bigl(I+\sum_{k=1}^\infty F_k^{(0)} (\lambda) z^k\Bigr)  z^{J^{(0)}},\quad J^{(0)}=\mathrm{diag}(\mu_1,\mu_2,\mu_3),
$$
and assume that  $A_{[2,2]}$ is  non-resonant (i.e $\pm \sqrt{A_{32}A_{23}} $ not a half integer), so that \eqref{21maggio2021-3} is 
$$
 Y_F(z,\lambda)= \mathcal{T}(x) \Bigl( I+\sum_{j=1}^\infty F_j(\lambda) z^{-j}\Bigr) z^{\mathrm{diag}(0,~\rho,-\rho)} e^{\Lambda z}.
$$
 Take $\mathcal{T}(x)$ in \eqref{13ottobre2021-2},   where $\mathcal{T}_0^{\rm particular}=\mathcal{T}_0$ has structure \eqref{13ottobre2021-4} and  
 $ \mathcal{B}(x)=\mathrm{diag}(1,x^\rho,x^{-\rho})
 $. 
  Then, $dG=\sum_{j=1}^2(\omega_j+\mathcal{D}_j)d\lambda_j G$ in  Theorem \ref{27marzo2021-5},  reduces to 
$$ \frac{dG}{dx}=\Bigl(\omega_1(x)+\frac{d\mathcal{T}(x)}{dx}\mathcal{T}(x)^{-1}\Bigr)G\equiv~ \frac{A}{x}~G, 
$$
so that 
$$ G^{(0)}(x)=G_0^{(0)} x^{J^{(0)}},\quad\quad \hbox{ with constant $G_0^{(0)}$ such that }  (G_0^{(0)})^{-1}AG_0^{(0)}=J^{(0)}.
$$
 Then, it is easy to check that the essential monodromy data  of system \eqref{12maggio2022-5} with the solutions $Y^{(0)}(z,\lambda)$, $ Y_F(z,u)$  above and an admissible direction $\arg z=\tau$ are the same data of the system  \eqref{12maggio2022-6} with admissible direction $\arg \zeta=\tau+\arg(\lambda_1-\lambda_2)$, relative to the fundamental matrix solutions 
$$
\widetilde{Y}^{(0)}(\zeta)=G_0^{(0)}\Bigl(I+\sum_{k=1}^\infty \widetilde{F}_k^{(0)} 
\zeta^k\Bigr) \zeta^{J^{(0)}},
\quad\quad
\widetilde{Y}_F(\zeta)=\mathcal{T}_0 \Bigl( I+\sum_{j=1}^\infty \widetilde{F}_j\zeta^{-j}\Bigr) \zeta^{\mathrm{diag}(0,~\rho,-\rho)} e^{ \zeta\cdot \mathrm{diag}(1,0,0)},
$$
where the matrices $\widetilde{F}_k^{(0)}$, $\widetilde{F}_j$ are constant.
}
\ere


\section{A few words on  the 4-dimensional case}
\label{20settembre2022-9}

The 3-dimensional case is rigid, namely one can reduce to a constant $A$. The first non-trivial isomonodromic deformations occur with dimension 4, and are already highly transcendental. Apart from the trivial situation when $\Lambda$ ony has one eigenvalue and $A$ is necessarily constant, the cases we need to consider are (up to permutation)
$$\begin{aligned}
\hbox{Case (1): }\quad& \Lambda=\mathrm{diag}(\underbrace{\lambda_1,\lambda_1},~\lambda_2,~\lambda_3)
\\
\hbox{Case (2): }\quad& \Lambda=\mathrm{diag}(\underbrace{\lambda_1,\lambda_1,\lambda_1},~\lambda_2)
\\
\hbox{Case (3): }\quad& \Lambda=\mathrm{diag}(\underbrace{\lambda_1,\lambda_1},~\underbrace{\lambda_2,\lambda_2})
\end{aligned}
$$

It can be proved analogously to the 3-dimensional case that cases (2) and (3) are rigid. Heuristically, we can see this by the gauge transformation $Y=e^{\lambda_1 z} \widetilde{Y} $ and the change of variable $\zeta=(\lambda_2-\lambda_1) z$, which respectively yield in the two cases 
$$ 
\frac{d\widetilde{Y}}{d\zeta}=\left(
\mathrm{diag}(0,0,0,1) +
\frac{A}{\zeta}
\right)\widetilde{Y}\quad \hbox{ and } 
\quad 
\frac{d\widetilde{Y}}{d\zeta}=\left(
\mathrm{diag}(0,0,1,1) +
\frac{A}{\zeta}
\right)\widetilde{Y}.
$$

 Case (1) is non-trivial. Heuristically, we understand this by the gauge transformation $Y=e^{\lambda_1 z} \widetilde{Y} $ 
and the change of variable $\zeta=(\lambda_3-\lambda_1) z$, which yield
\be
\label{20settembre2022-3}
\frac{d\widetilde{Y}}{d\zeta}=\left(
\mathrm{diag}(0,0,x,1) +
\frac{A}{\zeta}
\right)\widetilde{Y}
,\quad \quad x:= \frac{\lambda_2-\lambda_1}{\lambda_3-\lambda_1}.
\ee
The matrix $A(\lambda)$ must satisfy the isomonodromy  deformation equations \eqref{21agosto2021-1}, which we show below to be  highly transcendental.  

To simplify the treatment, we consider a case when  $A$ is  much simpler than the general  situation. Up to a gauge transformation,   we  can  assume that  
$$\widetilde{\omega}_j=\omega_j \quad \hbox{for all $j$}.
$$ 
In general, for every $n$ and $s$, we have seen in Section \ref{21settembre2022-1} that for an isomonodromic 
system the  relations  $[\Lambda,\omega_j]=[E_{p_j},A]$, $j=1,...,s$, hold. This, together with our assumption that $\omega^{(j)}_{[k,k]}=0$, $ \forall 1\leq j,k\leq s$, yields 
$$\sum_{j=1}^s \omega_j=0, \quad \quad 
 \sum_{j=1}^s \lambda_j \omega^{(j)}_{[k,\ell]}=A_{[k,\ell]}\delta_{k\ell},\quad \forall 1\leq k,\ell \leq s.
$$
Hence,  the isomonodromy deformation equations  \eqref{21agosto2021-1} imply
\be
\label{20settembre2022-1} 
\sum_{j=1}^s \frac{\partial A}{\partial \lambda_j}=0,
\quad\quad 
\sum_{j=1}^s \lambda_j\frac{\partial A_{[k,\ell]}}{\partial \lambda_j}=A_{[k,\ell]}A_{[\ell,\ell]}-A_{[k,k]}A_{[k,\ell]},  \quad \forall 1\leq k,\ell \leq s.
\ee
Successively, we assume that 
$$ 
A_{[j,j]}=0 \quad  \hbox{for all } j=1,...,s, 
$$
so that \eqref{20settembre2022-1} becomes 
\be
\label{20settembre2022-2} 
\sum_{j=1}^s \frac{\partial A}{\partial \lambda_j}=0,\quad
\sum_{j=1}^s \lambda_j\frac{\partial A_{[k,\ell]}}{\partial \lambda_j}=0\quad \forall 1\leq k,\ell \leq s.
\ee
The above implies the functional dependence 
$$ 
A=A\left(\left\{\frac{\lambda_j-\lambda_1}{\lambda_s-\lambda_1}\right\}_{j=2}^{s-1}\right).
$$
In case (1), $s=3$ and $n=4$, the functional dependence reduces to 
$$
A=A(x), \quad\quad x= \frac{\lambda_2-\lambda_1}{\lambda_3-\lambda_1}.
$$
The deformation equations \eqref{21agosto2021-1} reduce to 
\be
\label{20settembre2022-5}
\frac{dA}{dx}=[\hat{\omega}_2,A],
\ee
 where $\hat{\omega}_2=\hat{\omega}_2(x)$ is 
$$ 
\hat{\omega}_2=\left[
\begin{array}{ccc}
\boldsymbol{0}_2 & \hat{\omega}^{(2)}_{[1,2]} & \vec{0}
\\
\noalign{\medskip}
 \hat{\omega}^{(2)}_{[2,1]} & 0 &  \hat{\omega}^{(2)}_{[2,3]}
\\
\noalign{\medskip}
\vec{0}^{~\mathrm{T}} & \hat{\omega}^{(2)}_{[3,2]} & 0
\end{array}
\right],
$$
and 
$$
\vec{0}=\left[
\begin{array}{c}
0
\\
0
\end{array}
\right],\quad \vec{0}^{~\mathrm{T}} =[0~0],\quad \boldsymbol{0}_2:=\left[
\begin{array}{cc}
0& 0
\\
0& 0
\end{array}
\right],
$$
$$ 
 \hat{\omega}^{(2)}_{[1,2]}=\frac{A_{[1,2]}}{x},\quad  \hat{\omega}^{(2)}_{[2,1]}=\frac{A_{[2,1]}}{x},
 \quad
  \hat{\omega}^{(2)}_{[2,3]}=\frac{A_{[2,3]}}{x-1},\quad  \hat{\omega}^{(2)}_{[3,2]}=\frac{A_{[3,2]}}{x-1},
  $$
  More explicitly,  equations  \eqref{20settembre2022-5} are 
    \be
  \label{20settembre2022-7} 
  \left\{
  \begin{aligned}
&  \frac{d A_{[1,2]}}{d x}=\frac{A_{[1,3]}A_{[3,2]}}{1-x}, &   \quad \frac{d A_{[2,1]}}{d x}=\frac{A_{[3,1]}A_{[2,3]}}{x-1},
\\
\noalign{\medskip}
&  \frac{d A_{[2,3]}}{d x}=\frac{A_{[2,1]}A_{[1,3]}}{x} ,& \quad \frac{d A_{[3,2]}}{d x}=\frac{A_{[3,1]}A_{[1,2]}}{-x},
\\
\noalign{\medskip}
&  \frac{d A_{[1,3]}}{d x}=\frac{A_{[1,2]}A_{[2,3]}}{x(1-x)} ,& \quad \frac{d A_{[3,1]}}{d x}=\frac{A_{[2,1]}A_{[3,2]}}{x(x-1)}
  \end{aligned}
  \right.
  \ee
  where\footnote{Matrices are partitioned into the blocks inherited form $\Lambda=\mathrm{diag}(\lambda_1,\lambda_1,~\lambda_2,~\lambda_3)=\lambda_1 I_2\oplus \lambda_2 I_1\oplus \lambda_3 I_1$.} 
$$
\begin{aligned}
& 
A_{[1,2]}=\left[
\begin{array}{c}
A_{13}
\\
A_{23}
\end{array}
\right],
&
 A_{[1,3]}=\left[
\begin{array}{c}
A_{14}
\\
A_{24}
\end{array}
\right],
\\
\noalign{\medskip}
 &
 A_{[2,1]}=[A_{31}~A_{32}], & A_{[2,3]}=A_{34},
 \\
 &A_{[3,1]}=[A_{41}~A_{42}], &A_{[3,3]}=A_{43},
\end{aligned}
$$

  Suppose now that $A$ is skew-symmetric, so that we write it as 
  $$
  A
  =
  \begin{pmatrix}
  0 & 0 & \phi_1 & \phi_3
  \\
  0 & 0 & \phi_2 & \phi_4 
  \\
  0 &0  & 0 & \phi_5 
  \\
  0 & 0 & 0 & 0
  \end{pmatrix}- \begin{pmatrix}
  0 & 0 & \phi_1 & \phi_3
  \\
  0 & 0 & \phi_2 & \phi_4 
  \\
  0 &0  & 0 & \phi_5 
  \\
  0 & 0 & 0 & 0
  \end{pmatrix}^{\mathrm{T}}.
  $$
  Thus, system  \eqref{20settembre2022-7} becomes
  $$ 
  \left\{
  \begin{aligned}
  & \frac{d\phi_1}{dx}=\frac{\phi_3\phi_5}{1-x}, &  \frac{d\phi_3}{dx}=\frac{\phi_1\phi_5}{x(1-x)},
  \\
  \noalign{\medskip}
  & \frac{d\phi_2}{dx}=\frac{\phi_4\phi_5}{1-x}, &  \frac{d\phi_4}{dx}=\frac{\phi_2\phi_5}{x(1-x)},
  \\
  \noalign{\medskip}
  &  \frac{d\phi_5}{dx}=-\frac{\phi_1\phi_3+\phi_2\phi_4}{x}
  \end{aligned}
  \right.
  $$
  If we further restrict to the simpler case $\phi_1=\phi_2$ and $\phi_3=\phi_4$, we receive the three equations
  $$ 
  \left\{
  \begin{aligned}
  & \frac{d\phi_1}{dx}=\frac{\phi_3\phi_5}{1-x}, 
  &  \frac{d\phi_3}{dx}=\frac{\phi_1\phi_5}{x(1-x)},   \\
  \noalign{\medskip}
  &  \frac{d\phi_5}{dx}=-\frac{2\phi_1\phi_3}{x}.
  \end{aligned}
  \right.
  $$
Setting $\Omega_1:= -i \phi_5$, $\Omega_2:=\sqrt{2}\phi_1$, $\Omega_3:=i\sqrt{2} \phi_3$, the above system becomes
\be
\label{20settembre2022-8}
 \frac{d\Omega_1}{dx}=\frac{\Omega_2\Omega_3}{x}, 
\quad\quad
  \frac{d\Omega_2}{dx}=\frac{\Omega_1\Omega_3}{1-x},   
\quad\quad
  \frac{d\Omega_3}{dx}=\frac{\Omega_1\Omega_2}{x(x-1)}.
  \ee
 This system appears in \cite{Dub2,guzz2001}, where it is proved to be  equivalent to the sixth Painlev\'e equation\footnote{In classical form  $$
\frac{d^2y }{ dx^2}=\frac{1}{ 2}\left[ 
\frac{1}{ y}+\frac{1}{ y-1}+\frac{1}{ y-x}
\right]
           \left(\frac{dy}{ dx}\right)^2
-\left[
\frac{1}{ x}+\frac{1}{ x-1}+\frac{1}{ y-x}
\right]\frac{dy }{ dx}
$$
$$
+
\frac{y(y-1)(y-x)}{ x^2 (x-1)^2}
\left[
\alpha+\beta \frac{x}{ y^2} + \gamma \frac{x-1}{ (y-1)^2} +\delta
\frac{x(x-1)}{ (y-x)^2}
\right]
$$ 
} with parameters 
   $\alpha\in\mathbb{C}$, $\beta=\gamma=0$, $\delta=1/2$. 
   
   We conclude that already in the very simplified case with $A$ skew-symmetric depending on only three independent entries  $\phi_1,\phi_3,\phi_5$, the isomonodromy problem is as  transcendental as the Painlev\'e equations.   
  

\section{Example: the Caustic of a Dubrovin-Frobenius manifold in a generic case}
\label{3agosto2022-1}

 In this section, we give an important example where the theory of non generic  isomonodromy deformations along a stratum of a coalescence locus applies. It concerns the {\it caustic} of a {\it semisimple Dubrovin-Frobenius manifolds}. The result of this section is to  show  that  our Theorems \ref{30marzo2021-2} and  \ref{20agosto2021-3} are realized at generic points of the caustic.
 
  We cannot explain here  the details of  Dubrovin-Frobenius manifolds, introduced by B. Dubrovin to give a geometrical formulation of 2-D topological field theories. The reader is referred to \cite{Dub1,Dub2}. For the geometry  of the  caustic, we refer to \cite{Her,Str1,Str2} and in particular to \cite{Reyes}, in whose setting we will work. 
 
In a Dubrovin-Frobenius manifold $M$ of dimension $n$, the tangent spaces $T_pM$  at each point $p\in M$ is a Frobenius algebra with multiplication $\circ$, analytically depending on  $p$. If this algebra is  semisimple\footnote{A point $p\in M$ is a semisimple if  there are no {\it nilpotent} vectors  $\boldsymbol{v}$ in $T_pM$, namely no vectors  such that  $\boldsymbol{v}^{\circ m}=0$, for some integer $m$.} on an open dense subset of $M$, the manifold is called semisimple.  The subset  $\mathcal{K}\subset M$ where $\circ$ is not semisimple is called {\it caustic}. 
 This is empty or a hypersurface \cite{Her}. A flat metric $\eta$ is defined\footnote{We mean a symmetric non-degenerate bilinear form, non necessarily positive definite.} on $M$, compatible with the product: 
 \be
 \label{8agosto2022-5}
 \eta(\boldsymbol{u}\circ\boldsymbol{v},\boldsymbol{w})=
 \eta(\boldsymbol{u},\boldsymbol{v}\circ\boldsymbol{w}),
 \ee
 for any vector fields $\boldsymbol{u},\boldsymbol{v},\boldsymbol{w}$.  
 
  At semisimple points there is a basis  of idempotent vector fields $\pi_1,...,\pi_n$,  such that $\pi_i\circ \pi_j=\delta_{ij}\pi$, which are orthogonal  with respect to $\eta$.  
  They commute, $[\pi_i,\pi_j]=0$, so that locally there are coordinates $u=(u_1,...,u_n)$ such that each $\pi_j=\frac{\partial}{\partial u_j}$  (with abuse of notation). 
 They are called {\it canonical coordinates}, being  uniquely determined (up to permutation) as  eigenvalues of a multiplication operator $E\circ$, where $E$ is a preferred global {\it Euler} vector field of weight 1. This is a field satisfying $  \mathrm{Lie}_E(\boldsymbol{u}\circ\boldsymbol{v})-\mathrm{Lie}_E(\boldsymbol{u})\circ \boldsymbol{v}-\boldsymbol{u}\circ \mathrm{Lie}_E(\boldsymbol{v})
 =\boldsymbol{u}\circ\boldsymbol{v}$  (in short notation, $\mathrm{Lie}_E~\circ=\circ$) and $\mathrm{Lie}_E(\eta)=(2-d)\eta$, where $d\in\mathbb{C}$. 
 
 \vskip 0.2 cm 
 A sufficient condition for a point to be semisimple is that  $u_a\neq u_b$ for 
 all $1\leq a\neq b\leq n$. 
\vskip 0.2 cm

 A Frobenius manifold is essentially characterized by a  ``$z$-deformed'' connection $\widetilde{\nabla}$ defined by Dubrovin 
\cite{Dub1} as follows
\be
\label{5agosto2022-1}
 \widetilde{\nabla}_{\frac{d}{dz}}\boldsymbol{v}:=\frac{\partial \boldsymbol{v}}{\partial z} 
 +E\circ \boldsymbol{v}-\frac{1}{z}\hat{\mu}(\boldsymbol{v}),
   \quad
 \quad 
\widetilde{\nabla}_{\boldsymbol{u}}\boldsymbol{v} :={\nabla}_{\boldsymbol{u}}\boldsymbol{v} +z
 \boldsymbol{u}\circ \boldsymbol{z} 
 ,
 \quad\quad z\in\mathbb{C}^*:=\mathbb{C}\backslash\{0\}
,
\ee
 on the vector bundle $\pi^* TM\bigr|_{\mathbb{C}^*\times M}\longrightarrow \mathbb{C}^*\times M$, where $\pi: \mathbb{C}^*\times M \longrightarrow  M$ is the natural projection, and 
 $\boldsymbol{u},\boldsymbol{v}\in (\pi^*\mathscr T_M )(\mathbb{C}^*\times M)$.  
 Here $\hat{\mu}(\boldsymbol{v}):=(1-d/2)\boldsymbol{v}-\nabla_{\boldsymbol{v}}E$, and  $\nabla$ is the Levi-Civita connection of $\eta$.  
 
 Locally, at semisimple points where  all the $(u_1,...,u_n)$ are pairwise distinct, the flatness (the zero curvature condition) of the above connection  is  an integrable  Pfaffian system\footnote{ It is obtained considering the connection $\widetilde{\nabla}$  acting on the cotangent bundle, by looking at flat coordinates $\widetilde{t}$ satisfying 
 $\widetilde{\nabla}d\widetilde{t}=0$.
  System \eqref{4agosto2022-2} is the representation of $\widetilde{\nabla}d\widetilde{t}=0$ on the basis of normalized idempotents.},  which  Dubrovin writes as 
 \be
 \label{4agosto2022-2}
 d\mathscr{Y}=\left[\left(U+\frac{V(u)}{z}\right)dz+\sum_{j=1}^n \Bigl(z E_j +  V_j(u)\Bigr)du_j\right] \mathscr{Y},
 \ee 
  Here
  $$
  U=\mathrm{diag}(u_1,...,u_n),
  $$
  is the matrix representing $E\circ$ both on the basis $\pi_1,...,\pi_n$ and on the normalized basis 
 $$f_j:=\frac{\pi_j}{\eta_j},\quad \quad\eta_j:=\sqrt{\eta(\pi_j,\pi_j)}, \quad\quad j=1,...,n,
 $$
  (for a chosen sign of the square root), $V$ represents on the normalized basis the operator  $\hat{\mu}$ and is skew-symmetric
  $$\quad V^T=-V,
  $$
  $ E_j=\partial U/\partial u_j$ has all zero entries except for 1 in position $(j,j)$,  and $V_j$ has entries
   \be
 \label{4agosto2022-3} 
 V^{(j)}_{aa}(u)=0,\quad \quad\quad V^{(j)}_{ab}(u)
= 
\frac{V_{ab}(u)~(\delta_{aj}-\delta_{bj})}{u_a-u_b},\quad \quad a\neq b=1,...,n.
\ee
The above \eqref{4agosto2022-2} is a particular case of  \eqref{30marzo2021-1} when all eigenvalues remain pairwise distinct, and the matrices  $V_j$ are the analogous of our $\omega_j$. 
The  integrability of  \eqref{4agosto2022-2} is
 \be
 \label{4agosto2022-4} 
\frac{\partial V}{\partial u_j}=[V_j,V],\quad j=1,...,n,
\ee
   which is a particular case of the     isomonodromy  deformation equations \eqref{21maggio2021-2}. By  Corollary \ref{4agosto2022-1}, they are necessary and sufficient conditions  for the strong isomonodromy of the differential system
   \be
 \label{7giugno2022-2}
 \frac{d\mathscr{Y}}{dz}=\left(U+\frac{V(u)}{z}\right) \mathscr{Y}.
 \ee
 
 The local structure of the manifold can be explicitly constructed at semisimple points where all the $(u_1,...,u_n)$ are pairwise distinct, in terms of a fundamental matrix solution of \eqref{7giugno2022-2} in Levelt form at $z=0$ (see \cite{Dub1,Dub2,guzz2001}). 
 
  \vskip 0.2 cm
  
  In \cite{CDG1}, we have shown that the coalescence  of a pair of canonical coordinates $u_a-u_b\to 0$, for some $1\leq a\neq b\leq n$,  corresponds to a semisimple point of  $M$ 
  if and only if  $\lim_{u_a-u_b\to 0}V_{ab}(u)=0$ holomorphically.
    
  In general, for a pair $a\neq b\in\{1,...,n\}$, the condition $V_{ab}(u)\to 0$ is not satisfied along  solutions of \eqref{4agosto2022-4} when the corresponding $u_a-u_b\to 0$. Consequently,   the entries of  $V(u)$ may have a branching  at $u_a-u_b=0$, and may diverge along any direction $u_a-u_b\to 0$. Such coalescences of canonical coordinates may not correspond to a point of the manifold. In case they do, then the point must belong to the caustic. 

\vskip 0.3 cm 
Due to the singularity of $V(u)$ at coalescence points $u_a=u_b$ , the canonical coordinates are not the good ones to describe  the manifold close to the caustic.   
{\it The purpose of this section is to show that in  suitable local coordinates on $M$ the flatness of the connection \eqref{5agosto2022-1} is represented by a Pfaffian system which, restricted to the caustic, is a system  of type \eqref{30marzo2021-1}, satisfying  Theorem \ref{30marzo2021-2},  where $\Lambda$ has repeated eigenvalues and $\mathcal{T}$ is non trivial. Moreover, the above mentioned local coordinates, restricted at the caustic, are deformation parameters for a strong isomonodromy deformation described by our Theorems \ref{30marzo2021-2} and \ref{20agosto2021-3}.}
This purpose is realized in  Proposition \ref{6agosto2022-3} and Remark \ref{13agosto2022-2} below. 

\vskip 0.3 cm 

 A geometric study of the caustic of a Frobenius manifold has been done in \cite{Reyes}, under the assumptions that on an open dense subset of $\mathcal{K}$  there are are  $n-1$ idempotent vector fields, and the metric $\eta$ is non-degenerate on $\mathcal{K}$. These are the assumption we will also make here.  Such cases are realized in singularity theory, or equivalently for manifolds given by the orbit space of Coxeter groups (see \cite{Reyes}). 
 
 The points of $\mathcal{K}$ in the above dense  subset of \cite{Reyes} correspond to the coalescence of two canonical coordinates. Let $p\in\mathcal{K}$ belong to this subset and, without loss in generality, suppose that in a sufficiently small neighbourhood $\mathcal{B}\subset M$ of $p$ the caustic $\mathcal{K}$ is reached when 
 $$ 
 u_1-u_2\to 0.
 $$
 It has been shown in \cite{Reyes} that  theorems 2.11 and  4.7   (classification of 2-dimensional $F$ manifolds) of \cite{Her} allow to conclude that 
 in a neighbourhood of $p$ the germ of the manifold is (notations are borrowed  from and explained in \cite{Her})
 \be
 \label{5agosto2022-5}
 (M,p)=(I_2(m),p)\times \underbrace{(A_1,p)\times \dots \times (A_1,p)}_{n-2~\mathrm{times}}.
\ee 
These means the following. Local coordinates 
\be
\label{6agosto2022-1}
(t_1,t_2,u_3,...,u_n)
\ee
 are defined in a sufficiently small neighbourhood $\mathcal{B}\subset M$ of $p\in\mathcal{K}$, with  local basis of vector fields  $$ 
\underbrace{\frac{\partial}{\partial t_1},\quad \frac{\partial}{\partial t_2}}_{T_pI_2(m)},\quad 
\quad\underbrace{\pi_3=
 \frac{\partial}{\partial u_3},~\dots, ~ \pi_n=\frac{\partial}{\partial u_n}}_{\mathbb{C}^{n-2}},
 $$
 satisfying  the multiplication table
 \be
 \label{7giugno2022-1} 
 \left\{
 \begin{aligned}
 &
 \frac{\partial}{\partial t_1}\circ \frac{\partial}{\partial t_1}=\frac{\partial}{\partial t_1},\quad \frac{\partial}{\partial t_1}\circ \frac{\partial}{\partial t_2}=\frac{\partial}{\partial t_2},
\quad
 \frac{\partial}{\partial t_2}\circ\frac{\partial}{\partial t_2}=t_2^{m-2} \frac{\partial}{\partial t_1},
\\
 \noalign{\medskip}
&
\pi_i\circ \pi_j=\delta_{ij} \pi_i, 
\\
 \noalign{\medskip}
&
 \frac{\partial}{\partial t_1}\circ \pi_j= \frac{\partial}{\partial t_2}\circ \pi_j=0
 \end{aligned}
 \right. 
 \quad\quad 
 \left.
 \begin{aligned}
 & m\geq 2 ,\hbox{ integer}
 \\
 \\
 & i,j\geq 3.
 \end{aligned}
 \right.
 \ee
  The caustic $\mathcal{B}\cap\mathcal{K}$ around $p$ corresponds to 
$$t_2=0,
$$
and $\partial/\partial t_2$ is nilpotent at $\mathcal{B}\cap \mathcal{K}$. 

For points in $ \mathcal{B}\backslash \mathcal{K}$, the   idempotents $\pi_1,...,\pi_n$ and pairwise distinct canonical coordinates $(u_1,...,u_n)$ are well defined. The Euler vector field and the unit of $\circ$ on $\mathcal{B}\backslash\mathcal{K}$ are
 $$
 E=\sum_{j =1}^n u_j \frac{\partial}{\partial u_j},\quad \quad e= \pi_1+\pi_2+ \pi_3+\cdots+\pi_n.
 $$
  When $\mathcal{K}\cap \mathcal{B}$ is reached, then $u_1-u_2\to 0$ and  the good coordinates become \eqref{6agosto2022-1}, which  include $u_3,...,u_n$. On the whole $\mathcal{B}$, we can write  
  \cite{Reyes}
 \be
 \label{5agosto2022-3} 
 E= t_1\frac{\partial}{\partial t_1}+\frac{2}{m} t_2\frac{\partial}{\partial t_2}+\sum_{j\geq 3} u_j \frac{\partial}{\partial u_j},\quad \quad e= \frac{\partial}{\partial t_1}+ \pi_3+\cdots+\pi_n,
 \ee
 so that 
 $ \frac{\partial}{\partial t_1}=\pi_1+\pi_2
 $
  is well defined also at the caustic. Moreover,  $\pi_1+\pi_2,~\pi_3,...,\pi_n$ are the $n-1$ idempotents defined on $\mathcal{K}\cap \mathcal{B}$. Notice that $E\circ$ is diagonalizable on $\mathcal{K}\cap \mathcal{B}$, with a repeated eigenvalue
  $t_1\equiv u_1=u_2$ and pairwise distinct eigenvalues  $u_3,...,u_n$.
  
  \vskip 0.2 cm 
  So far we have reviewed the local  description of the manifold at a generic point of the caustic, according to \cite{Reyes}. We now show that the flatness of the Dubrovin deformed connection in the coordinates \eqref{6agosto2022-1} is realized at the caustic by a strongly isomonodromic theory obeying our Theorems \ref{30marzo2021-2} and  \ref{20agosto2021-3}. 
    \vskip 0.2 cm 
    
 Using  the multiplication table \eqref{7giugno2022-1}, $E\circ$ is  represented on the basis
 $$
\boldsymbol{w}_1:=\frac{\partial}{\partial t_1},\quad  \boldsymbol{w}_2:= \frac{\partial}{\partial t_2}, 
\quad \boldsymbol{w}_j:=f_j=\frac{\pi_j}{\sqrt{\eta_j}}\quad j\geq 3,
$$
 by a matrix
\be
\mathcal{U}=
 \begin{pmatrix}
 \widehat{\mathcal{U}} & 0 
 \\
 0 & U_{n-2}
 \end{pmatrix},
\ee
where
$$
  \widehat{\mathcal{U}} =
 \begin{pmatrix}
 t_1 & \dfrac{2}{m} t_2^{m-1}
 \\
 \noalign{\medskip}
  \dfrac{2}{m} t_2 & t_1
 \end{pmatrix},\quad\quad U_{n-2}=\mathrm{diag}(u_3,\dots,u_n).
$$
In $\mathcal{B}\backslash \mathcal{K}$, let  $\Psi$ be a matrix such that 
 $$ 
 \Psi \mathcal{U} \Psi^{-1}=U=\mathrm{diag}(u_1,u_2,\dots,u_n),
$$
 with normalization 
\be
\label{5agosto2022-2}
\Psi^\mathrm{T} \Psi=\Bigl(\eta(\boldsymbol{w}_\alpha,\boldsymbol{w}_\beta)\Bigr)_{\alpha,\beta=1}^n .
\ee
The eigenvalues $u_1,u_2$ of $\widehat{\mathcal{U}}$ are immediately computed:
\be
\label{8giugno2022-14} 
\left\{
\begin{aligned}
&u_1=t_1+\frac{2}{m} t_2 ^{m/2}
\\
\noalign{\medskip}
&u_2= t_1 -\frac{2}{m} t_2 ^{m/2}
\end{aligned}
\right.
\quad\Longrightarrow\quad
\left\{
\begin{aligned}
&t_1=\frac{u_1+u_2}{2},
\\
\noalign{\medskip}
&t_2= \left(\frac{m}{4}(u_1-u_2)\right)^{2/m}.
\end{aligned}
\right. 
\ee
Since system  \eqref{7giugno2022-2}  represents the $z$ component  of the deformed connection \eqref{5agosto2022-1} on the basis of normalized idempotents $f_1,...,f_n$ in $\mathcal{B}\backslash\mathcal{K}$, and $\mathcal{U}$ represents $E\circ$ on the basis $\boldsymbol{w}_1,...,\boldsymbol{w}_n$, then the gauge transformation 
 $$
 \mathscr{Y}=\Psi Y, 
 $$ 
  gives a differential system representing the $z$-component of \eqref{5agosto2022-1}  on the basis $\boldsymbol{w}_1,...,\boldsymbol{w}_n$:
 \be
 \label{6agosto2022-2} 
 \frac{d Y}{dz} =\left(
 \mathcal{U} +\frac{\mathcal{V}}{z}
 \right) Y, \quad \quad \mathcal{V}:=\Psi^{-1} V \Psi
 \ee
 In other words,  $\Psi$ gives the change of basis
\be
\label{7giugno2022-3} 
\Bigl[ \frac{\partial}{\partial t_1},\frac{\partial}{\partial t_2},f_3,\dots,f_n\Bigr]=
\bigl[f_1,f_2,f_3,\dots,f_n
\bigr]\Psi.
\ee
System \eqref{6agosto2022-2} is expressed in the good coordinates $(t_1,t_2,u_3,...,u_n)$.   The manifold structure is analytic at $\mathcal{K}$ in these coordinates. Therefore, the coefficients of  \eqref{6agosto2022-2} analytically extend from $\mathcal{B}\backslash \mathcal{K}$ to the whole
 $\mathcal{B}$, and in particular $\mathcal{V}(t_1,t_2,u_3,...,u_n)$  must be well defined and holomorphic  also at the caustic $t_2=0$. 

\vskip 0.2 cm 
 The matrix $\Psi$ diagonalizing $\mathcal{U}$ is partitioned into blocks as $\mathcal{U}$: 
  \be
 \label{5agosto2022-6} 
 \Psi=\begin{pmatrix}
 \widehat{\Psi}& 0 
 \\
 0 & I_{n-2}
 \end{pmatrix}
,\quad\quad  
\left\{\begin{aligned}
&\quad
 \widehat{\Psi} 
=
\begin{pmatrix}
\dfrac{a}{\sqrt{2}}t_2^{(2-m)/4} & \dfrac{a}{\sqrt{2}}t_2^{(m-2)/4}
\\
\noalign{\medskip}
\dfrac{ib}{\sqrt{2}} t_2^{(2-m)/4} & -\dfrac{ib}{\sqrt{2}}t_2^{(m-2)/4}
\end{pmatrix}
\\
\noalign{\medskip}
&\quad  I_{n-2} = \hbox{ $n-2$ dimensional identity matrix}
\\
\noalign{\medskip}
&
\hbox{ $\quad a$ and $b$ functions  of $(t_1,t_2,u_3,...,u_n)$}
\end{aligned}
\right.
\ee
The condition \eqref{5agosto2022-2} implies 
\be
\label{8giugno2022-1}
a^2= \tilde{\eta}_{12}+ t_2^{\frac{m-2}{2}}\tilde{\eta}_{11},
\quad\quad
b^2= \tilde{\eta}_{12}- t_2^{\frac{m-2}{2}}\tilde{\eta}_{11}.
\ee
where 
 $$\tilde{\eta}_{11}:=\eta\left(\frac{\partial }{\partial t_1},\frac{\partial }{\partial t_1} \right),
 \quad 
 \tilde{\eta}_{12}:=\eta\left(\frac{\partial }{\partial t_1},\frac{\partial }{\partial t_2} \right),
 \quad 
 \tilde{\eta}_{22}:=\eta\left(\frac{\partial }{\partial t_2},\frac{\partial }{\partial t_2} \right).
 $$
 Notice that $\tilde{\eta}_{11}$, $\tilde{\eta}_{22}$ are well defined at $t_2=0$, because so is the metric  $\eta$ and the manifold structure is analytic in the coordinates $(t_1,t_2,u_3,...,u_n)$. 
 
 By \eqref{5agosto2022-6}, 
$\mathcal{U}$ is not holomorphically diagonalizabe at $t_2=0$ for $m\geq 3$, in accordance with the fact that the change of coordinates $(u_1,u_2;~u_3,...,u_n)\mapsto(t_1,t_2;~u_3,...,u_n)$ is singular at $\mathcal{K}$  as in \eqref{8giugno2022-14}. For   $m=2$, $\mathcal{U}$ is  holomorphically diagonalizabe, which is an example of the  semisimple coalescence studied in \cite{CDG1}.

 \bre
 \label{5agosto2022-7}
 {\rm 
 By \eqref{8agosto2022-5},  \eqref{7giugno2022-1} and the unit $e$ in \eqref{5agosto2022-3}, we receive
 \be
 \label{8agosto2022-6}
 \tilde{\eta}_{11}=\eta_1+\eta_2,\quad \quad \tilde{\eta}_{22}=t_2^{m-2}\tilde{\eta}_{11}.
 \ee
 Let $m\geq 3$. The above \eqref{8agosto2022-6} implies that $
\tilde{\eta}_{22}\bigr|_{t_2=0}=0$, 
and thus
$$
\tilde{\eta}_{12}\bigr|_{t_2=0}\neq 0,
$$
otherwise the metric would be degenerate at $t_2=0$, which cannot be by assumption.  
Now let $m=2$. In case 
$ \tilde{\eta}_{12}\bigr|_{t_2=0}=0$, then $ \tilde{\eta}_{11}\bigr|_{t_2=0}=\tilde{\eta}_{22}\bigr|_{t_2=0}\neq 0$,  
otherwise the metric would be degenerate at $t_2=0$. 

 }
 \ere
 
Let  $\mathcal{V}_{[1,1]}$ be the $2\times2$  upper left block  of $\mathcal{V}$.
 A lengthy but elementary computation proves the implication 
 $$
 \left\{
 \begin{aligned}
 &V+V^T=0,
 \\
 \noalign{\medskip}
 &
 \mathcal{V}=\Psi^{-1} V \Psi 
 \quad\hbox{ and  \eqref{8giugno2022-1}}
 \end{aligned}
 \right.
 \quad\quad
 \Longrightarrow
 \quad 
 \mathcal{V}_{[1,1]}=
\frac{i V_{12}(u)}{\sqrt{ \tilde{\eta}_{12}^2-\tilde{\eta}_{11}^2~ t_2^{m-2}}}
\begin{pmatrix}
\tilde{\eta}_{12} & \tilde{\eta}_{11}~ t_2^{m-2}
\\
\noalign{\medskip}
-\tilde{\eta}_{11} & -\tilde{\eta}_{12}
\end{pmatrix}.
$$
The above matrix has eigenvalues $\pm i V_{12}(u)$. 
 As already explained, $\mathcal{V}_{[1,1]}$, $ \tilde{\eta}_{12}$, $ \tilde{\eta}_{11}$, $ \tilde{\eta}_{22}$ are well defined at $t_2=0$, namely for $u_1=u_2$, as analytic functions of $t_1(\equiv u_1=u_2)$, $u_3,...,u_n$. Hence,   the limit for $u_1-u_2\to 0$ of $V_{12}(u)$ is well defined as an  analytic function  of  $t_1, u_3,...,u_n$. Let
\be
\label{12agosto2022-1}
\mathring{V}_{12}:=\lim_{u_1-u_2\to 0}V_{12}(u) \quad \hbox{in }\mathcal{B}.
\ee
 In principle, it depends holomorphically on $t_1,u_3,...,u_n$, but we will show that $\mathring{V}_{12}$  is {\it constant}.
Keeping the above discussion into account, in $\mathcal{B}\cap\mathcal{K}$ we have 
\be
\label{6agosto2022-5}
\mathcal{V}_{[1,1]}\bigr|_{t_2=0}=
\left\{
\begin{aligned}
&\left.
\frac{i \mathring{V}_{12}}{\sqrt{\tilde{\eta}_{12}^2}} 
\begin{pmatrix}
\tilde{\eta}_{12} & 0 
\\
\noalign{\medskip}
-\tilde{\eta}_{11} & -\tilde{\eta}_{12}
\end{pmatrix}
\right|_{t_2=0},& m\geq 3.
\\
\noalign{\medskip}
& \frac{i \mathring{V}_{12}}{\sqrt{\tilde{\eta}_{12}^2-\tilde{\eta}_{11}^2}}
\left.
\begin{pmatrix}
\tilde{\eta}_{12} & \tilde{\eta}_{11}
\\
\noalign{\medskip}
-\tilde{\eta}_{11} & -\tilde{\eta}_{12}
\end{pmatrix}\right|_{t_2=0},&m=2.
\end{aligned}
\right.
\ee

Using \eqref{8giugno2022-14} and the chain rule  
$ 
\frac{\partial \mathscr{Y}}{\partial u_k}= \frac{\partial \mathscr{Y}}{\partial t_1}\frac{\partial t_1}{\partial u_k}+\frac{\partial \mathscr{Y}}{\partial t_2}\frac{\partial t_2}{\partial u_k}
$ for $k=1,2$, the $du_j$ components ($j\in\{1,...,n\}$) of \eqref{4agosto2022-2} become
\begin{align}
\label{27luglio2022-2}
&\frac{\partial  Y}{\partial t_1}=\left(
z\underbrace{\Psi^{-1} (E_1+E_2)\Psi}_{(E_1+E_2)}+\Psi^{-1}(V_1+V_2)\Psi-\Psi^{-1}
\frac{\partial \Psi}{\partial t_1}
\right)  Y.
\\
\noalign{\medskip}
\label{27luglio2022-3} 
&\frac{\partial  Y}{\partial t_2}=\left(
z~t_2^{\frac{m-2}{2}}\Psi^{-1} (E_1-E_2)\Psi+t_2^{\frac{m-2}{2}}\Psi^{-1}(V_1-V_2)\Psi
-
\Psi^{-1}
\frac{\partial \Psi}{\partial t_2}
\right)  Y,
\\
\noalign{\medskip}
\noalign{\medskip}
\label{27luglio2022-1}
 &
 \frac{\partial  Y}{\partial u_j}=\Bigl(z\underbrace{\Psi^{-1}E_j\Psi}_{E_j}+\Psi^{-1}V_j\Psi
 -
\Psi^{-1}
\frac{\partial \Psi}{\partial u_j}\Bigr) Y,\quad \quad j\geq 3.
 \end{align}
with 
$$ 
z~t_2^{\frac{m-2}{2}}\Psi^{-1} (E_1-E_2)\Psi=z~\begin{pmatrix}
0 & t_2^{m-2} & 
\\
1 & 0 &
\\
 & & I_{n-2}
\end{pmatrix}.
$$
By the same chain rule,  it is easily seen that from the deformation equations \eqref{4agosto2022-4}  we receive   
\begin{align}
\label{12agosto2022-5}
&\frac{\partial \mathcal{V}}{\partial t_1}=\left[
\Psi^{-1}(V_1+V_2)\Psi -\Psi^{-1}\frac{\partial \Psi}{\partial t_1}~,~\mathcal{V}\right],
\\
\noalign{\medskip}
\label{12agosto2022-6}
&\frac{\partial \mathcal{V}}{\partial t_2}=\left[
t_2^{\frac{m-2}{2}}\Psi^{-1}(V_1-V_2)\Psi -\Psi^{-1}\frac{\partial \Psi}{\partial t_2}~,~\mathcal{V}\right],
\\
\noalign{\medskip}
\label{12agosto2022-7}
&\frac{\partial \mathcal{V}}{\partial u_j} = \left[
\Psi^{-1}V_j\Psi -\Psi^{-1}\frac{\partial \Psi}{\partial t_2}~,~\mathcal{V}\right],\quad j\geq 3.
\end{align}

The Pfaffian system given by \eqref{6agosto2022-2} and \eqref{27luglio2022-2}-\eqref{27luglio2022-1} represents the flatness of the connection \eqref{5agosto2022-1} in the good coordinates $(t_1,t_2,~u_3,...,u_n)$ in $\mathcal{B}$. All the coefficients must be  holomorphic at $t_2=0$ in these variables.
 The flatness of the connection \eqref{5agosto2022-1} restricted at $\mathcal{B}\cap\mathcal{K}$ is then represented by \eqref{6agosto2022-2},  \eqref{27luglio2022-2} and \eqref{27luglio2022-1}  restricted at   $t_2=0$,  namely
\be
\label{29luglio2022-5}
\left\{
\begin{aligned}
&\frac{d Y}{dz} =\left(\mathcal{U}\bigr|_{t_2=0}
 +\frac{\mathcal{V}\bigr|_{t_2=0}}{z}
 \right) Y, 
 &&
 \mathcal{U}\bigr|_{t_2=0}=
\mathrm{diag}(t_1,t_1;~u_3,...,u_n), 
 \\
 \noalign{\medskip}
 &
 \frac{\partial  Y}{\partial t_1}=\left(z(E_1+E_2)+\mathcal{V}_2\bigr|_{t_2=0}
 -
 \left.\Psi^{-1}
\frac{\partial \Psi}{\partial t_1}\right|_{t_2=0}\right) Y,
&&
\quad\quad\quad
  \mathcal{V}_2:=\Psi^{-1}(V_1+V_2)\Psi,
 \\
 \noalign{\medskip}
&  \frac{\partial  Y}{\partial u_j}=\left(zE_j+
  \mathcal{V}_j\bigr|_{t_2=0}-
  \left.
\Psi^{-1}
\frac{\partial \Psi}{\partial u_j}\right|_{t_2=0}\right) Y,
  \quad
  \quad
   j\geq 3,
   && 
   \quad\quad\quad 
    \mathcal{V}_j:=\Psi^{-1}V_j\Psi.
 \end{aligned}
 \right.
 \ee
The deformation equation \eqref{12agosto2022-5} and \eqref{12agosto2022-7} become
\be
\label{12agosto2022-8}
\left\{\begin{aligned}
&\frac{\partial \mathcal{V}}{\partial t_1}=\left[
\mathcal{V}_2
-
 \left.\Psi^{-1}\frac{\partial \Psi}{\partial t_1}\right|_{t_2=0}~,~\mathcal{V}\right],
\\
\noalign{\medskip}
&\frac{\partial \mathcal{V}}{\partial u_j} = 
\left[
\mathcal{V}_j -
  \left.
\Psi^{-1}
\frac{\partial \Psi}{\partial u_j}\right|_{t_2=0}~,~\mathcal{V}\right],\quad j\geq 3.
\end{aligned}
\right.
\ee

\bpr
\label{6agosto2022-3}
The Pfaffian system \eqref{29luglio2022-5}, representing the flatness of the Dubrovin deformed connection \eqref{5agosto2022-1} restricted at caustic $\mathcal{K}\cap\mathcal{B}$,  is exactly a system of type \eqref{30marzo2021-1}, 
with 
$$ 
s=n-1,\quad p_1=2,~p_3=\cdots=p_{n-1}=1,\quad (\lambda_1,...,\lambda_{n-1}):=(t_1,u_3,...,u_n).
$$
and
$$
\Lambda:=\hbox{diag}(\lambda_1,\lambda_1,~\underbrace{\lambda_2,...,\lambda_{n-1}}_{n-2~distinct})\equiv \mathcal{U}\bigr|_{t_2=0},  \quad\quad A(\lambda):= \mathcal{V}\bigr|_{t_2=0} \quad \hbox{ as in \eqref{6agosto2022-5}}.
$$
With the above identification, system \eqref{12agosto2022-8} is exactly the isomonodromy deformation equations \eqref{20agosto2021-6}. 

The block-diagonal matrix  $ \mathcal{T}(\lambda)$ which  reduces to Jordan form the  block diagonal part $A_{[1,1]}\oplus A_{33}\oplus \cdots \oplus A_{nn}$ of $A(\lambda)$ is
\be
\label{8agosto2022-1} 
\mathcal{T}(\lambda)
=
 \underbrace{\mathcal{T}_1(\lambda)}_{2\times 2}\oplus ~ \mathrm{diag}(h_2,...,h_{n-1}),
\ee
where   $ \mathcal{T}_1(\lambda)$ bringings the upper left $2\times 2$ block $A_{[1,1]}$ to Jordan form, while the other $h_\ell$, $2\leq \ell\leq n-1$, are arbitrary scalar constants. 
In \eqref{29luglio2022-5}, one  exactly has
\be
\label{27luglio2022-5}
-\Psi^{-1}\frac{\partial \Psi}{\partial t_1}\Bigr|_{t_2=0}=\frac{\partial \mathcal{T}}{\partial t_1}\mathcal{T}^{-1},
\quad\quad
-\Psi^{-1}\frac{\partial \Psi}{\partial u_j}\Bigr|_{t_2=0}=\frac{\partial \mathcal{T}}{\partial u_j}\mathcal{T}^{-1},\quad j\geq 3,
\ee
and
$$
\mathcal{V}_2\bigr|_{t_2}=\omega_1,\quad \quad \mathcal{V}_j\bigr|_{t_2}=\omega_{j-1},\quad j=3,...,n.
$$

\epr

The proof will be given after Remark \ref{13agosto2022-2}. Notice that $ A(\lambda):= \mathcal{V}\bigr|_{t_2=0}$ is holomorphic in $\mathcal{B}\cap\mathcal{K}$, and $\mathcal{B}$ can be made  sufficiently small so that  Assumption 1 of the Introduction is satisfied.

\bre{\rm
According to the block structure $\Lambda=\lambda_1I_2\oplus\bigl( \lambda_2I_1\oplus \cdots\oplus \lambda_{n-1}I_1\bigr)$, we have 
$$ 
E_{p_1}=E_1+E_2,\quad \quad E_{p_j}= E_{j+1},\quad 2\leq j \leq n-1,
$$
and the  matrix $A$ must be  partitioned into the following blocks.  
$A_{[1,1]}$ is a $2\times 2$ block. Let $j,k\in\{2,...,n-1\}$. 
Then, the
 $A_{[j,1]}$ are $1\times 2$ blocks;  the $A_{[1,k]}$  are $2\times 1$ blocks; the $A_{[j,k]}=A_{j+1,k+1}$ are matrix entries (1 dimensional blocks!). The structure \eqref{8agosto2022-1}  is required by this block decomposition to fulfil Theorem \ref{30marzo2021-2}.

}
\ere

Proposition \ref{6agosto2022-3} not only tells us that the isomonodromy deformation theory developed in this paper applies at the caustics, but also allow us to predict properties of the caustic itself, as in the following 

\bcr
\label{12agosto2022-2}
The block $\mathcal{V}_{[1,1]}\bigr|_{t_2=0}$ is constant. In particular,  $
\mathring{V}_{12}=\lim_{u_1-u_2}V_{12}(u)
$ defined in \eqref{12agosto2022-1} is constant, and  $\tilde{\eta}_{11}\bigr|_{t_2=0}$ and  $\tilde{\eta}_{12}\bigr|_{t_2=0}$ are constant. 
\ecr

\vskip 0.3 cm 
\begin{proof}[Proof of Corollary \ref{12agosto2022-2}] 
It follows from PART II of Theorem \ref{30marzo2021-2}, which applies thanks to Proposition \ref{6agosto2022-3}, and from  \eqref{6agosto2022-5} .
\end{proof}

\bre[\bf on the value of  $\mathring{V}_{12}$]
\label{13agosto2022-1}
{\rm  Remarkably, the value of the constant $\mathring{V}_{12}$  can  be written explicitly:
$$
\mathring{V}_{12}=\frac{i(m-2)}{2m},
$$
(recall that $m\geq 2$). This result was first observed in \cite{Reyes}, and can be proved in three ways. 
\begin{itemize} 
\item[1)]  It is proved in \cite{Reyes} {\it by purely geometric arguments}, starting from the decomposition \eqref{5agosto2022-5}. 

\item[2)]   It is rather laborious  to show analytically that the isomonodromy  deformation equations \eqref{4agosto2022-4}  admit  a solution $V(u)$ with the property that  $\lim_{u_2-u_2\to 0} V_{12}(u)$ exists finite, if and only if this solution is such that  in a neighbourhood of $u_1-u_2=0$ 
$$ 
V_{12}(u)=\frac{i\sigma}{2} + \sum_{(h,k)\neq(0,0)}^{\infty} b_{hk}(x_3,...,x_{n-1})
~x_2^{h +(1-\sigma)k} + \sum_{(h,q)\neq(0,0)}^{\infty} a_{hk}(x_3,...,x_{n-1})
~x_2^{h +(1+\sigma)k},
$$
where 
$$ \sigma\in\mathbb{C},\quad 0\leq \Re\sigma<1; \quad\quad x_j=\frac{u_j-u_1}{u_n-u_1},\quad j=2,...,n-1,
$$
and the functions $a_{hk}(x_3,...,x_{n-1}),b_{hk}(x_3,...,x_{n-1})$ are analytic around $p$.  
Hence, 
$$
\mathring{V}_{12}=\frac{i\sigma}{2} \quad\hbox{ is  constant}.
$$
Then, using again  equations \eqref{4agosto2022-4} one can prove that $\mathring{V}_{12}=i(m-2)/2m$. This is more laborious than the geometric proof of  \cite{Reyes}. 

\item[3)] Using the approach of this paper. 
We have seen that the coefficients of  equations \eqref{27luglio2022-3} and \eqref{12agosto2022-6} must be holomorphic at $t_2=0$ in the variables $(t_1,t_2,u_3,...,u_n)$. We prove at the end of this Section that 
$$
t_2^{\frac{m-2}{2}}\Psi^{-1}(V_1-V_2)\Psi -\Psi^{-1}\frac{\partial \Psi}{\partial t_2},
$$
is holomorphic at $t_2=0$ (as it must be)  if and only if  $\mathring{V}_{12}=i(m-2)/2m$. 
\end{itemize}
}
\ere

\bre[\bf On the strong isomonodromy at the caustic]
\label{13agosto2022-2}
{\rm
From Remark \ref{13agosto2022-1} it follows that  $0\leq \Re(\mathring{V}_{12})<1/2$.  Thus, $\mathcal{V}_{[1,1]}\bigr|_{t_2=0}$ is non resonant, so that  by Theorem  \ref{20agosto2021-3} system \eqref{12agosto2022-8} is a {\it necessary and sufficient condition for strong isomonodromy}. Therefore, likewise the case of semisimple points, also at the generic points of the caustic here considered, the Dubrovin flat connection is realized by a strongly isomonodromic system.}
\ere

\vskip 0.3 cm 
\begin{proof}[Proof of Proposition \ref{6agosto2022-3}]  The  $z(E_1+E_2)$ and $zE_j$ terms in \eqref{29luglio2022-5}  correctly display $ 
E_{p_1}=E_1+E_2$ and $E_{p_j}= E_{j+1}$, $2\leq j \leq n-1$. 

\vskip 0.2 cm 
$\bullet$ First, we show that the matrices $\mathcal{V}_j$, $j= 2,...,n$, in \eqref{29luglio2022-5} are 
$$
\mathcal{V}_2\bigr|_{t_2=0}=\omega_1(\lambda),\quad \quad \mathcal{V}_j\bigr|_{t_2=0}=\omega_{j-1}(\lambda),\quad j=3,...,n.
$$
with the structure \eqref{31marzo2021-6}. 
We start by computing  $ \mathcal{V}_2$ in $\mathcal{B}$. 
\begin{align*}
 \mathcal{V}_2=\Psi^{-1}(V_1+V_2)\Psi
 =
 &~
 \Psi^{-1}
 \begin{pmatrix}
 0 & 0 & \cdots~ \frac{V_{1k}}{u_1-u_k}~ \cdots
 \\
 \noalign{\medskip}
  0 & 0 & \cdots~ \frac{V_{2k}}{u_2-u_k}~ \cdots
 \\
 \noalign{\medskip}
 \vdots & \vdots &  0~ \cdots~ 0 ~\cdots ~0
 \\
 \noalign{\medskip}
 \frac{V_{k1}}{u_1-u_k} & \frac{V_{k2}}{u_2-u_k} &0~ \cdots~ 0 ~\cdots ~0
  \\
 \noalign{\medskip}
 \vdots & \vdots & 0~ \cdots~ 0 ~\cdots ~0
 \end{pmatrix}
 \Psi
\\
\noalign{\medskip}
&
 =
\left( 
\begin{aligned}
& \begin{aligned}
 \quad  \quad\quad 0 && \quad \quad 0 
  \\
 \quad  \quad\quad 0 && \quad  \quad 0 
 \end{aligned}
  && 
   \widehat{\Psi}^{-1} 
 \left[\begin{aligned}
\cdots ~ \frac{V_{1k}}{u_1-u_k} ~\cdots
 \\
\cdots ~ \frac{V_{2k}}{u_2-u_k} ~\cdots
 \end{aligned}
 \right]
  \\
  &
  \left[\begin{aligned}
  \vdots\quad&\quad\quad \vdots
  \\
   \frac{V_{k1}}{u_1-u_k}  &\quad \frac{V_{k2}}{u_2-u_k}
   \\
    \vdots\quad&\quad\quad \vdots
   \end{aligned}
 \right]\widehat{\Psi} 
  & &
  \begin{aligned}
  \quad 0 \quad \cdots \quad 0 \quad \dots \quad 0
  \\
 \quad 0 \quad \cdots \quad 0 \quad \dots \quad 0
  \\
\quad 0 \quad \cdots \quad 0 \quad \dots \quad 0
\end{aligned}
  \end{aligned}
  \right)
\end{align*}
We want to take the limit for $t_2\to 0$, namely for $u_1-u_2\to 0$. We use (from \eqref{8giugno2022-14}  and $V=\Psi \mathcal{V}\Psi^{-1}$) the relations 
$$
u_1-u_2=\frac{4}{m}t_2^{m/2},\quad \quad \left\{
\begin{aligned}
V_{1k}=\frac{a}{\sqrt{2}}\left( t_2^{\frac{2-m}{4}}~\mathcal{V}_{1k} + t_2^{\frac{m-2}{4}}~\mathcal{V}_{2k} \right)
\\
\noalign{\medskip}
V_{2k}=\frac{ib}{\sqrt{2}}\left( t_2^{\frac{2-m}{4}}~\mathcal{V}_{1k} - t_2^{\frac{m-2}{4}}~\mathcal{V}_{2k} \right)
\end{aligned}
\right.\quad \quad  k= 3,...,n,
$$
and receive 
$$
\begin{aligned}
 \widehat{\Psi}^{-1} 
 \left[\begin{aligned}
\cdots ~ \frac{V_{1k}}{u_1-u_k} ~\cdots
 \\
\cdots ~ \frac{V_{2k}}{u_2-u_k} ~\cdots
 \end{aligned}
 \right]
 =
 \begin{pmatrix}
~ \cdots & \dfrac{\mathcal{V}_{1k}}{2}\dfrac{u_1+u_2-2u_k}{(u_1-u_k)(u_2-u_k)}
 -\dfrac{2~ t_2^{m-1} }{m(u_1-u_k)(u_2-u_k)}\mathcal{V}_{2k} & \cdots~
 \\
 \noalign{\medskip}
 ~\cdots & \dfrac{\mathcal{V}_{2k}}{2}\dfrac{u_1+u_2-2u_k}{(u_1-u_k)(u_2-u_k)}
 -\dfrac{2 ~t_2 }{m(u_1-u_k)(u_2-u_k)}\mathcal{V}_{1k} & \cdots~
 \end{pmatrix}
\\
\noalign{\medskip}
\underset{t_2\to 0, ~u_1-u_2\to 0}\longrightarrow 
\begin{pmatrix}
 ~\cdots & \dfrac{\mathcal{V}_{1k}\bigr|_{t_2=0}}{u_1-u_k}
 & \cdots~
 \\
 \noalign{\medskip}
~ \cdots & \dfrac{\mathcal{V}_{2k}\bigr|_{t_2=0}}{u_1-u_k}
 & \cdots~
 \end{pmatrix}
 \equiv
\begin{pmatrix}
 ~\cdots & \dfrac{A_{1k}}{\lambda_1-\lambda_{k-1}}
 & \cdots~
 \\
 \noalign{\medskip}
~ \cdots & \dfrac{A_{2k}}{\lambda_1-\lambda_{k-1}}
 & \cdots~
 \end{pmatrix},\quad \forall~m\geq 2.
\end{aligned}
$$
In the last step we have used $t_1=u_1=u_2$ at the caustic, the definition $\lambda_1:=t_1$, $\lambda_{k-1}:=u_k$, $k=3,...,n$, and $A(\lambda):=\mathcal{V}\bigr|_{t_2=0}$. 
Successively, we 
use 
$$ 
\left\{
\begin{aligned}
V_{k1}=\frac{1}{\sqrt{2}~ a} \left(
t_2^{\frac{2-m}{4}}\mathcal{V}_{k2}  + t_2^{\frac{m-2}{4}}\mathcal{V}_{k1} 
\right)
\\
\noalign{\medskip}
V_{k2}=\frac{i}{\sqrt{2}~ b} \left(
 t_2^{\frac{2-m}{4}}\mathcal{V}_{k2} -t_2^{\frac{m-2}{4}}\mathcal{V}_{k1}  
\right)
\end{aligned}
\right.\quad \quad  k= 3,...,n,
$$
and find
$$
  \left[\begin{aligned}
  \vdots\quad&\quad\quad \vdots
  \\
   \frac{V_{k1}}{u_1-u_k}  &\quad \frac{V_{k2}}{u_2-u_k}
   \\
    \vdots\quad&\quad\quad \vdots
   \end{aligned}
 \right]\widehat{\Psi} =
 $$
 $$
 =\begin{pmatrix}
 \vdots & \vdots
 \\
 \noalign{\medskip}
 \dfrac{\mathcal{V}_{k1}}{2}\dfrac{u_1+u_2-2u_k}{(u_1-u_k)(u_2-u_k)}
 -\dfrac{2 ~t_2 ~\mathcal{V}_{k2}}{m(u_1-u_k)(u_2-u_k)} 
 &
 \dfrac{\mathcal{V}_{k2}}{2}\dfrac{u_1+u_2-2u_k}{(u_1-u_k)(u_2-u_k)}
 -\dfrac{2~ t_2^{m-1} ~\mathcal{V}_{k1}}{m(u_1-u_k)(u_2-u_k)}
 \\
 \noalign{\medskip}
 \vdots & \vdots
 \end{pmatrix}
 $$
 $$
 \underset{t_2\to 0, ~u_1-u_2\to 0}\longrightarrow 
 \begin{pmatrix}
 \vdots & \vdots 
 \\
 \noalign{\medskip}
 \dfrac{\mathcal{V}_{k1}\bigr|_{t_2=0}}{u_1-u_k} & \dfrac{\mathcal{V}_{k2}\bigr|_{t_2=0}}{u_1-u_k}
  \\
 \noalign{\medskip}
  \vdots & \vdots 
 \end{pmatrix}\equiv \begin{pmatrix}
 \vdots & \vdots 
 \\
 \noalign{\medskip}
 \dfrac{A_{k1}}{\lambda_1-\lambda_{k-1}} & \dfrac{A_{k2}}{\lambda_1-\lambda_{k-1}}
  \\
 \noalign{\medskip}
  \vdots & \vdots 
 \end{pmatrix},\quad \forall~m\geq 2.
 $$ 
 Notice that  the above computations  confirm that $\mathcal{V}_2$ is holomorphic in a neighbourhood of $t_2=0$. 
 In conclusion, we have found that 
 $$
 \mathcal{V}_2\bigr|_{t_2=0}
 =
 \begin{pmatrix}
 0 & 0 & \cdots~ \dfrac{A_{1k}}{\lambda_1-\lambda_{k-1}}~ \cdots
 \\
 \noalign{\medskip}
  0 & 0 & \cdots~ \dfrac{A_{2k}}{\lambda_1-\lambda_{k-1}}~ \cdots
 \\
 \noalign{\medskip}
 \vdots & \vdots &  0~ \cdots~ 0 ~\cdots ~0
 \\
 \noalign{\medskip}
 \dfrac{A_{k1}}{\lambda_1-\lambda_{k-1}} & \dfrac{A_{k2}}{\lambda_1-\lambda_{k-1}} &0~ \cdots~ 0 ~\cdots ~0
  \\
 \noalign{\medskip}
 \vdots & \vdots & 0~ \cdots~ 0 ~\cdots ~0
 \end{pmatrix}.
 $$
 The above is the correct form $\omega_1(\lambda)$, corresponding to the left upper block $\mathrm{diag}(t_1,t_1)=t_1I_2\equiv \lambda_1I_2$ of $\Lambda=\mathcal{U}\bigr|_{t_2=0}$. 
 Analogous computations confirm that the $\mathcal{V}_j$ are holomorphic in a neighbourhood of $t_2=0$ and  
 $$ 
 \mathcal{V}_j\bigr|_{t_2=0}=\left(
 \frac{\mathcal{V}_{rs}\bigr|_{t_2=0}(\delta_{rj}-\delta_{sj})}{u_r-u_s}
 \right)_{r,s=1}^n
=\left(
 \frac{A_{rs}(\lambda)~(\delta_{r-1,j-1}-\delta_{s-1,j-1})}{\lambda_{r-1}-\lambda_{s-1}}
 \right)_{r,s=1}^n
,\quad j\geq 3.
$$
which is the correct form of $\omega_{j-1}(\lambda)$ corresponding to the other 1-dimensional diagonal blocks 
 of $\Lambda=\mathcal{U}\bigr|_{t_2=0}$.

$\bullet$ Next, we check if in \eqref{29luglio2022-5} the correct term $\frac{\partial \mathcal{T}(\lambda)}{\partial \lambda_j}\cdot  \mathcal{T}(\lambda)^{-1}$ appears, namely we prove  the equalities \eqref{27luglio2022-5}, with $\mathcal{T}(\lambda)\equiv \mathcal{T}(t_1,u_3,...,u_n)$ having structure \eqref{8agosto2022-1}. 
  Observe that in $\mathcal{B}$ 
\be
\label{8agosto2022-7}
\Psi^{-1}\frac{\partial \Psi}{\partial t_1}
=
\begin{pmatrix}   \widehat{\Psi}^{-1} \dfrac{\partial  \widehat{\Psi}}{\partial t_1}
&
\begin{aligned} 0 &\cdots & 0 
\\
0 &\cdots & 0
\end{aligned}
\\
\noalign{\medskip}
 \vdots \quad \vdots & 
 \\ 
 0 \quad  0 & \Large{\boldsymbol{0}}
 \\
   \vdots \quad \vdots &
\end{pmatrix},
\quad\quad 
\Psi^{-1}\frac{\partial \Psi}{\partial u_j}
=
\begin{pmatrix}  \widehat{\Psi}^{-1} \dfrac{\partial  \widehat{\Psi}}{\partial u_j}
&
\begin{aligned} 0 &\cdots & 0 
\\
0 &\cdots & 0
\end{aligned}
\\
\noalign{\medskip}
 \vdots \quad \vdots & 
 \\ 
 0 \quad  0 & \Large{\boldsymbol{0}}
 \\
   \vdots \quad \vdots &
\end{pmatrix}, \quad j\geq 3,
\ee
 The $h_\ell$ are a priori arbitrary functions, because they reduce the 1-dimensional blocks $A_{[\ell,\ell]}\equiv \mathcal{V}_{\ell+1,\ell+1}\bigr|_{t_2=0}$  to ``diagonal form''! Corresponding to the {\large $\boldsymbol{0}$} block in \eqref{8agosto2022-7}, equations  \eqref{27luglio2022-5}  are satisfied if the $h_\ell$  are non-zero arbitrary constants.   
$\mathcal{T}_1$ must reduce the block $A_{[1,1]}=\mathcal{V}_{[1,1]}\bigr|_{t_2=0}$ in \eqref{6agosto2022-5} to Jordan (diagonal, in our case) form.  Thus, we have to solve for $\mathcal{T}_1$ the system 
\be
\label{28luglio2022-1}
\left\{ 
\begin{aligned}
& \frac{\partial \mathcal{T}_1}{\partial t_1}\mathcal{T}_1^{-1}= - \widehat{\Psi}^{-1}\frac{\partial  \widehat{\Psi}}{\partial t_1}\Bigr|_{t_2=0}
\\
\noalign{\medskip}
& \frac{\partial \mathcal{T}_1}{\partial u_j}\mathcal{T}_1^{-1}= - \widehat{\Psi}^{-1}\frac{\partial  \widehat{\Psi}}{\partial u_j}\Bigr|_{t_2=0},\quad j\geq 3,
\\
\noalign{\medskip}
& 
\mathcal{T}_1^{-1} \mathcal{V}_{[1,1]}\bigr|_{t_2=0}\mathcal{T}_1= \begin{pmatrix}
i\mathring{V}_{12} & 0 
\\
0 & -i\mathring{V}_{12}\end{pmatrix}\quad \hbox{``constraint''}.
\end{aligned}
\right.
\ee
The "known terms" $ \widehat{\Psi}^{-1}\frac{\partial  \widehat{\Psi}}{\partial t_1}\bigr|_{t_2=0}$, $ \widehat{\Psi}^{-1}\frac{\partial  \widehat{\Psi}}{\partial u_j}\bigr|_{t_2=0}$ in \eqref{28luglio2022-1} all have the same structure. Indeed, let $\xi$ be one of the variables  $t_1, u_3,...,u_n$. Then, the corresponding "known term'' is  the value at $t_2=0$ of 
\be
\label{29luglio2022-1}
 \widehat{\Psi}^{-1}\frac{\partial  \widehat{\Psi}}{\partial \xi}\underset{\eqref{5agosto2022-6}}= 
 \frac{1}{2(\tilde{\eta}_{12}^2-t_2^{m - 2}\tilde{\eta}_{11}^2)}
\begin{pmatrix}
  \tilde{\eta}_{12}\dfrac{\partial \tilde{\eta}_{12}}{\partial 
\xi}
-
\tilde{\eta}_{11}\dfrac{\partial \tilde{\eta}_{11}}{\partial 
\xi}~t_2^{m - 2} 
&
t_2^{m-2}\left(
\tilde{\eta}_{12}\dfrac{\partial \tilde{\eta}_{11}}{\partial 
\xi}
-
\tilde{\eta}_{11}\dfrac{\partial \tilde{\eta}_{12}}{\partial 
\xi}
\right)
\\
\noalign{\medskip}
\tilde{\eta}_{12}\dfrac{\partial \tilde{\eta}_{11}}{\partial 
\xi} 
-
\tilde{\eta}_{11}\dfrac{\partial \tilde{\eta}_{12}}{\partial 
\xi}
& 
   \tilde{\eta}_{12}\dfrac{\partial \tilde{\eta}_{12}}{\partial 
\xi}
-
\tilde{\eta}_{11}\dfrac{\partial \tilde{\eta}_{11}}{\partial 
\xi}~t_2^{m - 2}
\end{pmatrix}.
\ee
This is analytic  in variables $(t_1,t_2,u_3,...,u_n)$ in a neighbourhood of $t_2=0$.

\vskip 0.3 cm 
 \underline{Case $m\geq 3$}. In this case $\tilde{\eta}_{12}\neq 0$ (see Remark \ref{5agosto2022-7}), so that 
 $$
\left. \widehat{\Psi}^{-1}\dfrac{\partial  \widehat{\Psi}}{\partial \xi}\right|_{t_2=0}=
\dfrac{1}{2} \left.\begin{pmatrix}
 \dfrac{1}{\tilde{\eta}_{12}}\dfrac{\partial \tilde{\eta}_{12}}{\partial \xi} & 0 
\\
\noalign{\medskip}
\left(
 \dfrac{1}{\tilde{\eta}_{12}}\dfrac{\partial \tilde{\eta}_{11}}{\partial \xi}- \dfrac{\tilde{\eta}_{11}}{\tilde{\eta}_{12}^2}\dfrac{\partial \tilde{\eta}_{12}}{\partial \xi}
\right)
& 
 \dfrac{1}{\tilde{\eta}_{12}}\dfrac{\partial \tilde{\eta}_{12}}{\partial \xi}
\end{pmatrix}\right|_{t_2=0}.
$$
The general solution  of \eqref{28luglio2022-1} is then
\be
\label{28luglio2022-2}
\mathcal{T}_1
=
\left.
\begin{pmatrix}
\dfrac{c_1}{\sqrt{\tilde{\eta}_{12}}} & 0 
\\
\noalign{\medskip}
-\dfrac{c_1}{2}\dfrac{\tilde{\eta}_{11}}{\tilde{\eta}_{12}^{3/2}} & \dfrac{c_2}{\sqrt{\tilde{\eta}_{12}}}
\end{pmatrix}\right|_{t_2=0},\quad\quad c_1,c_2\in\mathbb{C}\backslash\{0\}.
\ee

 \underline{Case $m=2$}.  By Remark \ref{13agosto2022-1}, $\mathcal{V}_{[1,1]}\bigr|_{t_2=0}=0$, so that we can take any solution  of $
\frac{\partial \mathcal{T}_1}{\partial \xi}\mathcal{T}_1^{-1}= -\left.  \widehat{\Psi}^{-1}\frac{\partial  \widehat{\Psi}}{\partial \xi}\right|_{t_2=0}$. 
 The general solution is, for an arbitrary constant invertible matrix $C$, 
\begin{align*} 
\mathcal{T}_1(t_1,u_3,...,u_n)&= \widehat{\Psi}^{-1}(t_1,0,u_3,...,u_n)~ C
\\
\noalign{\medskip}
&\underset{\eqref{5agosto2022-6},~ \eqref{8giugno2022-1}}=\frac{1}{\sqrt{2}}
\left. \begin{pmatrix}
\dfrac{1}{\sqrt{\tilde{\eta}_{12}+  \tilde{\eta}_{11}}}
&
\dfrac{-i }{\sqrt{\tilde{\eta}_{12}- \tilde{\eta}_{11}}}
\\
\noalign{\medskip}
\dfrac{1}{\sqrt{\tilde{\eta}_{12}+  \tilde{\eta}_{11}}}
&
\dfrac{i }{\sqrt{\tilde{\eta}_{12}-  \tilde{\eta}_{11}}}
\end{pmatrix}\right|_{t_2=0}\cdot C
.
\end{align*}
Notice that if hypothetically  $\mathring{V}_{12}\neq 0$, then the choice of $\mathcal{T}_1$ which diagonalizes 
$\mathcal{V}_{[1,1]}\bigr|_{t_2=0}$ in \eqref{6agosto2022-5} for $m=2$ is 
$$ 
C=\begin{pmatrix}
\alpha &\pm i\beta
\\
\pm i\alpha & \beta
\end{pmatrix}\quad\Longrightarrow 
\quad 
\mathcal{T}_1 \cdot \mathcal{V}_{[1,1]}^{-1}\bigr|_{t_2=0} \cdot\mathcal{T}_1
=
\hbox{diag} (\pm i \mathring{V}_{12},~ \mp i \mathring{V}_{12}).
$$
\end{proof}

\begin{proof}[Proof of point 3) of Remark \ref{13agosto2022-1}]
The coefficients   of equations \eqref{27luglio2022-3} and \eqref{12agosto2022-6} are holomorphic at $t_2=0$ in variables $(t_1,t_2,u_3,...,u_n)$ if and only if so is the term
$$
t_2^{\frac{m-2}{2}}\Psi^{-1}(V_1-V_2)\Psi -\Psi^{-1}\frac{\partial \Psi}{\partial t_2}.
$$
We have
$$
\Psi^{-1}(V_1-V_2)\Psi
 =
 \Psi^{-1}
 \begin{pmatrix}
 0 & \frac{-2V_{12}}{u_2-u_1} & \cdots~ \frac{V_{1k}}{u_1-u_k}~ \cdots
 \\
 \noalign{\medskip}
 \frac{2V_{21}}{u_1-u_2} & 0 & \cdots~ \frac{-V_{2k}}{u_2-u_k}~ \cdots
 \\
 \noalign{\medskip}
 \vdots & \vdots &  0~ \cdots~ 0 ~\cdots ~0
 \\
 \noalign{\medskip}
 \frac{V_{k1}}{u_1-u_k} & \frac{-V_{k2}}{u_2-u_k} &0~ \cdots~ 0 ~\cdots ~0
  \\
 \noalign{\medskip}
 \vdots & \vdots & 0~ \cdots~ 0 ~\cdots ~0
 \end{pmatrix}
 \Psi
$$
The analogous computations of the proof of Proposition \ref{6agosto2022-3}  lead to 
$$
t_2^{\frac{m-2}{2}}~ \Psi^{-1}(V_1-V_2)\Psi\underset{t_2\to 0}\sim\begin{pmatrix}
 \mathcal{O}(1/t_2) &  \mathcal{O}(t_2^{m-3}) & \cdots~ \dfrac{A_{1k}}{\lambda_1-\lambda_{k-1}}~ \cdots
 \\
 \noalign{\medskip}
    \mathcal{O}(1/t_2)&   \mathcal{O}(1/t_2) & \cdots~ \dfrac{-A_{2k}}{\lambda_1-\lambda_{k-1}}~ \cdots
 \\
 \noalign{\medskip}
 \vdots & \vdots &  0~ \cdots~ 0 ~\cdots ~0
 \\
 \noalign{\medskip}
 \dfrac{A_{k1}}{\lambda_1-\lambda_{k-1}} & \dfrac{-A_{k2}}{\lambda_1-\lambda_{k-1}} &0~ \cdots~ 0 ~\cdots ~0
  \\
 \noalign{\medskip}
 \vdots & \vdots & 0~ \cdots~ 0 ~\cdots ~0
 \end{pmatrix}.
 $$
 On the other hand, using \eqref{5agosto2022-6}  and \eqref{8giugno2022-1}  one finds 
 $$
 \Psi^{-1}\frac{\partial \Psi}{\partial t_2}
=
\begin{pmatrix}   \widehat{\Psi}^{-1} \dfrac{\partial  \widehat{\Psi}}{\partial t_2}
&
\begin{aligned} 0 &\cdots & 0 
\\
0 &\cdots & 0
\end{aligned}
\\
\noalign{\medskip}
 \vdots \quad \vdots & 
 \\ 
 0 \quad  0 & \Large{\boldsymbol{0}}
 \\
   \vdots \quad \vdots &
\end{pmatrix}
\underset{t_2\to 0}\sim\begin{pmatrix}
 \mathcal{O}(1/t_2) &  \mathcal{O}(t_2^{m-3}) & \cdots~0~ \cdots
 \\
 \noalign{\medskip}
    \mathcal{O}(1/t_2)&   \mathcal{O}(1/t_2) & \cdots~ 0~ \cdots
 \\
 \noalign{\medskip}
 \vdots & \vdots &  0~ \cdots~ 0 ~\cdots ~0
 \\
 \noalign{\medskip}
 0 & 0&0~ \cdots~ 0 ~\cdots ~0
  \\
 \noalign{\medskip}
 \vdots & \vdots & 0~ \cdots~ 0 ~\cdots ~0
 \end{pmatrix}.
 $$
All The $\mathcal{O}$ terms are explicitly computed in terms of $\tilde{\eta}_{11},\tilde{\eta}_{12}, V_{12}$. From this, one sees that in 
$$
t_2^{\frac{m-2}{2}}\Psi^{-1}(V_1-V_2)\Psi -\Psi^{-1}\frac{\partial \Psi}{\partial t_2}
$$
the divergences holomorphically cancel if and only if $\mathring{V}_{12}=i(m-2)/2m$, and in this case
$$
\left. t_2^{\frac{m-2}{2}}\Psi^{-1}(V_1-V_2)\Psi -\Psi^{-1}\frac{\partial \Psi}{\partial t_2}~\right|_{t_2=0}= \begin{pmatrix}
  0 & 0 & \cdots~ \dfrac{A_{1k}}{\lambda_1-\lambda_{k-1}}~ \cdots
 \\
 \noalign{\medskip}
  0&  0& \cdots~ \dfrac{-A_{2k}}{\lambda_1-\lambda_{k-1}}~ \cdots
 \\
 \noalign{\medskip}
 \vdots & \vdots &  0~ \cdots~ 0 ~\cdots ~0
 \\
 \noalign{\medskip}
 \dfrac{A_{k1}}{\lambda_1-\lambda_{k-1}} & \dfrac{-A_{k2}}{\lambda_1-\lambda_{k-1}} &0~ \cdots~ 0 ~\cdots ~0
  \\
 \noalign{\medskip}
 \vdots & \vdots & 0~ \cdots~ 0 ~\cdots ~0
 \end{pmatrix}.
 $$

\end{proof}

{\bf In conclusion}, Proposition \ref{6agosto2022-3} shows that the isomonodromy deformation theory developed in this paper, summarized in  Theorems \ref{30marzo2021-2} and  \ref{20agosto2021-3}, applies to  the caustic of a Dubrovin-Frobenius manifold of the class studied in \cite{Reyes}. The representation of the Dubrovin flat connection on the basis $\partial/\partial t_1,\partial/\partial t_2, \pi_3,...,\pi_n$,  and  restricted to the caustic, is exactly a Pfaffian system of type \eqref{30marzo2021-1}, and the $z$-component  is strongly isomonodromic (Remark \ref{13agosto2022-2}). Moreover,  the isomonodromy deformation theory developed in this paper allows us to predict some properties at the caustic, as  Corollary \ref{12agosto2022-2} and Remark \ref{13agosto2022-1}.


\section{Appendix. Proof of Lemma \ref{1april2021-1}}
We consider the Jordan form  $J=J_1\oplus\cdots \oplus J_s$ of $A_{[1,1]}\oplus \cdots \oplus A_{[s,s]}$ and the corresponding  $L=L_1\oplus\cdots\oplus L_s$. 
 We prove that if the deformation is strongly isomonodromic, then
 $$  
 \Bigl[\mathcal{T}_k^{-1} \frac{\partial \mathcal{T}_k}{\partial \lambda_j}~,~ L_k\Bigr]=\Bigl[\mathcal{T}_k^{-1} \frac{\partial \mathcal{T}_k}{\partial \lambda_j}~,~ J_k\Bigr]=0,\quad \forall k=1,...,s.
 $$
 For a strong isomonodromy deformation, $J$ and $R$, and equivalently $D$ and $L$, are constant.  The factor $z^D z^L$ in each 
 $$ 
 Y_\nu(z,\lambda)=\mathcal{T}(\lambda) \widehat{Y}_\nu(z,\lambda) z^D z^L e^{z\Lambda}
 $$ 
 corresponds to a fundamental solution in  Levelt form for the  system (4.1) of \cite{CDG}, which has  a Fuchsian singularity in $z=\infty$.  Up to a permutation, we can always  assume that the matrices $Y_\nu(z,\lambda)$ are taken so that $z^{D} z^{L}$ satisfies the properties of Section \ref{2maggio2021-1} (to which we refer for notations).\footnote{One can take a permutation matrix $P$, which does not change $\Lambda$ because it permutes indexes inside the same block, which yields $\mathcal{T}P(P^{-1} \widehat{Y}_\nu P)  z^{P^{-1}DP} z^{P^{-1}LP}e^{z\Lambda}$ with the desired properties of Section \ref{2maggio2021-1}.} Therefore, following Section \ref{2maggio2021-1}, we can write  
 $$ 
 Y_\nu(z,\lambda)=\mathcal{T}(\lambda) \widehat{Y}_\nu(z,\lambda)~ z^{D+\Sigma} z^N e^{z\Lambda}.
 $$ 
 
 {\bf Step 1.}  First, we show that 
 \be
 \label{1maggio2021-1} 
 [\mathcal{T}(\lambda)^{-1} d \mathcal{T}(\lambda), D+\Sigma]=0,
 \ee
 and
 \be
 \label{1maggio2021-2} 
 [\mathcal{T}(\lambda)^{-1} d \mathcal{T}(\lambda),N]=0.
 \ee
 In order to prove \eqref{1maggio2021-1}-\eqref{1maggio2021-2},  recall \eqref{21maggio2021-4}-\eqref{22maggio2021-2} and let  
\begin{align*}
 & \mathcal{Y}_\nu(z,\lambda):= Y_\nu(z,\lambda)\mathcal{T}(\lambda)^{-1},& \mathfrak{D}(\lambda):= \mathcal{T}(\lambda) (D+\Sigma)\mathcal{T}(\lambda)^{-1}
\\
\noalign{\medskip}
&  \widehat{\mathcal{Y}}_\nu(z,\lambda):=\mathcal{T}(\lambda) \widehat{Y}_\nu(z,\lambda)\mathcal{T}(\lambda)^{-1},&
\mathfrak{N}(\lambda):=\mathcal{T}(\lambda) N \mathcal{T}(\lambda)^{-1}
\end{align*}
 Then, we have
\begin{align*}
 Y_\nu(z,\lambda)
=&\widehat{\mathcal{Y}}_\nu (z,\lambda)~z^{\mathfrak{D}(\lambda)} z^{\mathfrak{N}(\lambda)}e^{z\Lambda}~\mathcal{T}(\lambda)=
\mathcal{Y}_\nu (z,\lambda)~\mathcal{T}(\lambda),
\\
\noindent{\medskip}
& \widehat{\mathcal{Y}}_\nu(z,\lambda)
 \sim 
 I+\sum_{j=1}^\infty 
  \mathcal{T}(\lambda) F_j(\lambda) \mathcal{T}(\lambda)^{-1} ~z^{-j},\quad \quad z\to\infty \hbox{ in } \mathcal{S}_\nu.
  \end{align*}
 Let $d_\lambda$ be the differential w.r.t. $\lambda_1,...,\lambda_s$. From \eqref{31marzo2021-5},
 $$
\sum_{j=1}^s \Bigl(z E_{p_j} + \omega_j(\lambda)\Bigr)d\lambda_j + d\mathcal{T}\cdot \mathcal{T}^{-1}=d_\lambda Y_\nu(z,\lambda) \cdot Y_\nu(z,\lambda)^{-1}
$$
The right hand-side is  
\be
\label{1maggio2021-3}
\begin{aligned} 
d_\lambda\mathcal{Y}_\nu \cdot \mathcal{Y}_\nu^{-1}+ \mathcal{Y}_\nu ~d\mathcal{T}\cdot \mathcal{T}^{-1}  \mathcal{Y}_\nu^{-1}=
\\
\noalign{\medskip}
= d_\lambda\widehat{\mathcal{Y}}_\nu \cdot\widehat{\mathcal{Y}}_\nu^{-1}+\widehat{\mathcal{Y}}_\nu \sum_{m=1}^\infty \frac{d(\mathfrak{D}^m)}{m!}(\ln z)^m 
  ~z^{-\mathfrak{D}}\widehat{\mathcal{Y}}_\nu^{-1} 
  +
  \widehat{\mathcal{Y}}_\nu  z^{\mathfrak{D}} \sum_{k=1}^{\overline{k}} \frac{d(\mathfrak{N}^k)}{k!}(\ln z)^k 
  ~z^{-\mathfrak{N}}z^{-\mathfrak{D}}\widehat{\mathcal{Y}}_\nu^{-1}
  +
\\
 \noalign{\medskip}
 + z \widehat{\mathcal{Y}}_\nu 
 z^{\mathfrak{D}}
  z^{\mathfrak{N}}
 d\Lambda 
 z^{-\mathfrak{N}} 
 z^{-\mathfrak{D}}\widehat{\mathcal{Y}}_\nu^{-1} 
  +
  \widehat{\mathcal{Y}}_\nu z^{\mathfrak{D}} z^{\mathfrak{N}} e^{\Lambda z} 
  d\mathcal{T}\cdot \mathcal{T}^{-1}   e^{-\Lambda z}  z^{-\mathfrak{N}} 
  z^{-\mathfrak{D}}\widehat{\mathcal{Y}}_\nu^{-1} 
\\
\noalign{\medskip}
= \underbrace{d_\lambda\widehat{\mathcal{Y}}_\nu \cdot\widehat{\mathcal{Y}}_\nu^{-1}}_{O(1/z)}+\widehat{\mathcal{Y}}_\nu \sum_{m=1}^\infty \frac{d(\mathfrak{D}^m)}{m!}(\ln z)^m 
  ~z^{-\mathfrak{D}}\widehat{\mathcal{Y}}_\nu^{-1} 
  +
  \widehat{\mathcal{Y}}_\nu  z^{\mathfrak{D}} \sum_{k=1}^{\overline{k}} \frac{d(\mathfrak{N}^k)}{k!}(\ln z)^k 
  ~z^{-\mathfrak{N}}z^{-\mathfrak{D}}\widehat{\mathcal{Y}}_\nu^{-1}
  +
\\
 \noalign{\medskip}
 + z \widehat{\mathcal{Y}}_\nu 
 d\Lambda 
 \widehat{\mathcal{Y}}_\nu^{-1} 
  +
  \widehat{\mathcal{Y}}_\nu z^{\mathfrak{D}} z^{\mathfrak{N}} 
  d\mathcal{T}\cdot \mathcal{T}^{-1}    z^{-\mathfrak{N}} 
  z^{-\mathfrak{D}}\widehat{\mathcal{Y}}_\nu^{-1}.
  \end{aligned}
\ee
In the last step, we have used the fact that $  e^{\Lambda z}$  commutes with 
$ d\mathcal{T}\cdot \mathcal{T}^{-1}$,  $\mathfrak{D}$ and $\mathfrak{N}$, due to the block structure. The absence of logarithmic singularities in $\sum_{j=1}^s (z E_{p_j} + \omega_j(\lambda))d\lambda_j+ d\mathcal{T}\cdot \mathcal{T}^{-1}$ requires that 
\be
\label{13giugno2022-2} 
d\mathfrak{D}=d\Bigl(\mathcal{T}(\lambda) (D+\Sigma) \mathcal{T}(\lambda)^{-1}\Bigr)=0 ,
\quad\quad 
d\mathfrak{N}=d\Bigl(\mathcal{T}(\lambda) N \mathcal{T}(\lambda)^{-1}\Bigr)=0 
 \ee
 Since $d(D+\Sigma) =dN=0$ for the strong isomonodromy deformation, the above conditions are satisfied if and only if \eqref{1maggio2021-1} and \eqref{1maggio2021-2} respectively hold. 
 In this way, also the last term in \eqref{1maggio2021-3}, namely 
 $$ \widehat{\mathcal{Y}}_\nu z^{\mathfrak{D}} z^{\mathfrak{N}} 
  d\mathcal{T}\cdot \mathcal{T}^{-1}    z^{-\mathfrak{N}} 
  z^{-\mathfrak{D}}\widehat{\mathcal{Y}}_\nu^{-1} \equiv \widehat{\mathcal{Y}}_\nu z^{D+\Sigma} z^{N} 
 \mathcal{T}^{-1} d\mathcal{T}   z^{-N} 
  z^{-D-\Sigma}\widehat{\mathcal{Y}}_\nu^{-1}, 
  $$
 does not contain logarithmic singularities when \eqref{1maggio2021-1}-\eqref{1maggio2021-2} hold, because it reduces  to 
 $$ 
 \widehat{\mathcal{Y}}_\nu d\mathcal{T}\cdot \mathcal{T}^{-1} \widehat{\mathcal{Y}}_\nu^{-1}=
 d\mathcal{T}\cdot \mathcal{T}^{-1}+O(1/z).
 $$
 Notice that the above behaviour is in agreement  with  $\sum_{j=1}^s (z E_{p_j} + \omega_j(\lambda))d\lambda_j+ d\mathcal{T}\cdot \mathcal{T}^{-1}$. 
 
 Notice that if $J$ is diagonal, $\mathcal{T}(\lambda) (D+\Sigma) \mathcal{T}(\lambda)^{-1}=\mathcal{T}(\lambda) ~J~ \mathcal{T}(\lambda)^{-1}$ in the first equation in  \eqref{13giugno2022-2}  
 \vskip 0.2 cm 

 {\bf Step 2.} 
 The relations \eqref{1maggio2021-1}-\eqref{1maggio2021-2} can be written for the individual 
blocks  inherited from  $L=L_1\oplus\cdots \oplus L_s$  and $\mathcal{T}= \mathcal{T}_1\oplus\cdots \oplus \mathcal{T}_s$.  It suffices to  consider a single block with label $k$. 

\begin{figure}
\centerline{\includegraphics[width=0.8\textwidth]{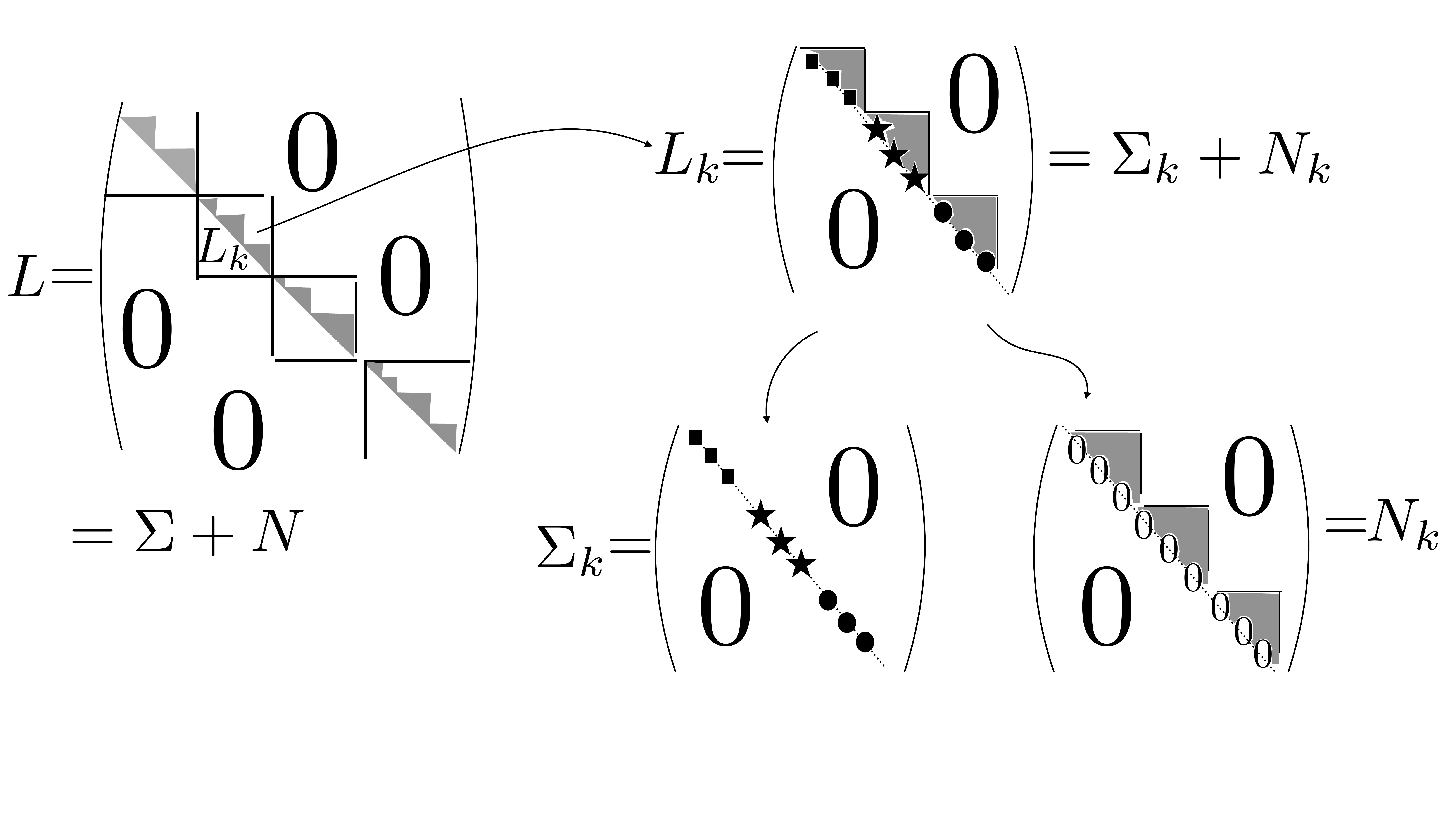}}
\caption{Structure of the blocks $L_k$ of $L$. Here, $\Sigma_k$ is diagonal and $N_k$ is nilpotent}
\label{2maggio2021-2}
\end{figure}

As already explained, up to a $Y_\nu \longmapsto Y_\nu P$ given by a suitable permutation matrix $P$, we  assume that for each block $L_k$, $k=1,...,s$, the Levelt structure explained in Section \ref{2maggio2021-1} applies. See figure \ref{2maggio2021-2}.

\begin{figure}
\centerline{\includegraphics[width=0.8\textwidth]{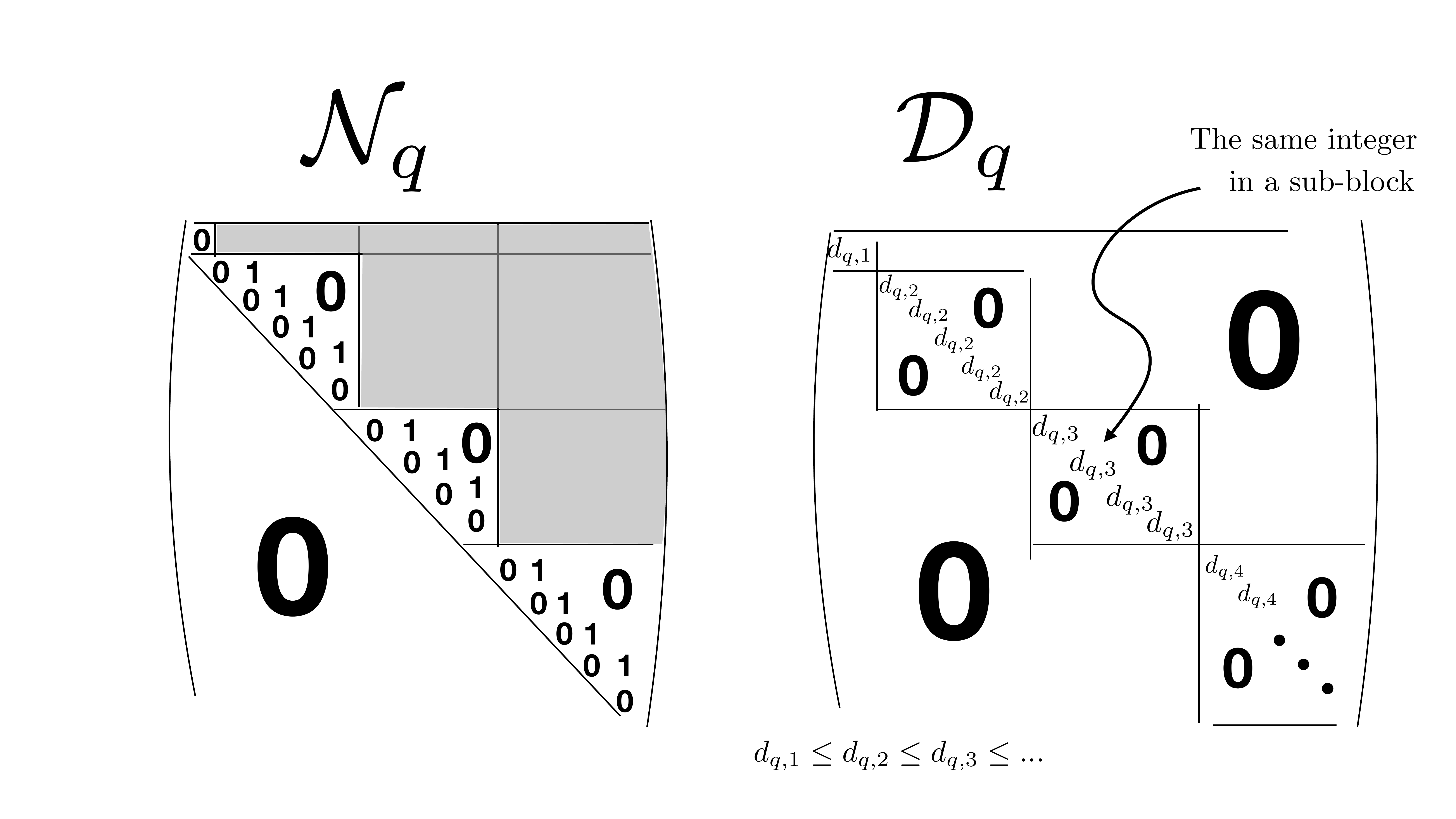}}
\caption{The structure of the sub-blocks $\mathcal{N}_q$ and $\mathcal{D}_q$ of $\mathcal{N}$ and $\mathcal{D}$ (that is, of a certain  $N_k$ and $D_k$), corresponding to $\Sigma_q=\sigma_q I_q$.}
\label{2maggio2021-4}
\end{figure}

 We consider the problem at the level of a single block  with label $k$. In order to avoid a proliferation of indices, from now on $\mathcal{J}$, $\mathcal{L}$, $\mathcal{R}$, $\mathcal{S}$, $\mathcal{D}$, $\Sigma$ and $\mathcal{N}$ will respectively stand for  $J_k$,  $L_k$, $R_k$, $S_k$, $D_k$, $\Sigma_k$ and $N_k$. We will take the label $k$ only for $\mathcal{T}_k$, in order not to confuse it with the full $\mathcal{T}$. 
 To them, the structures of  Section \ref{2maggio2021-1} apply.  Now,  we have $\mathcal{L}=\mathcal{L}_1\oplus\cdots\oplus \mathcal{L}_\ell$ for some $\ell$,  and 
$$ 
\Sigma= \sigma_1 I_1\oplus\cdots  \oplus  \sigma_\ell I_\ell, 
$$
with  eigenvalues $\sigma_q $ (with real part in $[0,1)$). 
Hence, since $ \mathcal{D}$ is diagonal,  \eqref{1maggio2021-1}  for the block $k$ of $\mathcal{T}^{-1} d \mathcal{T}$ gives 
$$
 [\mathcal{T}_k^{-1} d \mathcal{T}_k, \mathcal{D}+\Sigma]=0\quad \Longrightarrow \quad \mathcal{T}_k=
 \mathcal{T}_1^{(k)} \oplus\cdots  \oplus  \mathcal{T}_\ell^{(k)}.
  $$
  Now, $\mathcal{D}=\mathcal{D}_1\oplus\cdots\oplus \mathcal{D}_\ell$, and $\mathcal{N}=\mathcal{N}_1\oplus\cdots\oplus \mathcal{N}_\ell$. 
  Notice that 
  \be
\label{2maggio2021-5}
 [(\mathcal{T}_q^{(k)})^{-1} d \mathcal{T}_q^{(k)}, \mathcal{D}_q+\Sigma_q]=0.
  \ee
  Each $\mathcal{N}_q$ ($q=1,...,\ell$) is upper triangular, it has zeros on the diagonal, and its diagonal blocks are elementary Jordan sub-blocks with 1's on the second upper diagonal, as in figure \ref{2maggio2021-4}. Accordingly,  $\mathcal{D}_q$, which is diagonal  with a non decreasing sequence of integer eigenvalues, has sub-blocks with  the same eigenvalue  corresponding to a Jordan sub-block in $\mathcal{N}_q$,  as in figure \ref{2maggio2021-4}.

\begin{figure}
\centerline{\includegraphics[width=0.8\textwidth]{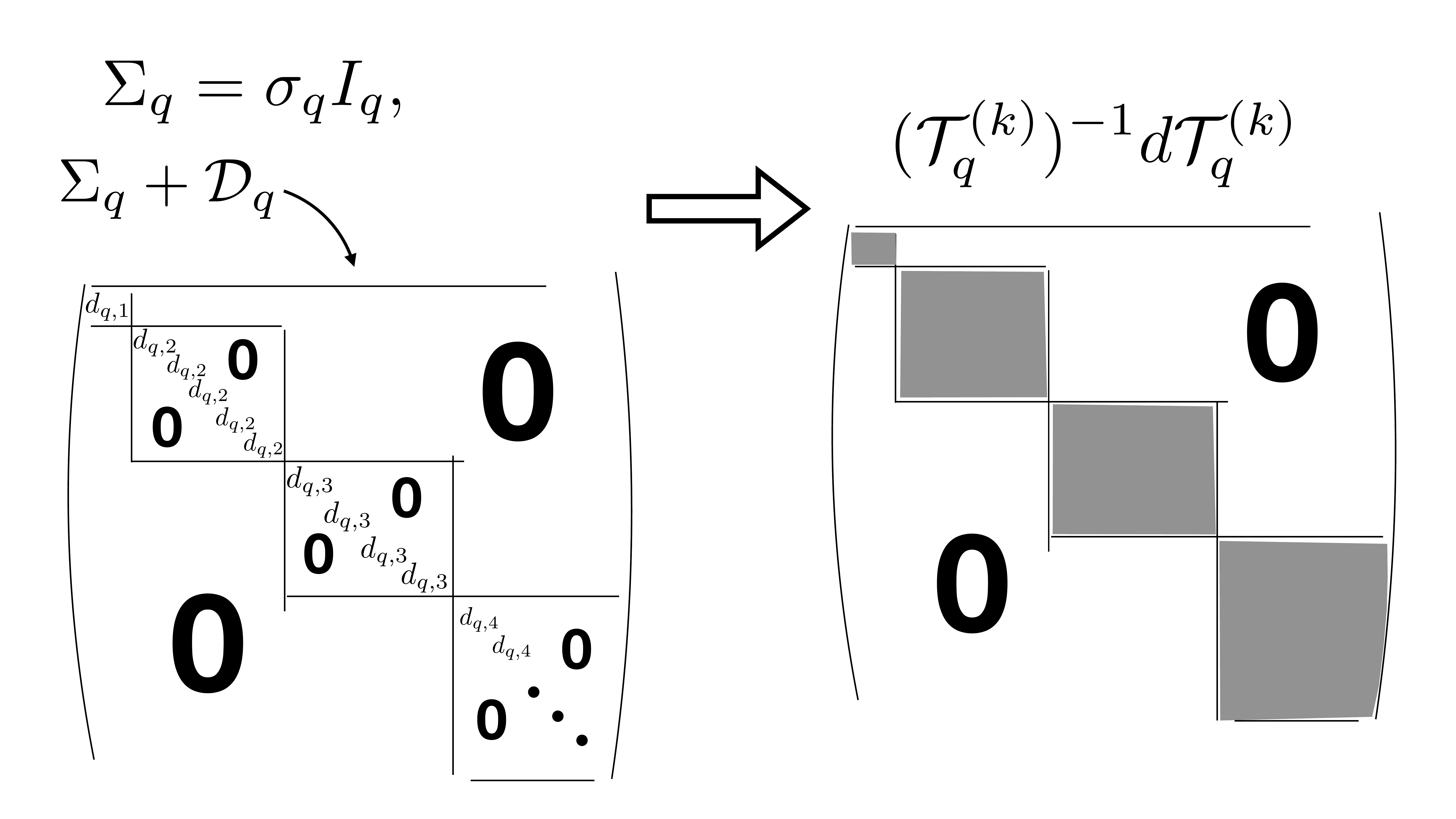}}
\caption{The sub-blocks-structure of the block $(\mathcal{T}_q^{(k)})^{-1} d \mathcal{T}_q^{(k)}$.}
\label{2maggio2021-6}
\end{figure}

  The above facts and \eqref{2maggio2021-5} imply that 
  $(\mathcal{T}_q^{(k)})^{-1} d \mathcal{T}_q^{(k)}$ is divided into sub-blocks as $\Sigma_q+\mathcal{D}_q$, where the only non-zero sub-blocks are the block-diagonal part, as in Figure \ref{2maggio2021-6} .

\noindent 
If follows that $[\mathcal{T}_k(\lambda)^{-1} d \mathcal{T}_k(\lambda),\mathcal{N}]=0$ (from  \eqref{1maggio2021-2}) reduces to 
\be
\label{2maggio2021-7}
[(\mathcal{T}_q^{(k)})^{-1} d \mathcal{T}_q^{(k)},\mathcal{N}_q]=0,
 \ee
Let $\mathcal{M}_q:=(\mathcal{T}_q^{(k)})^{-1} d \mathcal{T}_q^{(k)}$, and let $\mathcal{M}^{(q)}_{[a,b]}=\mathcal{M}_{[a,a]}^{(q)}\delta_{ab}$  be a sub-block. From  \eqref{2maggio2021-7} we receive
$$ 
\mathcal{M}_{[a,a]}^{(q)} \mathcal{N}_{[a,b]}^{(q)}-  \mathcal{N}_{[a,b]}^{(q)} \mathcal{M}_{[b,b]}^{(q)} =0.
$$ 
In particular
 $$ 
\mathcal{M}_{[a,a]}^{(q)} \mathcal{N}_{[a,a]}^{(q)}-  \mathcal{N}_{[a,a]}^{(q)} \mathcal{M}_{[a,a]}^{(q)} =0,
$$ 
which means that  $\mathcal{M}_q$ commutes with the block-diagonal matrix $ \mathcal{N}_{[1,1]}^{(q)}\oplus  \mathcal{N}_{[2,2]}^{(q)}\oplus \cdots \oplus  \mathcal{N}_{[\ell,\ell]}^{(q)}$ ($=$ Jordan  matrix obtained from the diagonal sub-blocks of $\mathcal{N}_q$  in the left part of figure \ref{2maggio2021-4}). 

Now, observe that $\mathcal{D}_q+\Sigma_q+ (\mathcal{N}_{[1,1]}^{(q)}\oplus  \mathcal{N}_{[2,2]}^{(q)}\oplus \cdots)=\mathcal{J}_q$. Since  $\mathcal{M}_q$ also commutes with $\mathcal{D}_q+\Sigma_q$, we conclude that it commutes with $\mathcal{J}_q$, namely 
$$ 
[(\mathcal{T}_q^{(k)})^{-1} d \mathcal{T}_q^{(k)},\mathcal{J}_q]=0.
$$
Therefore,
$$ 
[(\mathcal{T}_k)^{-1} d \mathcal{T}_k,\mathcal{J}]=0.
$$
Coming back to the original notations, the above is $ [(\mathcal{T}_k)^{-1} d \mathcal{T}_k,J_k]=0$. This is what we wanted to prove. 
$\Box$ 

\begin{small}

\end{small}
\end{document}